\documentclass[%
  reprint,
  superscriptaddress,
  amsmath,amssymb,
  aps,
  prb,
]{revtex4-2}

\usepackage{silence}
\allowdisplaybreaks

\usepackage{graphicx}
\graphicspath{{./}{figure}{figure/sdos}{figure/sdos-Zeeman}{figure/spin-sus}{figure/spin-sus-site}}
\usepackage{url}
\usepackage[dvipsnames]{xcolor}
\usepackage{tikz}
\usepackage[]{subcaption}
\usepackage{hyperref}
\hypersetup{
	setpagesize=false,
	bookmarksnumbered=true,%
	bookmarksopen=true,%
	colorlinks=true,%
	linkcolor=magenta,
	citecolor=magenta,
}

\usepackage{bbm,mathrsfs}
\usepackage{makecell}
\usepackage{dcolumn}
\usepackage{multirow}
\usepackage{booktabs}
\usepackage{braket}
\usepackage[capitalise]{cleveref}
\crefname{section}{Sect.}{Sect.}

\usepackage[version=4]{mhchem}
\usepackage{physics}
\usepackage{siunitx}
\usepackage{ulem}

\usepackage[whole]{bxcjkjatype}
\input{unicode-math-pdflatex.sty}

\newcommand{\rocomment}[1]{#1}

\usepackage{natbib}

\begin{document}
\preprint{APS/123-QED}

\title{
  Anisotropic paramagnetic response of topological Majorana surface states in the superconductor \texorpdfstring{\ce{UTe2}}{UTe2}
}
\author{Ryoi Ohashi}
\affiliation{Department of Materials Engineering Science, Osaka University, Toyonaka, Osaka 560-8531, Japan}
\author{Jushin Tei}
\affiliation{Department of Materials Engineering Science, Osaka University, Toyonaka, Osaka 560-8531, Japan}
\author{Yukio Tanaka}
\affiliation{Department of Applied Physics, Nagoya University, Nagoya 464-8603, Japan}
\affiliation{Research Center for Crystalline Materials Engineering, Nagoya University, Nagoya 464-8603, Japan}
\author{Takeshi Mizushima}
\affiliation{Department of Materials Engineering Science, Osaka University, Toyonaka, Osaka 560-8531, Japan}
\author{Satoshi Fujimoto}
\affiliation{Department of Materials Engineering Science, Osaka University, Toyonaka, Osaka 560-8531, Japan}
\affiliation{Center for Quantum Information and Quantum Biology, Osaka University, Toyonaka 560-8531, Japan}
\date{\today}

\begin{abstract}
  Identifying the superconducting gap symmetry and topological signatures in the putative spin-triplet superconductor \ce{UTe2} is an important issue. 
  Especially, a smoking-gun detection scheme for Majorana surface states hallmarking topological superconductivity in \ce{UTe2} is still lacking.
  In this study, we examine the surface spin susceptibility of \ce{UTe2} with a particular focus on the contribution of the surface states. 
  We find that Majorana surface states contribute significantly to the surface spin susceptibility, and give rise to an Ising-like anisotropy and anomalous enhancement in the surface spin susceptibility.
  We calculate the surface spin susceptibility as well as the local density of states using the recursive Green's function method and examine the anisotropy of the surface spin susceptibility in terms of the topological surface states and symmetry for all irreducible representations of odd-parity pairing states. 
  Our results indicate that the Ising anisotropy and the anomalous enhancement are attributed to the Majorana surface state protected by the crystalline symmetry. 
  These findings suggest the possibility of detecting the Majorana surface state via magnetic measurements.
\end{abstract}
\maketitle

\section{Introduction}

\ce{UTe2} is a heavy fermion superconductor that has garnered significant attention in the condensed matter physics community due to its unique and unconventional superconducting properties~\cite{ran_2019,aoki_2022_rev}. 
This material has a superconducting transition temperature ($T_\mathrm{c}$) between 1.6 and \SI{2.1}{K}~\cite{sakai_2022_Single} and is strongly anticipated to exhibit spin-triplet and topological superconductivity.

One of the most remarkable features of \ce{UTe2} is its distinct behavior 
under a magnetic field. 
Recent studies have employed a variety of experimental probes, including thermal, electrical, and magnetic measurements, to elucidate further the complex phase diagram of \ce{UTe2} under high magnetic fields~\cite{ran_2019a,rousel_2023,braithwaite_2019_Multiple,aoki_2020,thomas_2020_Evidence,knebel_2020,lin_2020_Tuning,ran_2020_Enhancement,aoki_2021_FieldInduced, lewin_2024_pre}.
These investigations have revealed the existence of a field-induced state with low or vanishing electrical resistance and provided the definitive evidence for bulk field-induced superconductivity.
This unique case among superconductors provides an opportunity to develop the theory of the mechanism of unconventional pairing~\cite{ishizuka_2021_Periodic,shishidou_2021,tei_2024_Pairing,hakuno_2024_Magnetism} and enhance our understanding of the conditions that allow for the emergence of spin-triplet superconductivity.

Determining the symmetry of the gap function still remains challenging, even at low magnetic fields and ambient pressure. 
Previous studies have reported the spontaneous breaking time-reversal symmetry (TRS) from scanning tunneling microscopy (STM) measurements~\cite{jiao_2020_Chiral} and Kerr effect measurements~\cite{hayes_2021_Multicomponent}.
However, recent Kerr effect measurements on high-quality samples have shown no evidence of spontaneous Kerr signals~\cite{ajeesh_2023}, leaving no support for the TRS breaking.
The identification of the irreducible representations (IRs) of the superconducting order parameter in \ce{UTe2} is crucial for understanding the underlying pairing mechanism. 
The IRs contain the superconducting orders with a full gap and point nodes, which may be probed experimentally through thermal and magnetic measurements.
The specific heat measurement in a high-quality sample displaying an optimal transition temperature at \SI{2}{K} exhibits a single transition and small residual heat capacity~\cite{rosa_2022_Single}.
Specific heat and magnetic penetration depth measurements support the existence of point nodes, indicating a chiral $B_{3u}+iA_{u}$ non-unitary state with broken TRS~\cite{ishihara_2023a}.
There is conflicting evidence about thermal conductivity measurements. 
While one group supports a fully gapped $A_u$ state~\cite{suetsugu_2024_Fully}, another group observes evidence for the existence of point nodes~\cite{hayes_2024}.
Furthermore, nuclear magnetic resonance (NMR) measurements~\cite{matsumura_2023} show a decrease in the Knight shift for all directions, indicating compatibility with a fully gapped $A_{u}$ pairing state.
The recent remarkable improvement in sample quality~\cite{sakai_2022_Single} provides promising prospects for further experimental studies.

In addition to unconventional superconductivity in the bulk, the existence of Majorana surface states (MSSs) in \ce{UTe2} has been predicted in previous studies~\cite{ishizuka_2019,tei_2023}.
\ce{UTe2} is a strong candidate for a spin-triplet superconductor and may harbor MSSs which is protected by the topological invariant defined by bulk Hamiltonian~\cite{sato_2010,Sato2011}. 
However, detecting these states is an urgent issue that needs to be clarified theoretically.
It is known that MSSs are special type of surface Andreev bound states (ABS)~\cite{tanaka_2024_progress}. 
From the study of unconventional superconductor junctions, it is clarified that so-called zero-bias conductance peak (ZBCP) appears in the tunneling spectroscopy~\cite{TK95,Kashiwaya00} in the presence zero energy surface ABSs 
\cite{Hu94,Sato2011} protected by 
topological invariant. 
Actually, there are many experimental reports of ZBCP in high $T_{c}$ cuprate junctions, where spin-singlet $d$-wave pairing is realized~\cite{Experiment1,Experiment2,Experiment3,Experiment4,Experiment5,Experiment6,Experiment7,Iguchi2000,Deutscher}. 
ZBCPs  have been observed also in other unconvnetional superconductors like \ce{CeCoIn5} \cite{Rourke}, \ce{PtCoGa5} \cite{Daghero2012}, \ce{Sr2RuO4} \cite{Laube2000,Kashiwaya2011}  and \ce{UBe13} \cite{Ott2000}.  
It is also noted that if the MSSs have a dispersion and the contribution of Majorana zero mode to tunneling conductance is not large ZBCP does not appear~\cite{Yamakage2012}. 
As regards tunneling experiment of \ce{UTe2}, there are several experimental reports. 
ZBCPs have not been observed in STM measurements on cleavage surfaces of a sample~\cite{gu_2023}.
On the other hand,  ZBCP has been reported in the experiment of point contact~\cite{yoon_2024_probing}.  
Thus, it is very timely to study MSSs of \ce{UTe2} and its influence on the various physical quantities. 

Motivated by these backgrounds, in this paper, we study the surface spin susceptibility of \ce{UTe2} with a particular focus on the contribution of the MSSs. 
We begin by presenting the local density of states (LDOS) of the MSSs and explore the possibility of detecting their signatures from STM measurements. 
We will also calculate the properties of the local spin susceptibility (LSS) and attempt to study it with NMR measurements in mind, thereby proposing a multifaceted method of MSS observation.
In addition, we organize the Majorana Ising anisotropy and the paramagnetic response in terms of discrete symmetry.
By summarizing these features, we clarify an anisotropic magnetic response depending on the pairing symmetry and also suggest a valuable scheme for determining the symmetry of the gap function.

This paper is organized as follows.
In \cref{sec:model_and_symmetry}, we formulate the theoretical model of superconductivity in \ce{UTe2} and briefly describe the symmetry that the model holds.
In \cref{sec:LDOS}, we numerically calculate LDOS as the first observable and characterize the MSSs.
In \cref{sec:local_spin_susceptibility}, we numerically calculate LSS as a second observable to reveal the magnetic anisotropy of the MSSs in \ce{UTe2}. 
We also explain how the anisotropy is related to the crystalline symmetry.
Lastly, in \cref{sec: summary_discussion}, we summarize our results and discuss the possibility of signal enhancement by attaching a diffusive metal with the superconductor \ce{UTe2}. 
\cref{app:sdos_zeeman} summarizes the relation between the gap-out of the MSS and the orientation of an external magnetic field. The numerical method for the recursive Green's function is described in \cref{app:sdos}.

\section{Model and Symmetry}
\label{sec:model_and_symmetry}

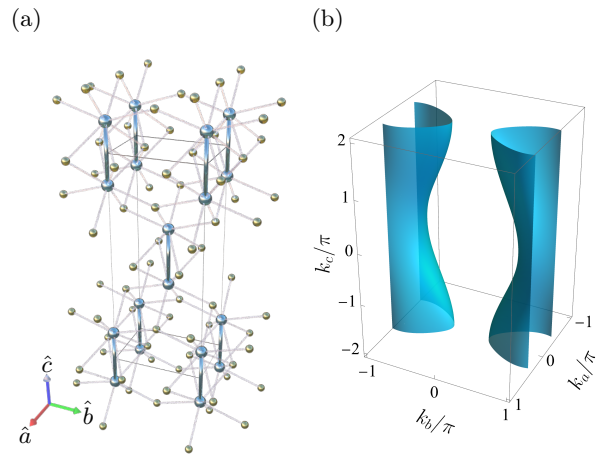
\begin{figure}[htbp]
	\centering
  \begin{minipage}[t]{0.45\linewidth}
    \subcaption{}\vspace{-5mm}
    \label{sfig: crys_str_UTe2_unit}
    \begin{tikzpicture}
      \node[anchor=south west,inner sep=0] (image) at (0,0) {\includegraphics[width=\linewidth]{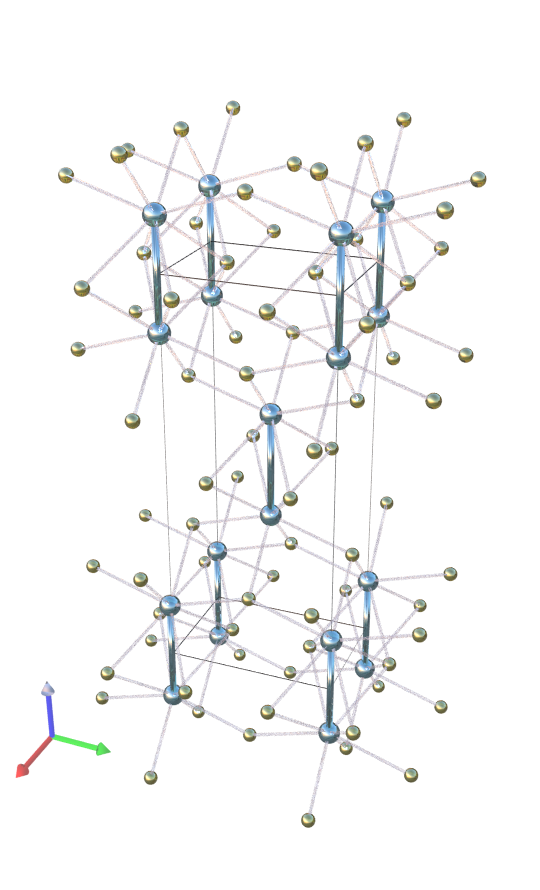}};
      \node [anchor=west] (note) at (-0.15,0.7) {$\hat{a}$};
      \node [anchor=west] (note) at (0.7,1.0) {$\hat{b}$};
      \node [anchor=west] (note) at (0.15,1.65) {$\hat{c}$};
    \end{tikzpicture}%
  \end{minipage}
  \begin{minipage}[t]{0.45\linewidth}
    \subcaption{}\vspace{5mm}
    \label{sfig: fermi_surface}
    \includegraphics[width=\linewidth]{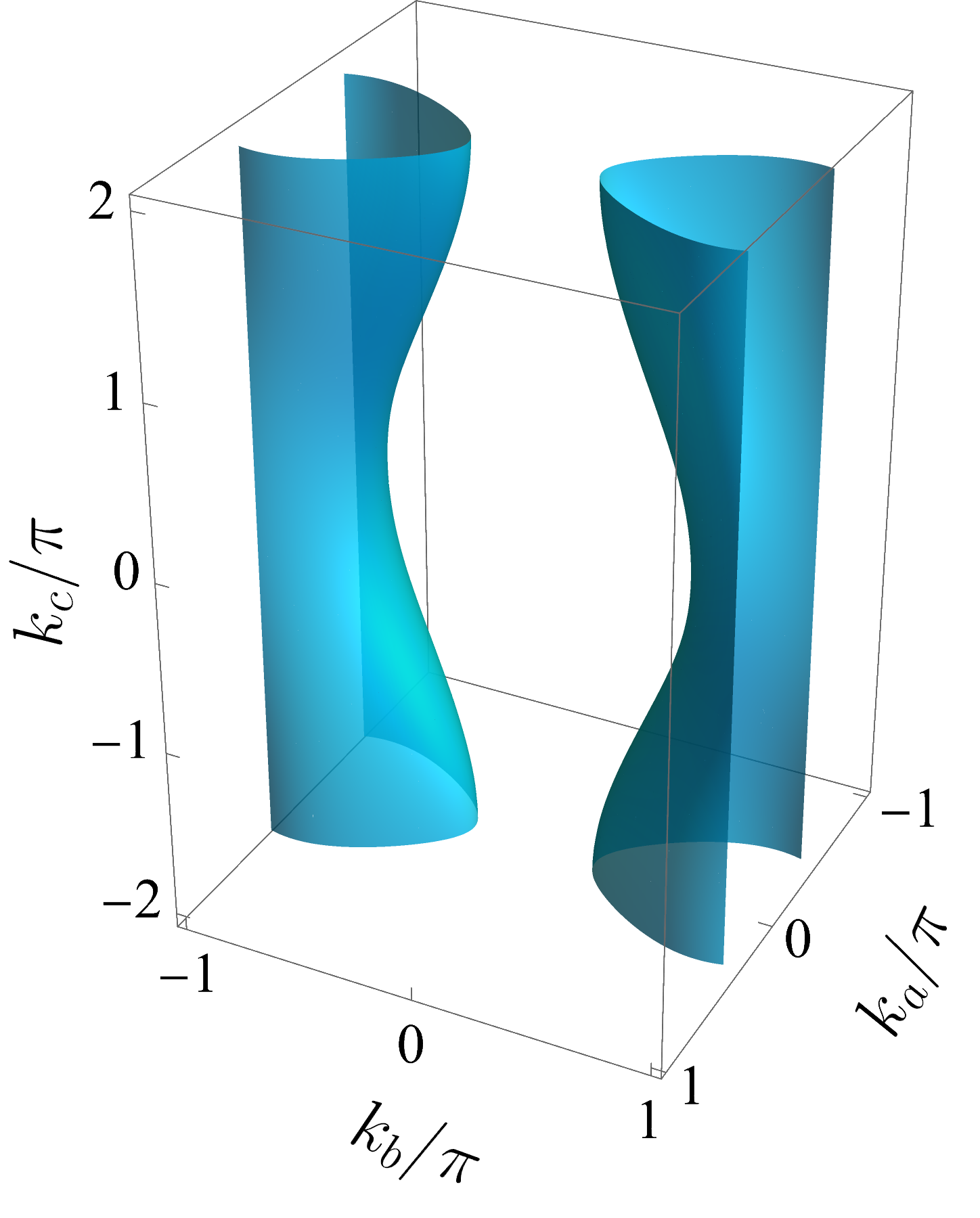}
  \end{minipage}
  \caption{
    (a) Crystal structure of \ce{UTe2}. 
    A blue (yellow) sphere describes the \ce{U} (\ce{Te}) atom.
    (b) Cylindrical electron Fermi surface of \ce{UTe2}.
  }
  \label{fig: crys_str_UTe2}
\end{figure}

First of all, in the numerical calculations, we construct the Hamiltonian that takes into account the symmetry of \ce{UTe2}.
Since \ce{UTe2} has a body-centred orthorhombic lattice structure and the space group symmetry is \rocomment{$Immm(\#71, D^{25}_{2h})$} (show \cref{sfig: crys_str_UTe2_unit}), the associated point group is $D_{2h}$ and has eight one-dimensional (1D) irreducible representations;
the possible IRs for the $\vb*{d}$-vector are $A_{u}$, $B_{1u}$, $B_{2u}$ and $B_{3u}$.

\subsection{Normal state}
\label{ssec:normal_state}
In this paper, the system is assumed to have TRS.
We consider the two-orbital $f$-electron Hamiltonian in the normal state~\cite{shishidou_2021,tei_2023} as
\begin{align}
  H_\mathrm{N}(\vb*{k}) = ϵ_{0}(\vb*{k}) - μ + f_{x}(\vb*{k})τ_{x}+f_{y}(\vb*{k})τ_{y}
  \label{eq:normal_hamiltonian}
\end{align}
with
\begin{subequations}
\begin{align}
  ϵ_{0}(\vb*{k}) &= 2t_{1}\cos k_{a} +2t_{2}\cos k_{b},\\
  f_{x}(\vb*{k})&=m_{0}+t_{3}\cos(k_{a}/2)\cos(k_{b}/2)\cos(k_{c}/2),\\
  f_{y}(\vb*{k})&=t_{4}\cos(k_{a}/2)\cos(k_{b}/2)\sin(k_{c}/2),
\end{align}
\end{subequations}
where $τ_{\mu}$ are the Pauli matrices for orbital degrees of freedom.
\rocomment{
  First-principles calculations \cite{xu_2019_Quasi, ishizuka_2021_Periodic} indicate that the cylindrical-hole Fermi surface, composed of \ce{U} site $d$-electrons and \ce{Te} site $p$-electrons, and the cylindrical-electron Fermi surface, composed of \ce{U} site $f$-electrons, are the Fermi surfaces observed in the de Haas–van Alphen experiments~\cite{aoki_2022}.
  In this paper, only the cylindrical-electron Fermi surface (see \cref{sfig: fermi_surface}) is included in the calculations, based on the assumption that the strong Coulomb repulsion in $f$-electrons induces magnetic fluctuations, which mediate spin-triplet odd-parity pairing.
}
We choose the parameters as follows \cite{shishidou_2021,tei_2023}:
$μ = -1.8$, $t₁ = -0.5$, $t₂ = 0.375$, $t₃ = 0.65$, $t₄ = -0.65$ and $m₀ = -0.7$.

\subsection{Superconducting state}
\label{ssec:superconducting_state}
The superconducting phases are classified by the crystalline point group symmetry~\cite{sigrist_1991}.
The Bogoliubov-de Gennes (BdG) Hamiltonian $\mathcal{H}$ in the superconducting state of \ce{UTe2} is described as
\begin{align}
  \mathcal{H} = 
    \frac{1}{2}∑_{\vb*{k},σ,σ^{\prime}}\hat{c}^{†}_{\vb*{k},σ}H_\mathrm{BdG}(\vb*{k})\hat{c}_{\vb*{k},σ^{\prime}}
\end{align}
with
\begin{align}
  H_\mathrm{BdG}(\vb*{k}) &= 
    \mqty(
      H_\mathrm{N}(\vb*{k}) & Δ(\vb*{k}) \\
      Δ^{†}(\vb*{k}) & -\qty(H_\mathrm{N}(-\vb*{k}))^\mathrm{t}
    )\qc
  \hat{c}_{\vb*{k},σ} = \mqty(c_{\vb*{k},σ} \\ c^{†}_{-\vb*{k},σ}),
  \label{eq:Hbdg}
\end{align}
where $c_{\vb*{k},σ}$ and $c_{\vb*{k},σ}^{†}$ are the annihilation and creation operator of the electron with momentum $\vb*{k}$, $σ$ labels spin and orbital degrees of freedom, and $Δ(\vb*{k})$ is the gap function, respectively.
In this paper, we only consider the inter-orbital pairing states which consist of orbital-triplet pairings~\cite{tei_2023},
\begin{align}
  Δ_\mathrm{IR}(\vb*{k}) = 
  \mqty(
    0 & Δ_\mathrm{IR}^\mathrm{inter}(\vb*{k})\\
    Δ_\mathrm{IR}^\mathrm{inter}(\vb*{k}) & 0
  ),
\end{align}
\begin{align}
  Δ_\mathrm{IR}^\mathrm{inter}(\vb*{k}) = i\vb*{d}_\mathrm{IR}\cdot\vb*{σ}σ_{y},
\end{align}
where $σ_{μ}$ are the Pauli matrices for spin degrees of freedom, and the $\vb*{d}$-vector is expressed for each IR as follows
\begin{subequations}
\begin{align}
  \vb*{d}_{A_{u}}(\vb*{k})  &=
    Δ_{0}\mqty(
      \sin k_{a}\\
      \sin k_{b}\\
      \sin k_{c}
    ), \label{eq:d-vector_IR_Au}\\
  \vb*{d}_{B_{1u}}(\vb*{k}) &= 
    Δ_{0}\mqty(
      \sin k_{b}\\
      \sin k_{a}\\
      0
    ), \label{eq:d-vector_IR_B1u}\\
  \vb*{d}_{B_{2u}}(\vb*{k}) &= 
    Δ_{0}\mqty(
      \sin k_{c}\\
      0\\
      \sin k_{a}
    ),\label{eq:d-vector_IR_B2u} \\
  \vb*{d}_{B_{3u}}(\vb*{k}) &= 
    Δ_{0}\mqty(
      0\\
      \sin k_{c}\\
      \sin k_{b}
    ).\label{eq:d-vector_IR_B3u}
\end{align}
\label{eq:d-vector_IR}
\end{subequations}
It has been predicted that including intraorbital pairing does not alter \rocomment{the  qualitative features of the results, such as the topological protection of the surface ABS~\cite{tei_2023}.}
In this paper, we choose the gap amplitude as $Δ₀ = 0.1$.

\rocomment{
Here we need to comment on the validity of our model.
Firstly, we have not considered the spin-orbit coupling (SOC) in this study.
In Ref.~\cite{shishidou_2021}, it is pointed out that \ce{UTe2} has staggered Rashba-type SOC due to local inversion symmetry breaking.
In general, with the inclusion of staggered Rashba-type SOC, there may be some mixing with spin-singlet pairing. 
However, unless the pairing interaction in the spin-singlet channel is comparable to the pairing interaction in the spin-triplet channel, the magnitude of mixing is suppressed by $E_\mathrm{SOC}/E_\mathrm{F}$ ($E_\mathrm{SOC}$ and $E_\mathrm{F}$ are, respectively, the energy scale of the spin-orbit coupling, and the Fermi energy) \cite{fujimoto_2007_Electron}.
Therefore, there is no qualitative effect on the results.
}
\rocomment{
Besides, we have to comment that the tight-binding model has some limitations for quantitative discussions because of the complexity of the electronic structure, particularly, near surfaces~\cite{sarma_2015_Substrateinduced, csire_2018_Relativistic, nyari_2023_Topological}.
However, these points neglected in our tight-binding model do not affect qualitatively the results shown below. Thus, this simplified model is suitable for our purpose of elucidating distinct qualitative features of magnetic responses of MSS.
}

\subsection{Symmetry}
\label{sec:symmetry}

The BdG Hamiltonian preserves TRS and particle-hole symmetry (PHS),
\begin{align}
  Θ H_\mathrm{BdG}(\vb*{k})Θ^{-1} &= H_\mathrm{BdG}(-\vb*{k}),\\
  C H_\mathrm{BdG}(\vb*{k})C^{-1} &= -H_\mathrm{BdG}(-\vb*{k}).
\end{align}
Using a chiral symmetry operator $Γ = iΘC$, one can introduce the topological invariant, i.e.,  the three-dimensional (3D) winding number. 
However, the 3D winding number vanishes when the Fermi surface is a cylindrical shape.
Therefore, we discuss chiral symmetry that takes into account crystalline symmetry.
For the point-group symmetry $D_{2h}$ of \ce{UTe2}, two types of crystalline symmetries exist: a $\pi$ rotation symmetry $\mathcal{C}_{\mu}$ and a mirror symmetry $\mathcal{M}_{\mu\nu}$ with $\mu,\nu=a,b,c$~\cite{tei_2023}.
From these crystalline symmetries, we can introduce chiral symmetry as follows:
\begin{align}
  Γ_{U} = e^{iϕ_{U}}\hat{U}Γ,
  \label{eq:gamma}
\end{align}
where $U = \mathcal{C}_{μ} \; \text{or} \; \mathcal{M}_{μν}$, $e^{iϕ_{U}}$ is a phase factor which ensures $Γ_{U}² = 1$\cite{Yamakage2017}. 
The unitary matrix, $\hat{U}$, is defined in the particle-hole space as
\begin{align}
  \hat{U} = \mqty(\dmat{U, sU^{\ast}} ),
\end{align}
where $s=±1$ is the sign produced when the gap function is unitarily transformed by $U$.

\cref{tb: Majorana_states_and_chiral_operator} summarizes the MSSs and the chiral symmetries that topologically protect the MSSs on each surface.
Recently, the existence and the dispersion of these surface Majorana states have been confirmed through the diagonalization of the lattice Hamiltonian~\cite{tei_2023}. 
The chiral symmetry $Γ_{U}$ can be used to define the 1D crystal winding number $w_{U}$\cite{sato11,mizushima_2012,shi14}. 
This may take nontrivial values when the Brillouin zone of the 1D subspace crosses a Fermi surface and is in one-to-one correspondence with the surface Majorana zero modes protected by the crystal symmetry.

In the next section of this paper, we will attempt to evaluate the surface density of states, which can be measured through tunneling spectroscopy \cite{Kashiwaya00,TK95,Deutscher}. 
\begin{table}[htbp]
  \caption{
    Summary of the MSSs and the chiral symmetries $Γ_{U}$ that protect the MSSs on each surface. 
    The Majorana cones (flat Fermi arc) imply that the MSSs have a cone-like gapless dispersion (zero-energy flat band), and a Fermi point (Fermi arc) exists in the surface Brillouin zone.
    }
  \label{tb: Majorana_states_and_chiral_operator}
\begin{tabular}{cccc}
  \toprule
  IR       & (100)  & (010) & (001) \\
  \midrule\midrule
  \multirow{2}{*}{$A_{u}$} & Majorana cones  & Majorana cones & \multirow{2}{*}{--}  \\
    & $Γ_{\mathcal{C}_{a}}$  & $Γ_{\mathcal{C}_{b}}$ & \\
  \midrule
  \multirow{2}{*}{$B_{1u}$} & flat Fermi arc  & flat Fermi arc & \multirow{2}{*}{--}  \\
    & $Γ_{\mathcal{M}_{ca}}$  & $Γ_{\mathcal{M}_{bc}}$ & \\
  \midrule
  \multirow{2}{*}{$B_{2u}$} & flat Fermi arcs & \multirow{2}{*}{--} & flat Fermi arcs \\
    & $Γ_{\mathcal{M}_{ab}}$ &  & $Γ_{\mathcal{M}_{bc}}$ \\
  \midrule
  \multirow{2}{*}{$B_{3u}$} & \multirow{2}{*}{--} & flat Fermi arcs & \multirow{2}{*}{--} \\
    & & $Γ_{\mathcal{M}_{ab}}$ & \\
  \bottomrule
\end{tabular}
\end{table}


\section{Local density of states}
\label{sec:LDOS}

In this section, we show the results of the LDOS for each IR. 
MSSs lead to a distinct LDOS structure on the surface as compared to the bulk.
This is particularly specific to the case of creating zero-energy peak (ZEP) structures~\cite{hara_1986_Polar,matsumoto_1999_Quasiparticle,tanuma_2001_Theoretical}.
The \ce{UTe2} has zero-energy surface ABS, which are protected by crystalline symmetries~\cite{tei_2023}.
Therefore, we evaluate the surface LDOS numerically to provide the detection of MSS on the surface of \ce{UTe2}.
The LDOS is calculated by numerically computing the recursive Green's function (see \cref{app:sdos})~\cite{umerski_1997,ohashi_2021,fukaya_2022}.
We introduce the $\vb*{k}_{∥}$-resolved LDOS from the recursive Green's function as
\begin{align}
  ρ_{ℓ}^{(x_{⟂})}(E,\vb*{k}_{∥}) = \frac{1}{π}\Im \tr[G_{ℓ}^\mathrm{R}(E,\vb*{k}_{∥},x_{⟂})],
  \label{eq: ARDOS_formula}
\end{align}
and the LDOS is also obtained by summing $ρ_{ℓ}^{(x_{⟂})}(E,\vb*{k}_{∥})$ over $\vb*{k}_{∥}$ as
\begin{align}
  ρ_{ℓ}^{(x_{⟂})}(E) = ∑_{\vb*{k}_{∥}}ρ_{ℓ}^{(x_{⟂})}(E,\vb*{k}_{∥}),
\end{align}
which $G^\mathrm{R}$ is the retarded Green's function, $\vb*{k}_∥$ is the wave numbers parallel to the cut-out plane (perpendicular to the open direction), $x_{⟂}$ is the site index perpendicular to the cut-out plane (parallel to the open direction), and $ℓ = \mathrm{S\; or\; N}$ is the label for the superconducting or normal state, respectively.

\textit{Overview}:
\cref{tb: Majorana_states_and_chiral_operator} also summarizes the dispersion of the surface ABS on the (100), (010), and (001) planes of \ce{UTe2} in each IR.
We find that the existence of the zero-energy states in each IR depends on the orientation of the cut-out plane:
In the $A_{u}$ pairing state, the structure of the full-gap LDOS in the bulk changes due to the contribution of the surface Majorana cone, resulting in a V-shaped structure similar to case of the pairing symmetry discussed in the context of \ce{Cu}-doped \ce{Bi2Se3}~\cite{Yamakage2012}
In the $B_{1u}$ pairing state, the LDOS has a full-gap structure in the bulk, but a ZEP appears due to the contribution of the surface flat Fermi arc 
(flat band zero energy surface ABS). 
In the $B_{2u}$ and $B_{3u}$ pairing states, the LDOS has a nodal structure in the bulk, but the surface flat Fermi arcs connecting the nodal points leads to the ZEP in the surface LDOS.
Therefore, we find that in all cases where MSS is present, the dispersion of the MSSs produces distinct LDOS structures from the bulk.
In the following subsections, we will examine the results in detail for each of the IRs.

\subsection{\texorpdfstring{$A_{u}$}{Au} pairing state}
\label{ssec:sdos_Au}
\begin{figure}[htbp]
  \begin{tabular}{ccc}
    \scalebox{1.2}{\fbox{$A_{u}$}} & $\bm{k}_{∥}$-resolved LDOS & LDOS\\
    \rotatebox[origin=c]{0}{(100)}&
    \begin{minipage}[c]{0.35\linewidth}
        \subcaption{}\vspace{-1mm}
        \label{sfig: sdos_Au_a_dos3d_1}
          \includegraphics[trim={0 20px 0 150px},clip,width=\linewidth]{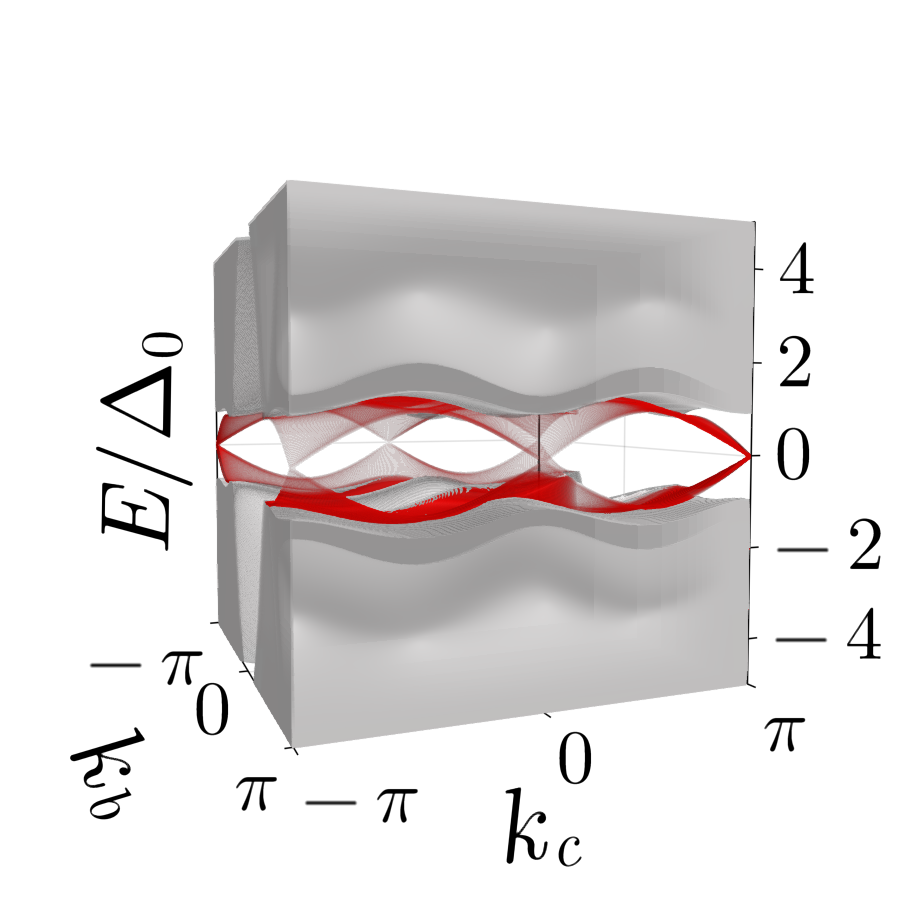}
    \end{minipage}&
    \begin{minipage}[c]{0.45\linewidth}
        \subcaption{}\vspace{-1mm}
        \label{sfig: sdos_Au_a_sdos}
          \includegraphics[width=\linewidth]{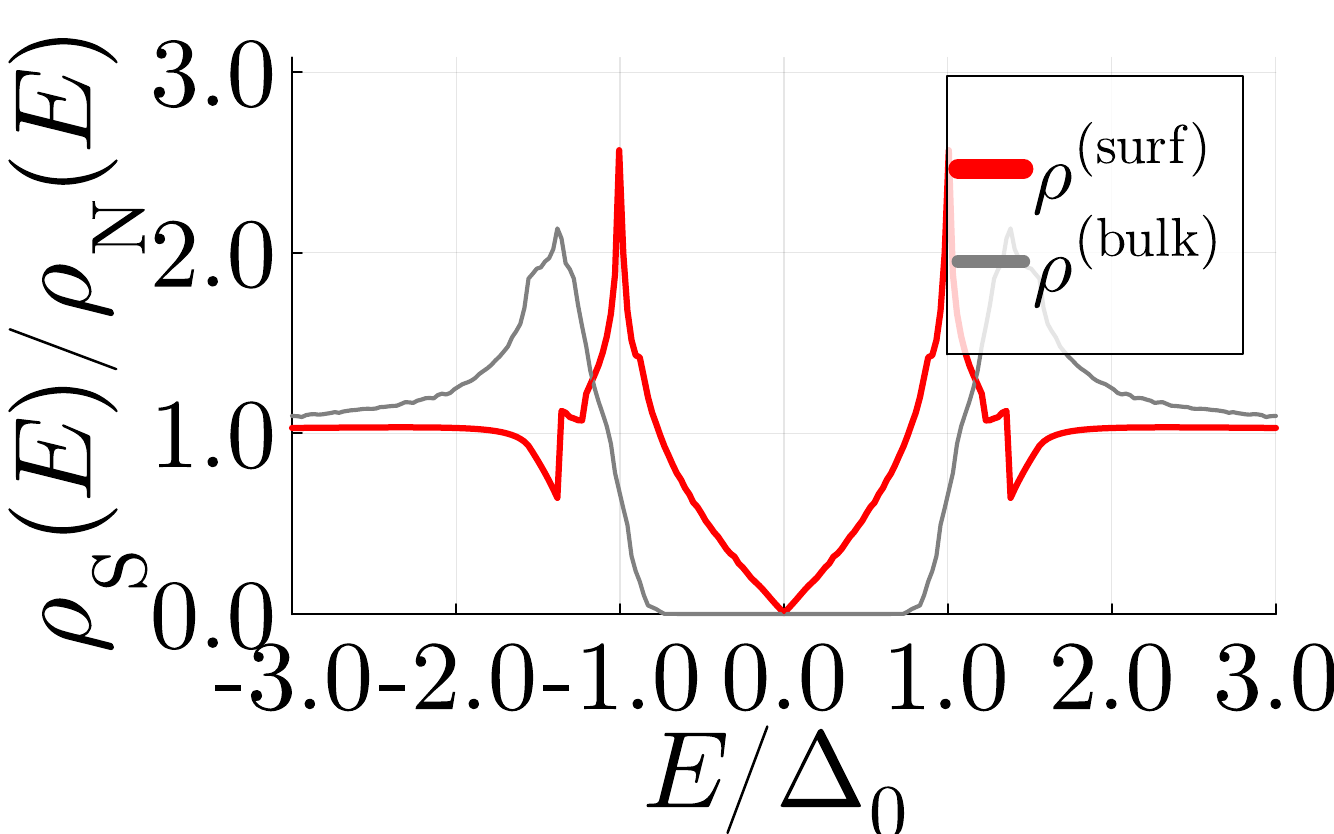}
    \end{minipage}\\
    \rotatebox[origin=c]{0}{(010)}&
    \begin{minipage}[c]{0.35\linewidth}
        \subcaption{}\vspace{-1mm}
        \label{sfig: sdos_Au_b_dos3d_1}
          \includegraphics[trim={0 20px 0 150px},clip,width=\linewidth]{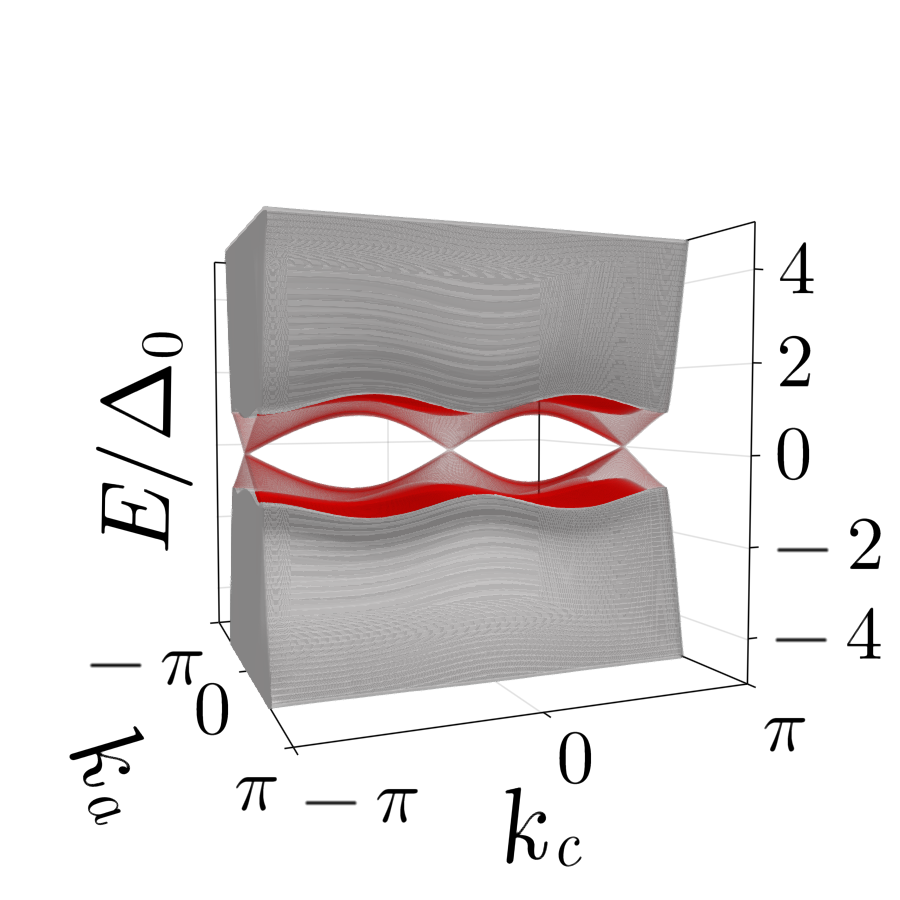}
    \end{minipage}&
    \begin{minipage}[c]{0.45\linewidth}
        \subcaption{}\vspace{-1mm}
        \label{sfig: sdos_Au_b_sdos}
          \includegraphics[width=\linewidth]{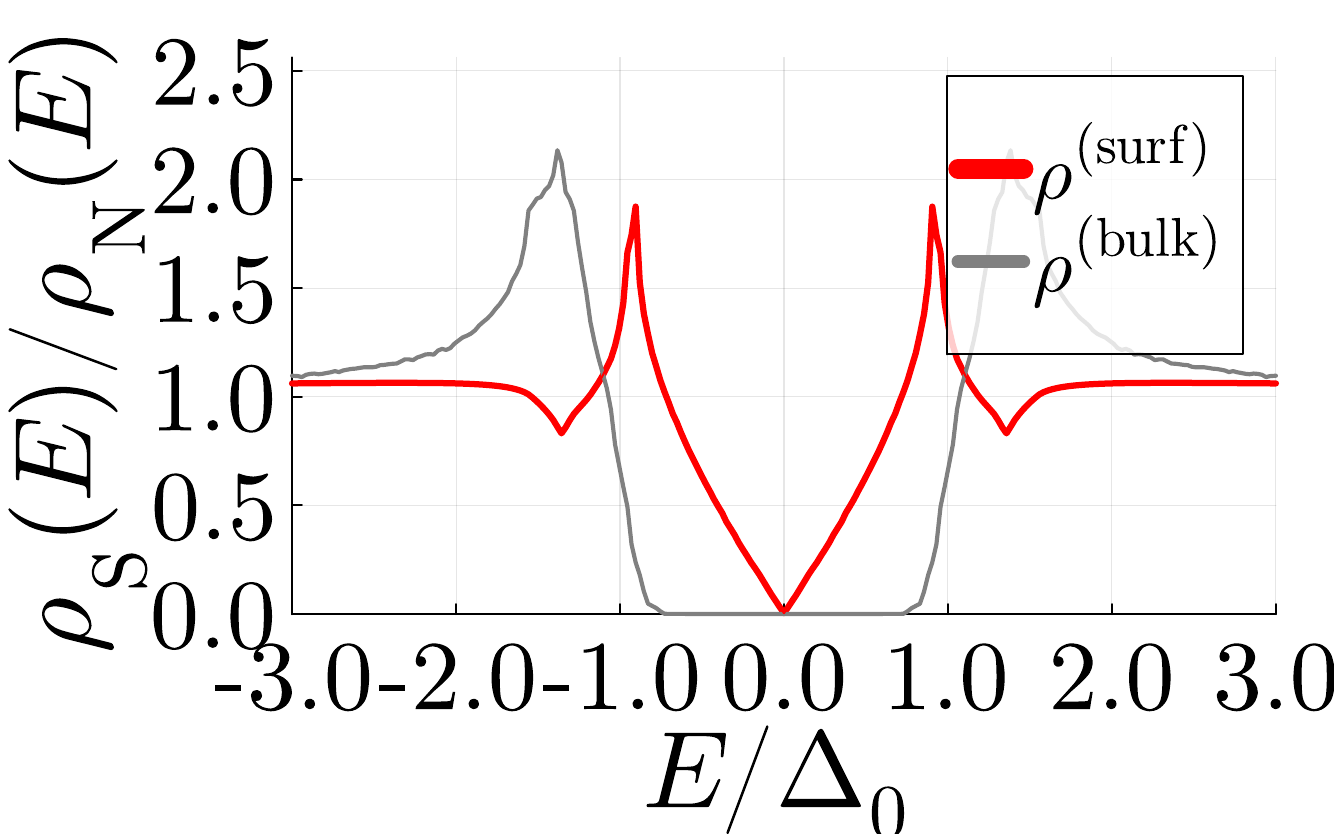}
    \end{minipage}\\
    \rotatebox[origin=c]{0}{(001)}&
    \begin{minipage}[c]{0.35\linewidth}
        \subcaption{}\vspace{-1mm}
        \label{sfig: sdos_Au_c_dos3d_1}
          \includegraphics[trim={0 20px 0 150px},clip,width=\linewidth]{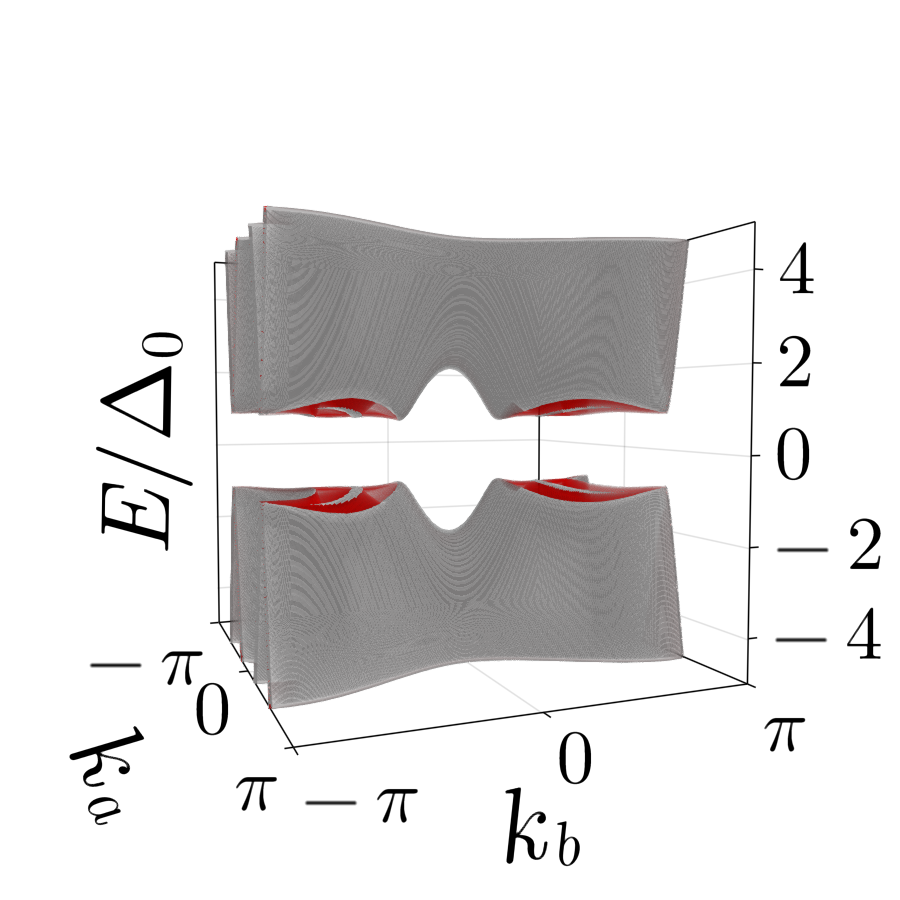}
    \end{minipage}&
    \begin{minipage}[c]{0.45\linewidth}
        \subcaption{}\vspace{-1mm}
        \label{sfig: sdos_Au_c_sdos}
          \includegraphics[width=\linewidth]{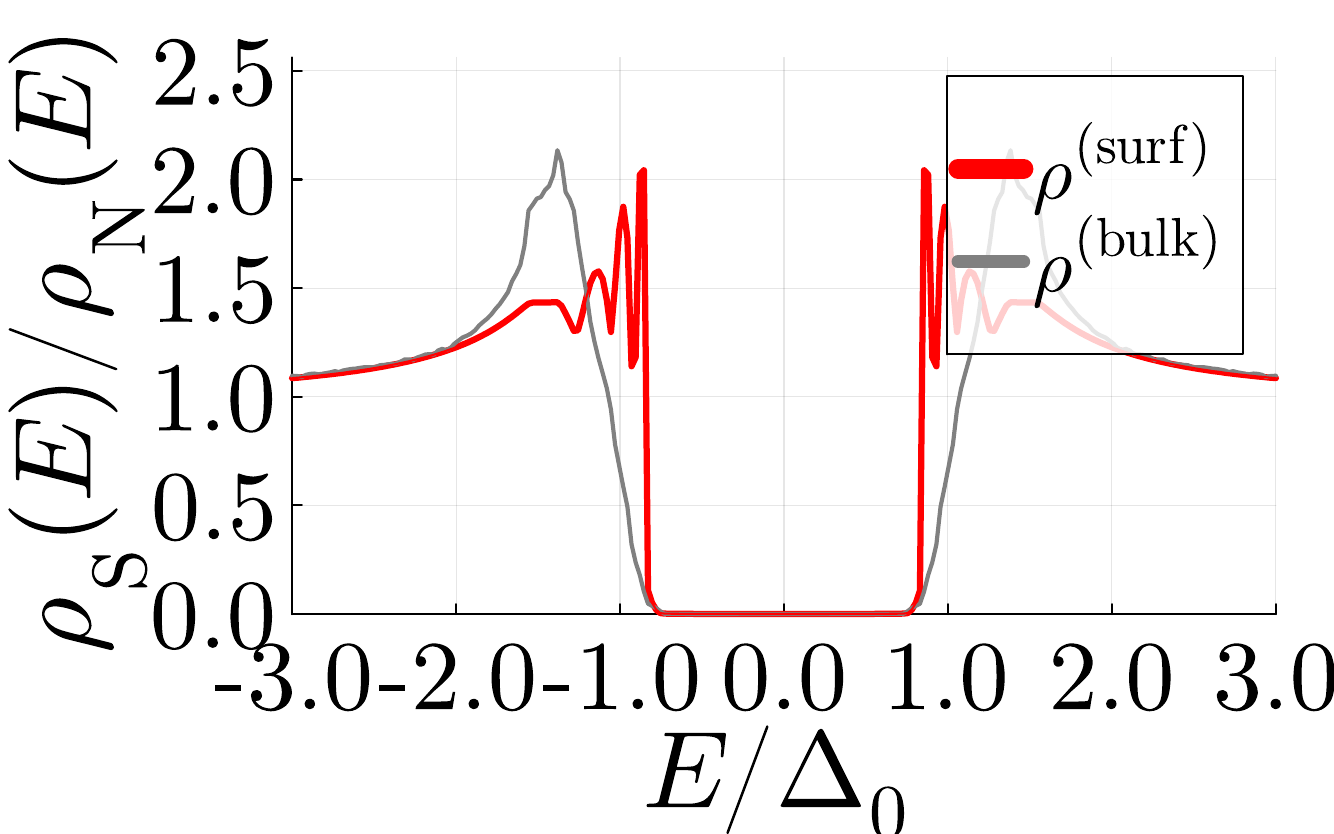}
    \end{minipage}
  \end{tabular}
  \caption{
    Local density of states for the $A_{u}$ pairing state.
    In all of these figures, the red (grey) color shows the surface (bulk) LDOS.
    Each row represents the results on the $(100)$, $(010)$ and $(001)$ surface, respectively.
    The first column represents $\vb*{k}_{∥}$-resolved LDOS by coloring in points with finite $ρ_\mathrm{S}(\vb*{k}_{∥})$ values.
    The bulk $A_{u}$ state has a full gap structure, while the in-gap (V-shaped) structure appears due to the contribution of the surface Majorana cone states to the LDOS as shown in (a), (b), (c) and (b).
  }
  \label{fig: sdos_Au}
\end{figure}

First, we consider the $A_{u}$ pairing state, where the $\vb*{d}$-vector is given by \cref{eq:d-vector_IR_Au}. 
In this state, there are Majorana cones protected by crystalline symmetry in the (100) and (010) planes when the Fermi surface is a cylindrical shape~\cite{tei_2023}.
The results of the LDOS calculations are summarized in \cref{fig: sdos_Au}. 
Each row of \cref{fig: sdos_Au} corresponds to the results on the (100), (010), and (001) planes, respectively. 
The first column of \cref{fig: sdos_Au} describe the structure of the $\vb*{k}_{∥}$-resolved LDOS, while the second column of \cref{fig: sdos_Au} describes the surface (red) and bulk (gray) LDOS.
We note that \cref{fig: sdos_B1u,fig: sdos_B2u,fig: sdos_B3u}, which represent the results in different IRs, are also equivalently arranged.

As shown in \cref{sfig: sdos_Au_a_dos3d_1}, which are the (100) plane cases, \ce{UTe2} has the surface-localized state and forms Majorana cones at $\vb*{k}_∥ = (k_{b},k_{c}) = (±π,0)\;\mathrm{and}\;(±π,±π)$. 
The Majorana cones are protected by the topological invariants associated with $Γ_{\mathcal{M}_{ab}}$ and $Γ_{\mathcal{C}_{a}}$, respectively (see \cref{tb: Majorana_states_and_chiral_operator} and Ref.~\citenum{tei_2023}). 
In terms of the LDOS, the bulk has a full-gap structure, whereas the surface LDOS has a V-shaped structure in the in-gap region due to the contribution of the Majorana cones, as shown in \cref{sfig: sdos_Au_a_sdos}.
\Cref{sfig: sdos_Au_b_dos3d_1}, which are the (010) plane cases, also show the surface-localized state and form Majorana cones at $\vb*{k}_∥ = (k_{a}, k_{c}) = (0,0)\;\mathrm{and}\;(0,±π)$. 
The Majorana cones at $\vb*{k}_∥ = (0,0)$ is protected by the topological invariant $Γ_{\mathcal{C}_{b}}$ (see \cref{sec:model_and_symmetry}). 
\rocomment{The MZM at $\vb*{k}_∥ = (0,±π)$, on the other hand, is topologically trivial, arising accidentally from the simplification of the model Hamiltonian \cite{tei_2023}. This MZM can be gapped out by adding symmetry-allowed terms to the Hamiltonian.}
\Cref{sfig: sdos_Au_b_sdos} also shows the existence of the in-gap states, which is a consequence of the contribution from the surface Majorana cones.
\Cref{sfig: sdos_Au_c_dos3d_1,sfig: sdos_Au_c_sdos}, which are the (001) plane case, shows the surface-localized state, but there is no zero-energy state. 
Therefore, a full-gap structure is seen for the surface LDOS as well as for the bulk structure.

In summary, there exist surface Majorana cones on the (100) and (010) planes of the $A_{u}$ pairing state, which cause the surface LDOS to have a V-shaped structure.

\subsection{\texorpdfstring{$B_{1u}$}{B1u} pairing state}
\label{ssec:sdos_B1u}
\begin{figure}[htbp]
  \begin{tabular}{ccc}
    \scalebox{1.2}{\fbox{$B_{1u}$}} & $\vb*{k}_{∥}$-resolved LDOS & LDOS\\
    \rotatebox[origin=c]{0}{(100)}&
    \begin{minipage}[c]{0.35\linewidth}
        \subcaption{}\vspace{-1mm}
        \label{sfig: sdos_B1u_a_dos3d_2}
          \includegraphics[trim={0 20px 0 150px},clip,width=\linewidth]{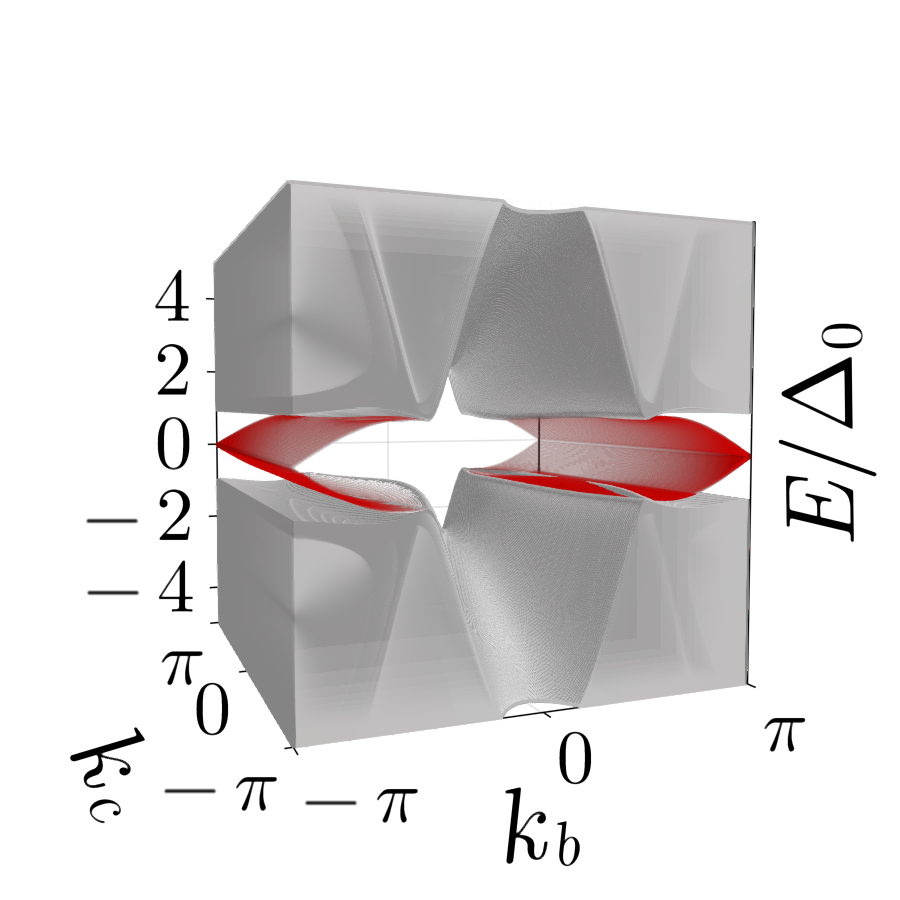}
    \end{minipage}&
    \begin{minipage}[c]{0.45\linewidth}
        \subcaption{}\vspace{-1mm}
        \label{sfig: sdos_B1u_a_sdos}
          \includegraphics[width=\linewidth]{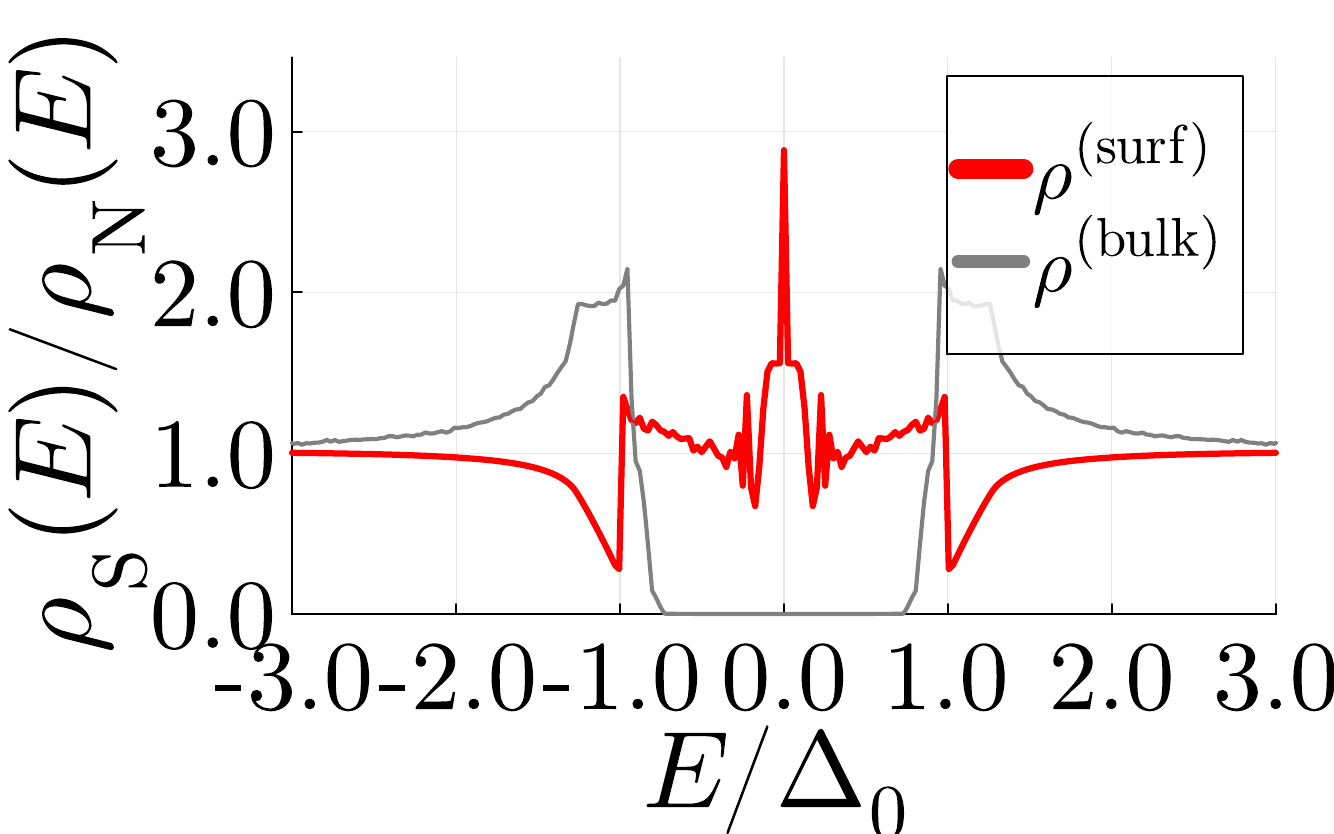}
    \end{minipage}\\
    \rotatebox[origin=c]{0}{(010)}&
    \begin{minipage}[c]{0.35\linewidth}
        \subcaption{}\vspace{-1mm}
        \label{sfig: sdos_B1u_b_dos3d_2}
          \includegraphics[trim={0 20px 0 150px},clip,width=\linewidth]{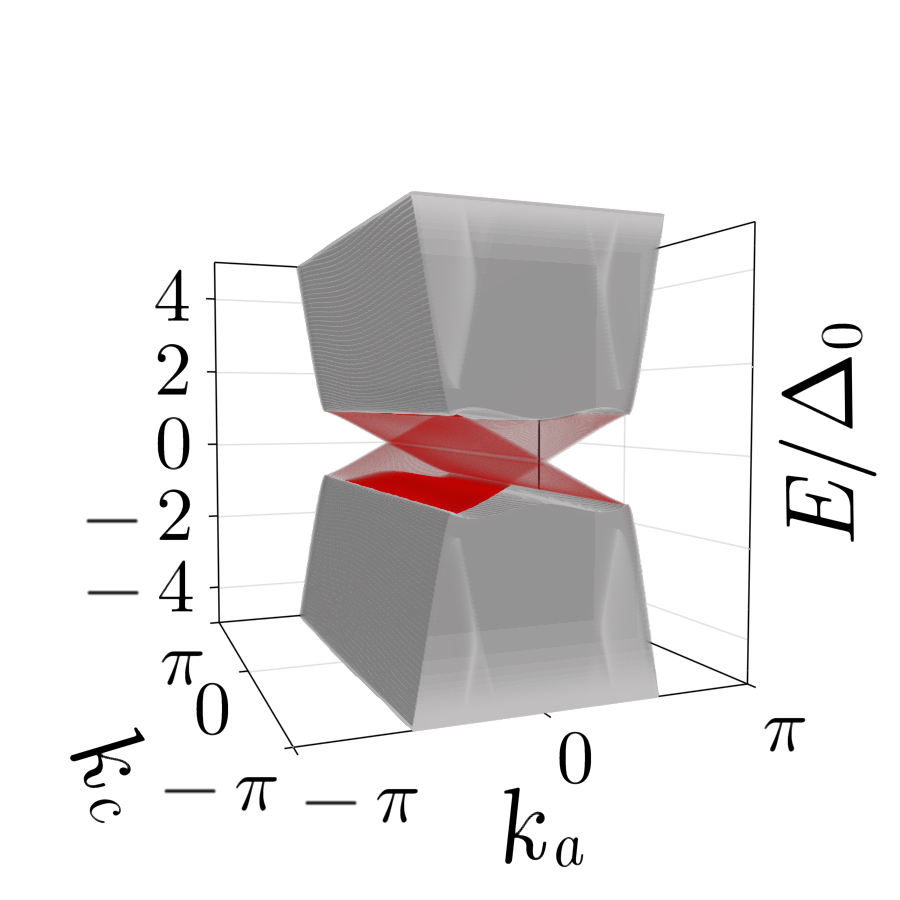}
    \end{minipage}&
    \begin{minipage}[c]{0.45\linewidth}
        \subcaption{}\vspace{-1mm}
        \label{sfig: sdos_B1u_b_sdos}
          \includegraphics[width=\linewidth]{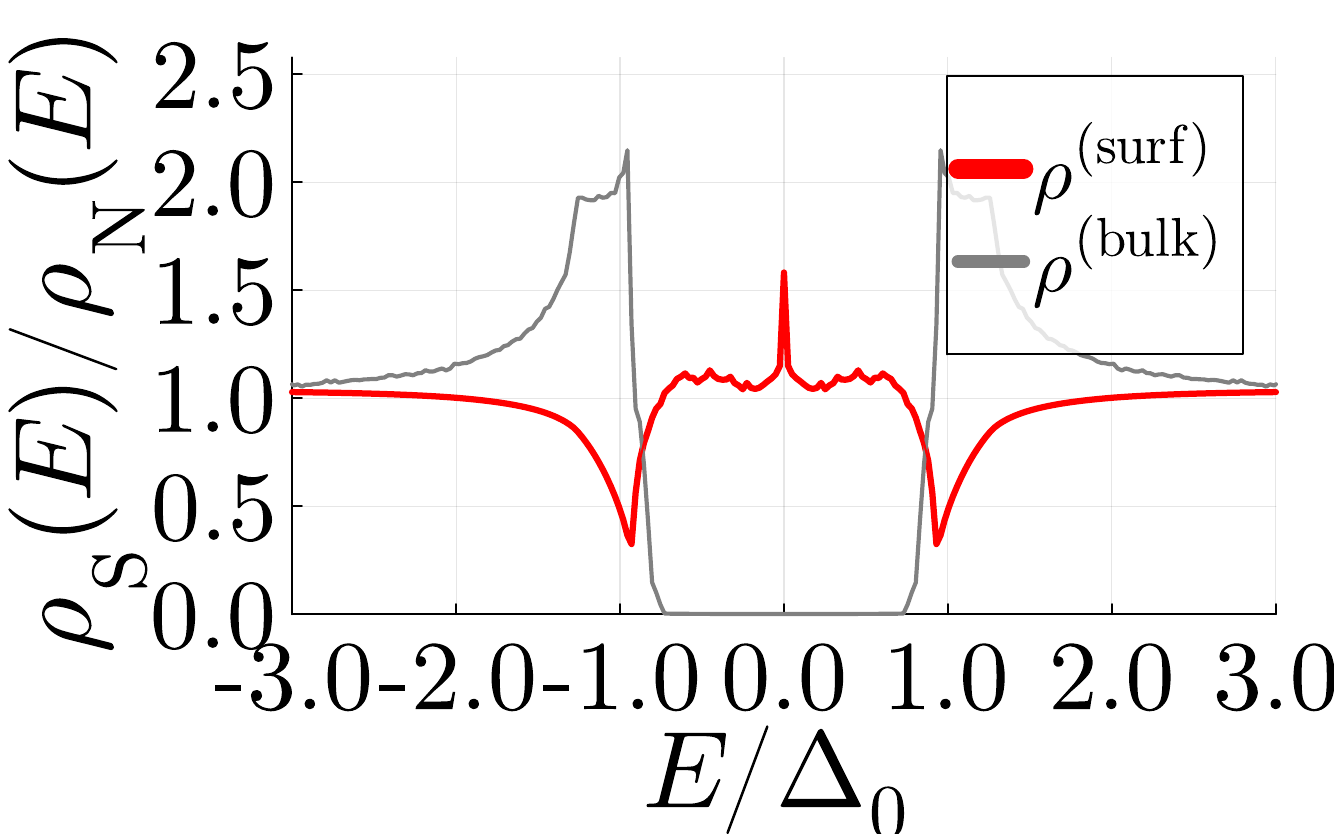}
    \end{minipage}\\
    \rotatebox[origin=c]{0}{(001)}&
    \begin{minipage}[c]{0.35\linewidth}
        \subcaption{}\vspace{-1mm}
        \label{sfig: sdos_B1u_c_dos3d_2}
          \includegraphics[trim={0 20px 0 150px},clip,width=\linewidth]{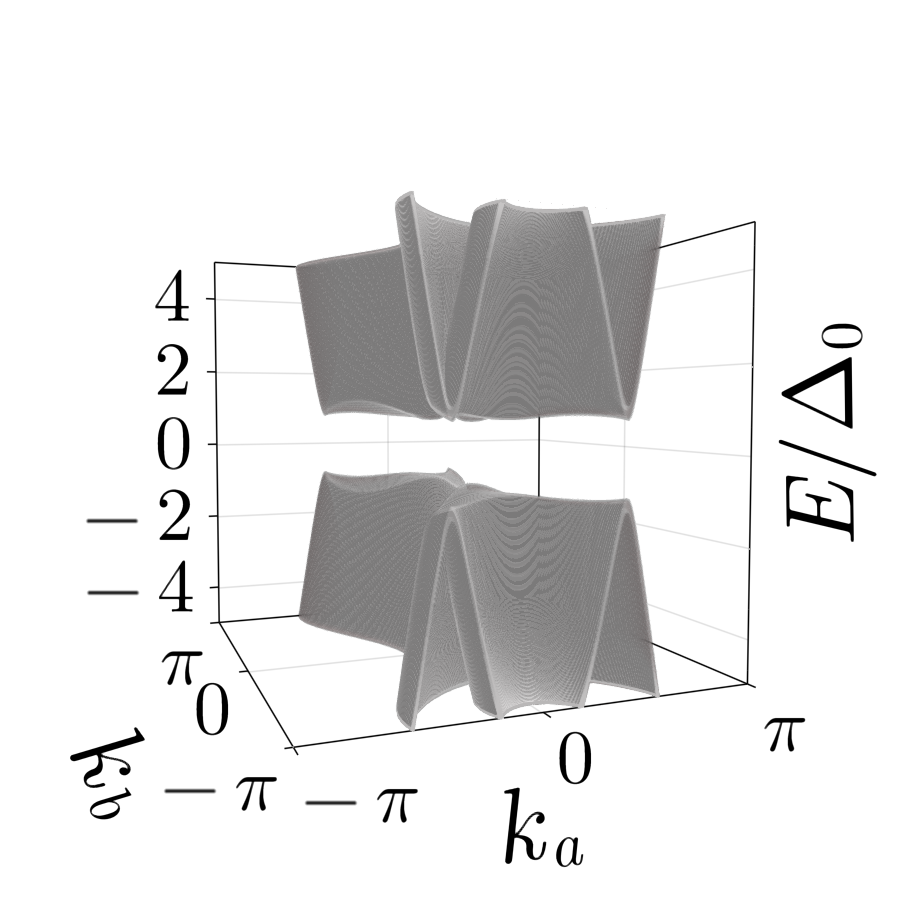}
    \end{minipage}&
    \begin{minipage}[c]{0.45\linewidth}
        \subcaption{}\vspace{-1mm}
        \label{sfig: sdos_B1u_c_sdos}
          \includegraphics[width=\linewidth]{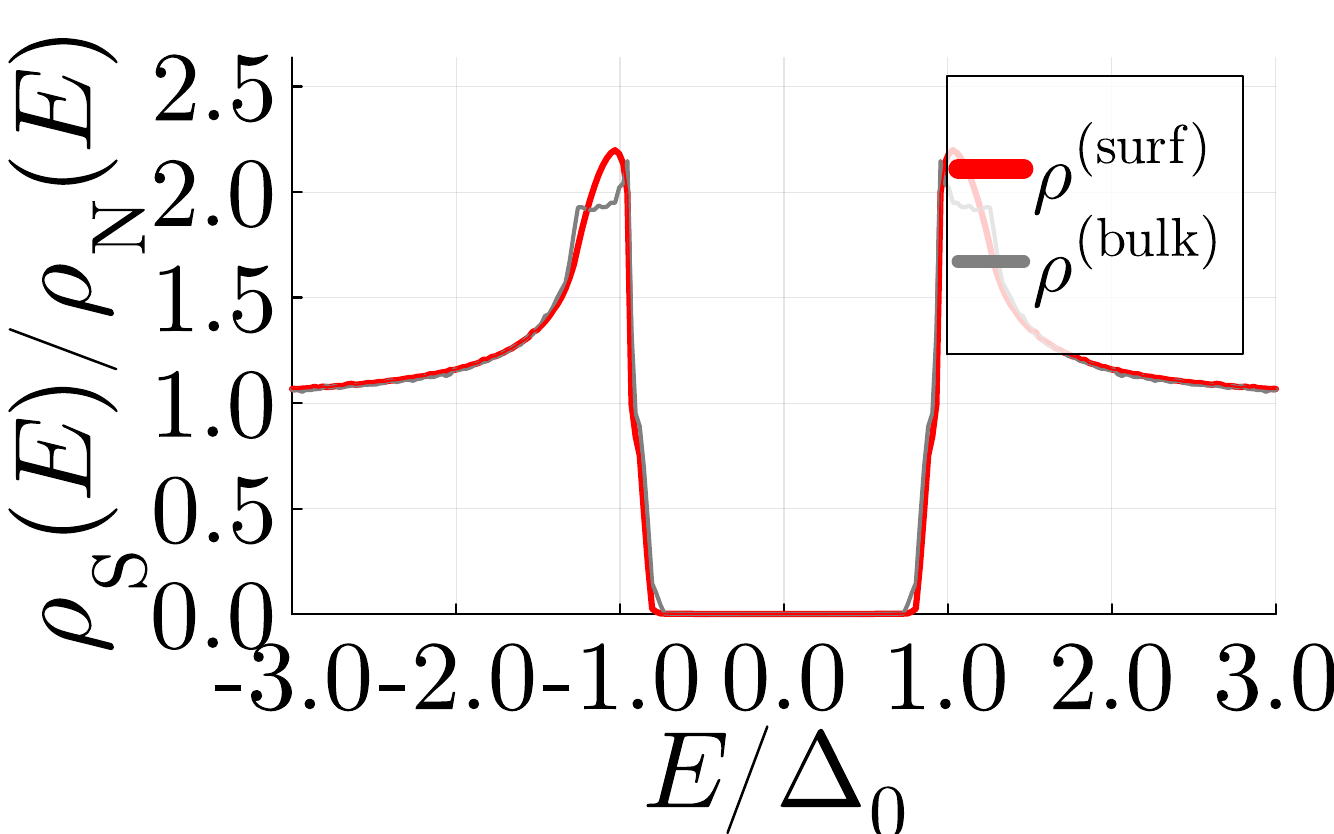}
    \end{minipage}
  \end{tabular}
  \caption{
    Local density of states for the $B_{1u}$ pairing state.
    In all of these figures, the red (grey) color shows the surface (bulk) LDOS. 
    Each row represents the results of the $(100)$, $(010)$ and $(001)$ surfaces, respectively.
    The first column represents $\vb*{k}_{∥}$-resolved LDOS by coloring in points with finite $ρ_\mathrm{S}(\vb*{k}_{∥})$ values.
    The bulk $B_{1u}$ state has a full gap structure in the cylindrical Fermi surface, while the ZEP structure appears due to the contribution of the surface flat Fermi arc in the MSS to the LDOS as shown in (a) and (b).
  }
  \label{fig: sdos_B1u}
\end{figure}

Next, we consider the $B_{1u}$ state, where the $\vb*{d}$-vector is given by \cref{eq:d-vector_IR_B1u}.
In this state, the bulk excitation is fully gapped in the cylindrical Fermi surface and there is a flat Fermi arc state~\cite{schnyder_2012} protected by crystalline symmetry on the (100) and (010) planes~\cite{tei_2023}.
The results of the LDOS are summarized in \cref{fig: sdos_B1u}.
As shown in \cref{sfig: sdos_B1u_a_dos3d_2}, which are the $\vb*{k}_{∥}$-resolved LDOS on the (100) plane, \ce{UTe2} has the surface-localized state and forms a flat Fermi arc at $k_{b} = ±π$. 
The flat Fermi arc is protected by the topological invariant defined by $Γ_{\mathcal{M}_{ab}}$ on $\vb*{k}_{∥} = (±π,0)$ (see \cref{sec:model_and_symmetry}). 
In terms of LDOS, the bulk has a full-gap structure, whereas the surface LDOS has a prominent ZEP structure in the in-gap region due to the flat Fermi arc contribution, as shown in \cref{sfig: sdos_B1u_a_sdos}.

\Cref{sfig: sdos_B1u_b_dos3d_2}, which shows the $\vb*{k}_{∥}$-resolved LDOS on the (010) plane, also indicates the existence of the surface-localized state and the formation of a flat Fermi arc on $k_{a} = 0$. The flat Fermi arc is protected by the topological invariant defined by $Γ_{\mathcal{M}_{bc}}$ (see \cref{tb: Majorana_states_and_chiral_operator}). 
\Cref{sfig: sdos_B1u_b_sdos} also shows that the ZEP structure of the LDOS appears as a consequence of the contribution from the flat Fermi arc.

The results on the (001) plane are displayed in \cref{sfig: sdos_B1u_c_dos3d_2,sfig: sdos_B1u_c_sdos}. 
This indicates that there is no surface-localized state on the surface. 
Therefore, both the surface LDOS and the bulk structure exhibit a full-gap structure.

In summary, for the $B_{1u}$ pairing state, in the (100) and (010) planes, there is a flat Fermi arc, which causes the surface LDOS to have a ZEP structure.

\subsection{\texorpdfstring{$B_{2u}$}{B2u} pairing state}
\label{ssec:sdos_B2u}

\begin{figure}[htbp]
  \begin{tabular}{ccc}
    \scalebox{1.2}{\fbox{$B_{2u}$}} & $\bm{k}_{∥}$-resolved LDOS & LDOS\\
    \rotatebox[origin=c]{0}{(100)}&
    \begin{minipage}[c]{0.35\linewidth}
        \subcaption{}\vspace{-1mm}
        \label{sfig: sdos_B2u_a_dos3d_1}
          \includegraphics[trim={0 20px 0 150px},clip,width=\linewidth]{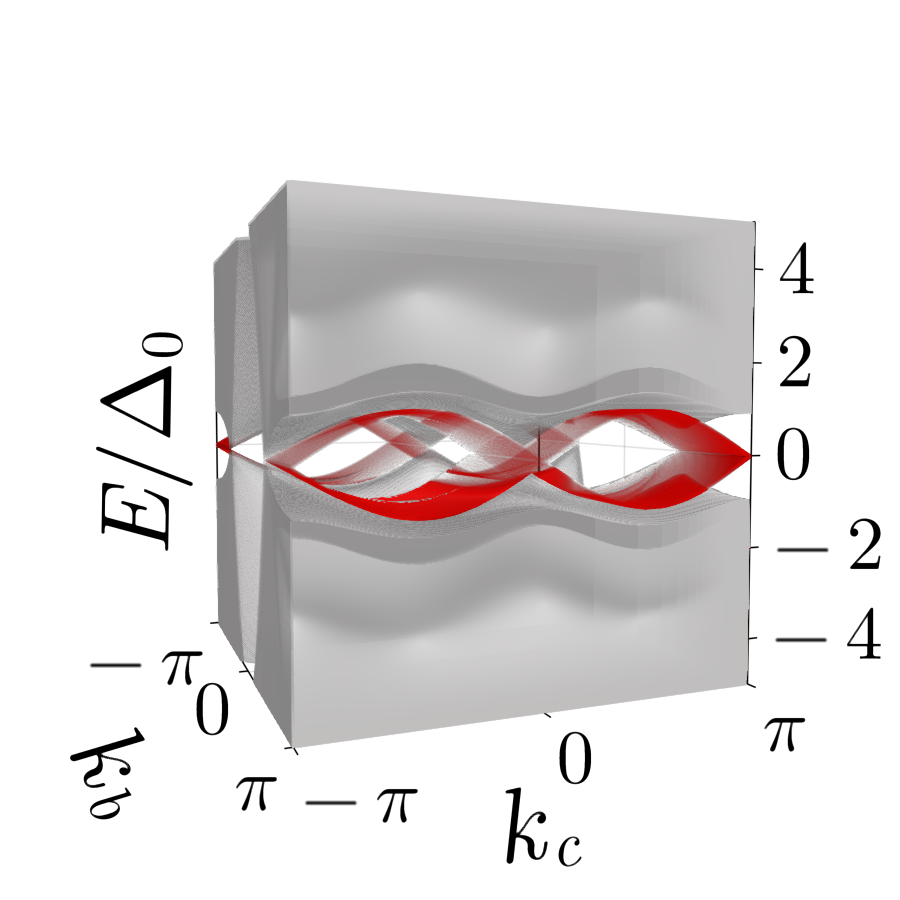}
    \end{minipage}&
    \begin{minipage}[c]{0.45\linewidth}
        \subcaption{}\vspace{-1mm}
        \label{sfig: sdos_B2u_a_sdos}
          \includegraphics[width=\linewidth]{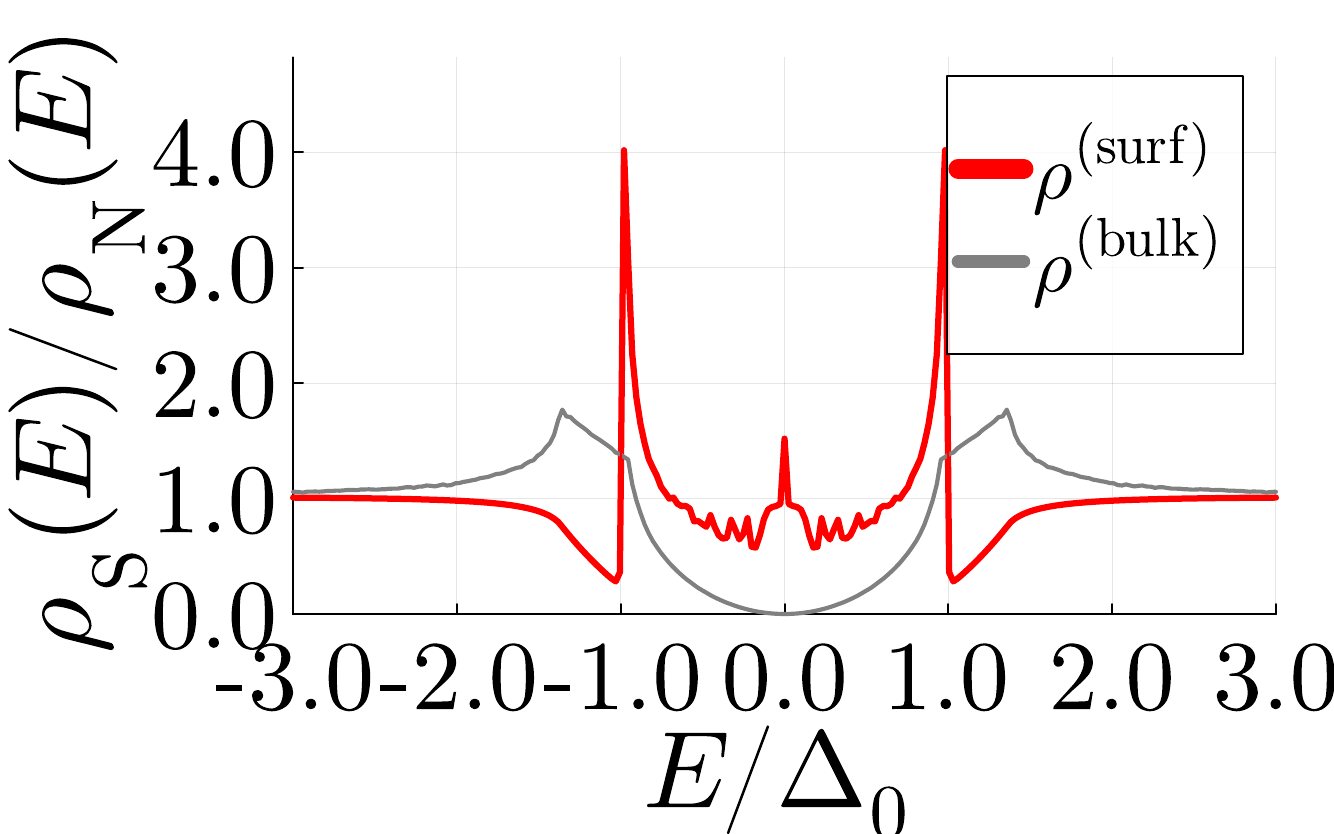}
    \end{minipage}\\
    \rotatebox[origin=c]{0}{(010)}&
    \begin{minipage}[c]{0.35\linewidth}
        \subcaption{}\vspace{-1mm}
        \label{sfig: sdos_B2u_b_dos3d_1}
          \includegraphics[trim={0 20px 0 150px},clip,width=\linewidth]{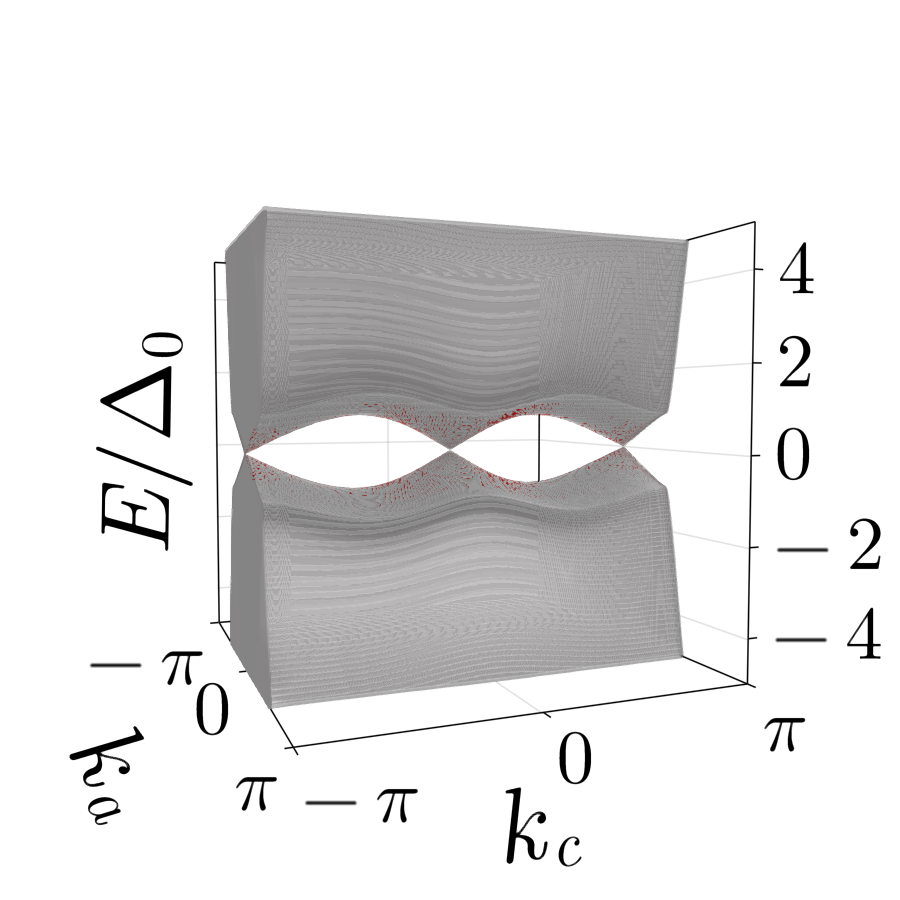}
    \end{minipage}&
    \begin{minipage}[c]{0.45\linewidth}
        \subcaption{}\vspace{-1mm}
        \label{sfig: sdos_B2u_b_sdos}
          \includegraphics[width=\linewidth]{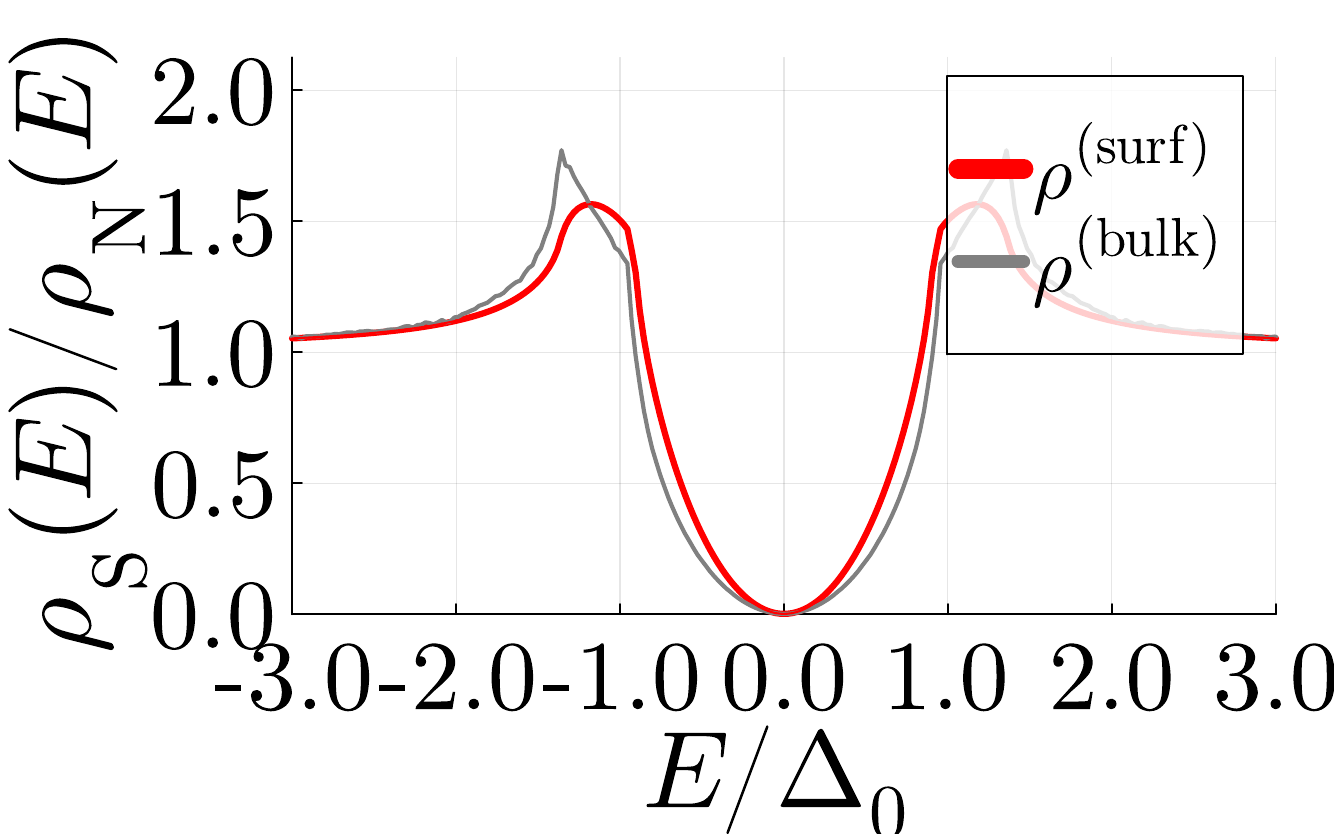}
    \end{minipage}\\
    \rotatebox[origin=c]{0}{(001)}&
    \begin{minipage}[c]{0.35\linewidth}
        \subcaption{}\vspace{-1mm}
        \label{sfig: sdos_B2u_c_dos3d_1}
          \includegraphics[trim={0 20px 0 150px},clip,width=\linewidth]{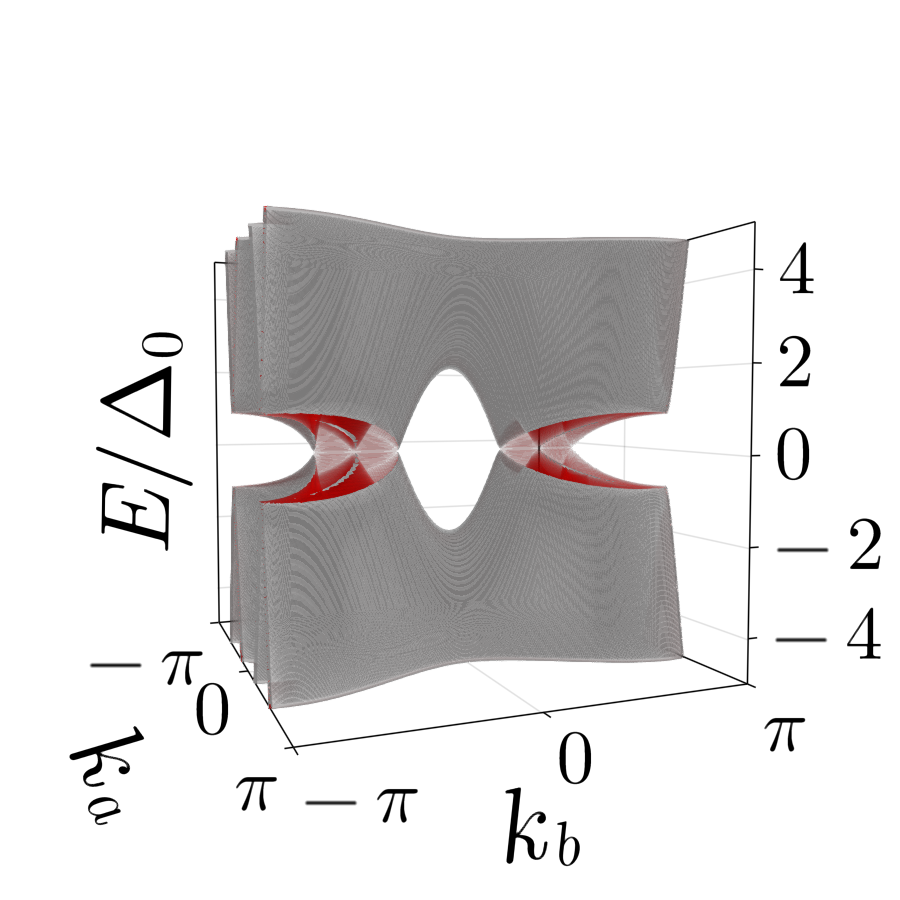}
    \end{minipage}&
    \begin{minipage}[c]{0.45\linewidth}
        \subcaption{}\vspace{-1mm}
        \label{sfig: sdos_B2u_c_sdos}
          \includegraphics[width=\linewidth]{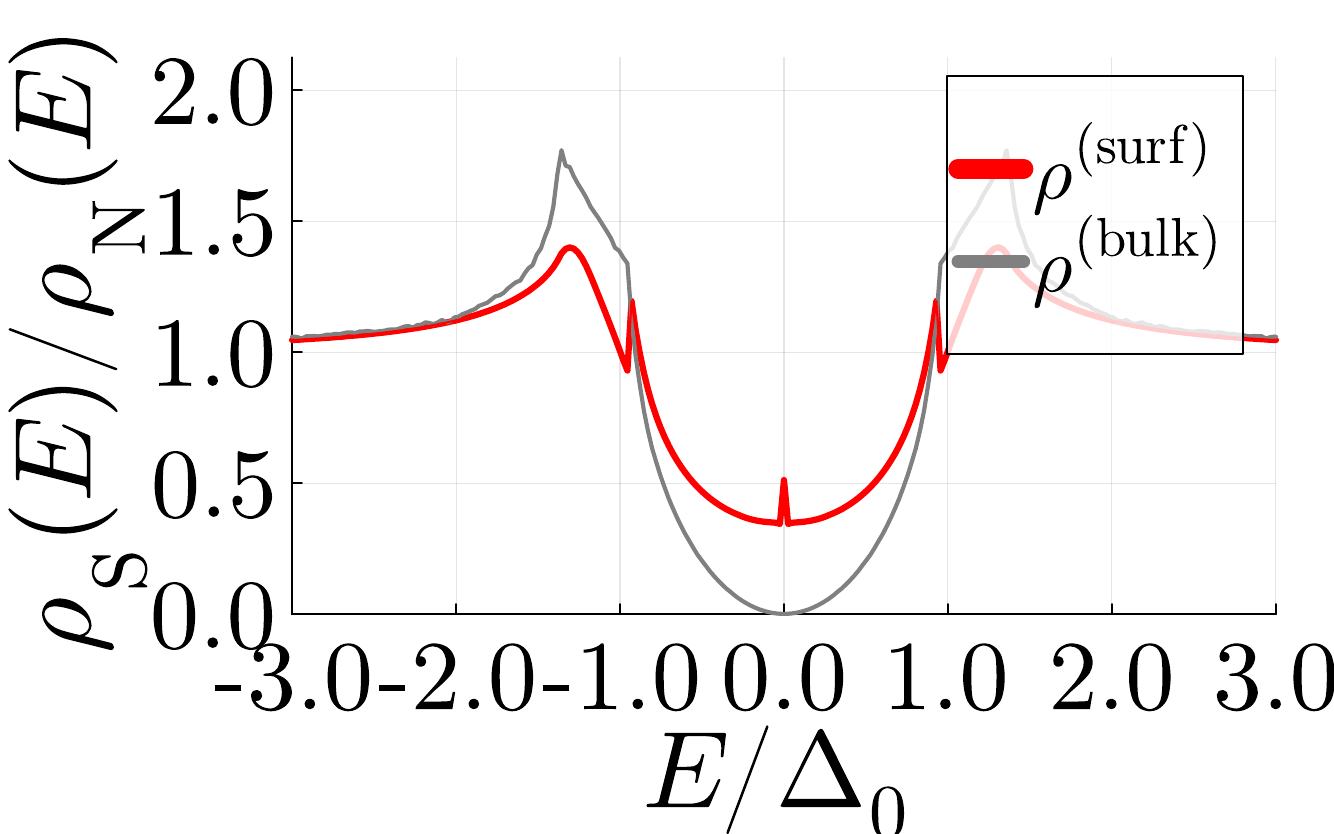}
    \end{minipage}
  \end{tabular}
  \caption{
    Local density of states for the $B_{2u}$ pairing state.
    In all of these figures, the red (grey) color shows the surface (bulk) LDOS. 
    Each row represents the results on the $(100)$, $(010)$ and $(001)$ surface, respectively. 
    The first column represents $\vb*{k}_{∥}$-resolved LDOS by coloring in points with finite $ρ_\mathrm{S}(\vb*{k}_{∥})$ values.
    The bulk $B_{2u}$ state has point nodes, while the ZEP structure appears due to the contribution of the surface flat Fermi arcs in the MSS to the LDOS as shown in (a) and (b).
  }
  \label{fig: sdos_B2u}
\end{figure}

We also consider the $B_{2u}$ pairing state, where the $\vb*{d}$-vector is given by \cref{eq:d-vector_IR_B2u}.
In this state, there is a flat Fermi arc state protected by crystalline symmetry on the (100) and (001) planes~\cite{tei_2023}.
The results of the LDOS calculations are summarized in \cref{fig: sdos_B2u}.

As shown in \cref{sfig: sdos_B2u_a_dos3d_1}, on the (100) plane, the $\vb*{k}_{∥}$-resolved LDOS exhibits the existence of the surface-localized state and the formation of the flat Fermi arcs on $k_{c} = 0 \mathrm{\; and \;}±π$. 
The flat Fermi arcs are protected by the topological invariant defined by $Γ_{\mathcal{M}_{ab}}$ (see \cref{tb: Majorana_states_and_chiral_operator}). 
The bulk DOS has a nodal structure, whereas the surface LDOS has a ZEP structure in the in-gap region due to the flat Fermi arcs contribution, as shown in \cref{sfig: sdos_B1u_a_sdos}.

\Cref{sfig: sdos_B2u_a_dos3d_1,sfig: sdos_B2u_a_sdos} shows the results on the (010) plane, indicating the absence of the surface-localized state. 
Therefore, both the surface LDOS and the bulk structure exhibit the spectral profile that reflects the existence of nodal points.

The LDOSs on the (001) plane in \cref{sfig: sdos_B2u_c_dos3d_1} show the existence of the surface-localized state and the formation of the flat Fermi arcs on $k_{a} = 0$.
The flat Fermi arcs are protected by the topological invariant defined by $Γ_{\mathcal{M}_{bc}}$ (see \cref{tb: Majorana_states_and_chiral_operator}). 
\Cref{sfig: sdos_B1u_c_sdos} also shows the ZEP structure of the LDOS, which is a consequence of the contribution from the flat Fermi arcs.

In summary, in the $B_{2u}$ pairing state, the flat Fermi arcs appear on the (100) and (001) planes, which causes the surface LDOS to have a ZEP structure. 
However, the height of the ZEP is smaller than that of the $B_{1u}$ pairing state and 
may be smeared out by the nonzero temperature in the actual tunneling spectroscopy.
The weak signature of the ZEP stems from the length of the flat Fermi arcs. In the $B_{1u}$ case, the flat Fermi arc spans the entire region of the surface Brillouin zone along $k_{c}\in [-π,π]$, while in the $B_{2u}$ case, the flat Fermi arcs only lie in a restricted region of $k_{c}$.

\subsection{\texorpdfstring{$B_{3u}$}{B3u} pairing state}
\label{ssec:sdos_B3u}

\begin{figure}[htbp]
  \begin{tabular}{ccc}
    \scalebox{1.2}{\fbox{$B_{3u}$}} & $\bm{k}_{∥}$-resolved LDOS & LDOS\\
    \rotatebox[origin=c]{0}{(100)}&
    \begin{minipage}[c]{0.35\linewidth}
        \subcaption{}\vspace{-1mm}
        \label{sfig: sdos_B3u_a_dos3d_1}
          \includegraphics[trim={0 20px 0 150px},clip,width=\linewidth]{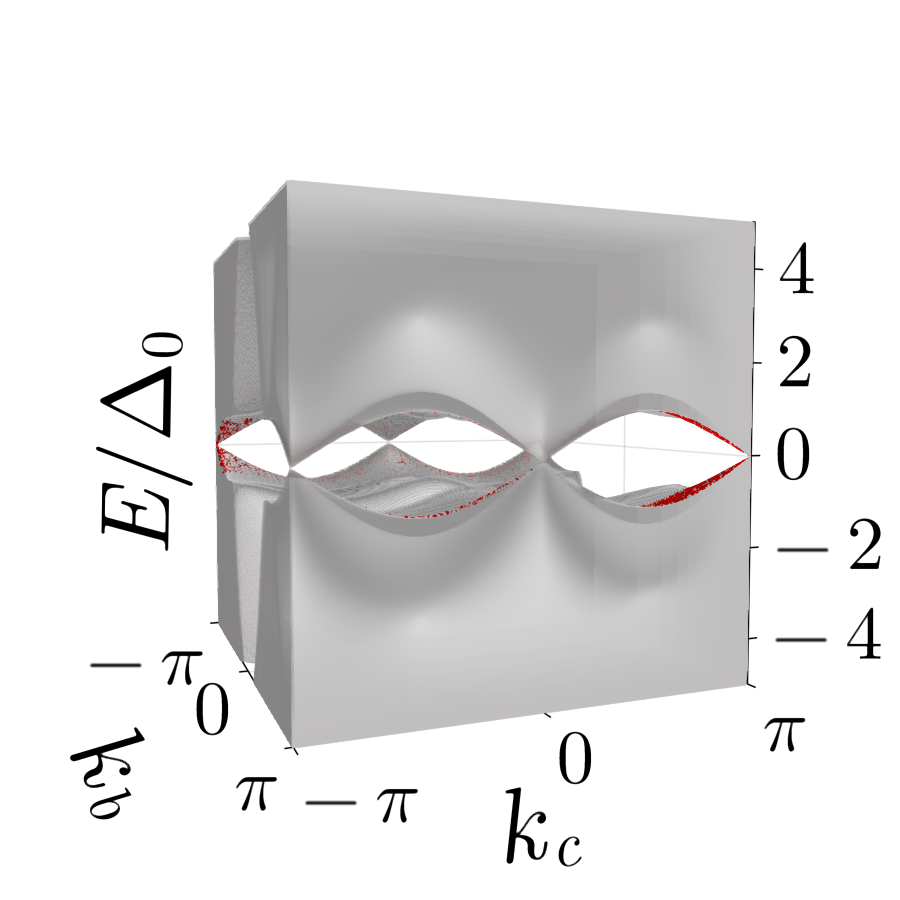}
    \end{minipage}&
    \begin{minipage}[c]{0.45\linewidth}
        \subcaption{}\vspace{-1mm}
        \label{sfig: sdos_B3u_a_sdos}
          \includegraphics[width=\linewidth]{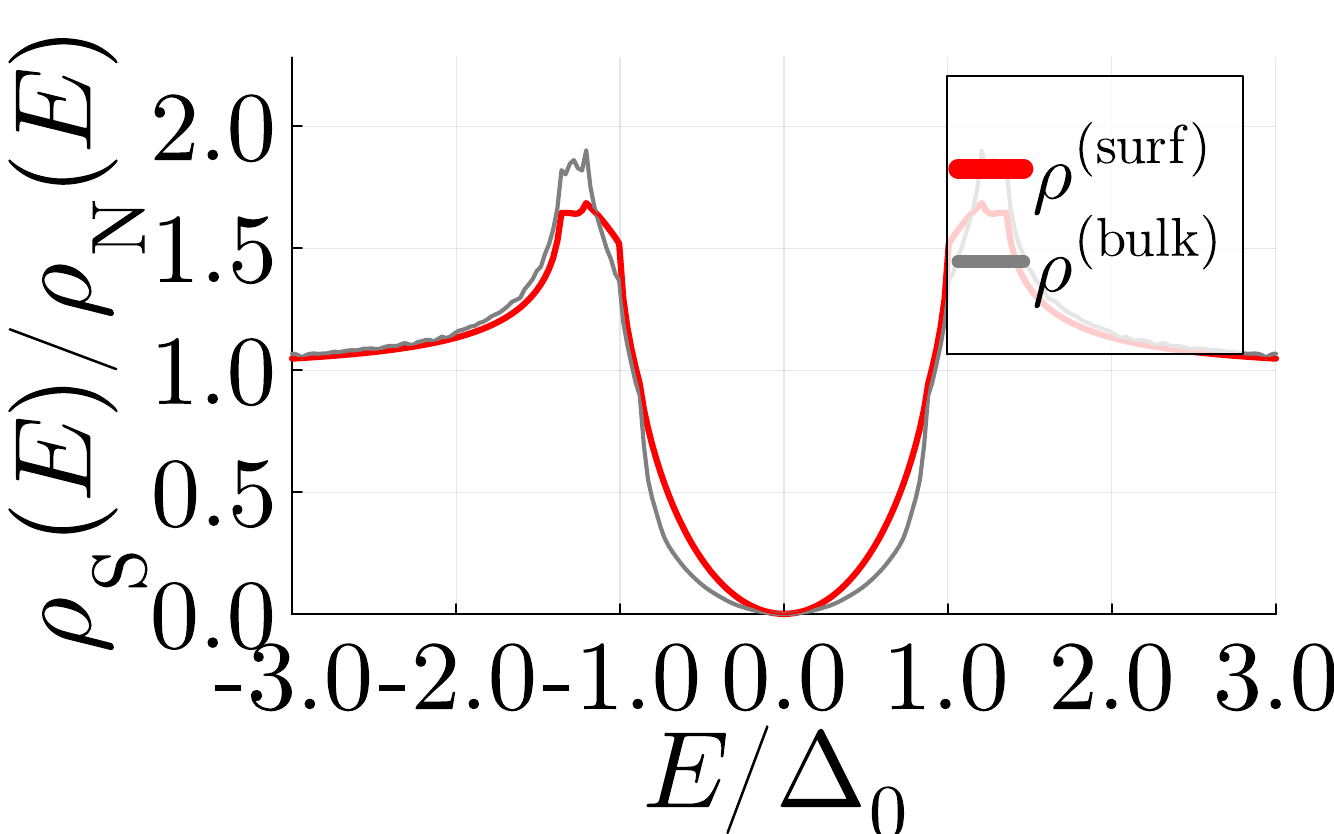}
    \end{minipage}\\
    \rotatebox[origin=c]{0}{(010)}&
    \begin{minipage}[c]{0.35\linewidth}
        \subcaption{}\vspace{-1mm}
        \label{sfig: sdos_B3u_b_dos3d_1}
          \includegraphics[trim={0 20px 0 150px},clip,width=\linewidth]{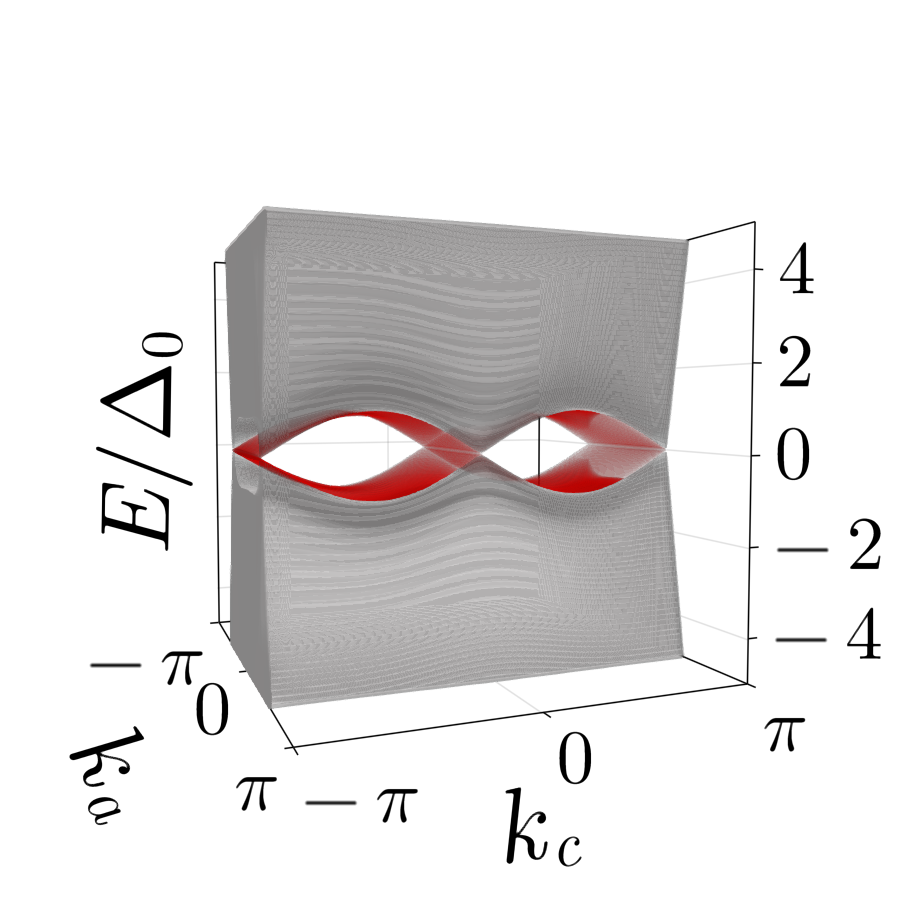}
    \end{minipage}&
    \begin{minipage}[c]{0.45\linewidth}
        \subcaption{}\vspace{-1mm}
        \label{sfig: sdos_B3u_b_sdos}
          \includegraphics[width=\linewidth]{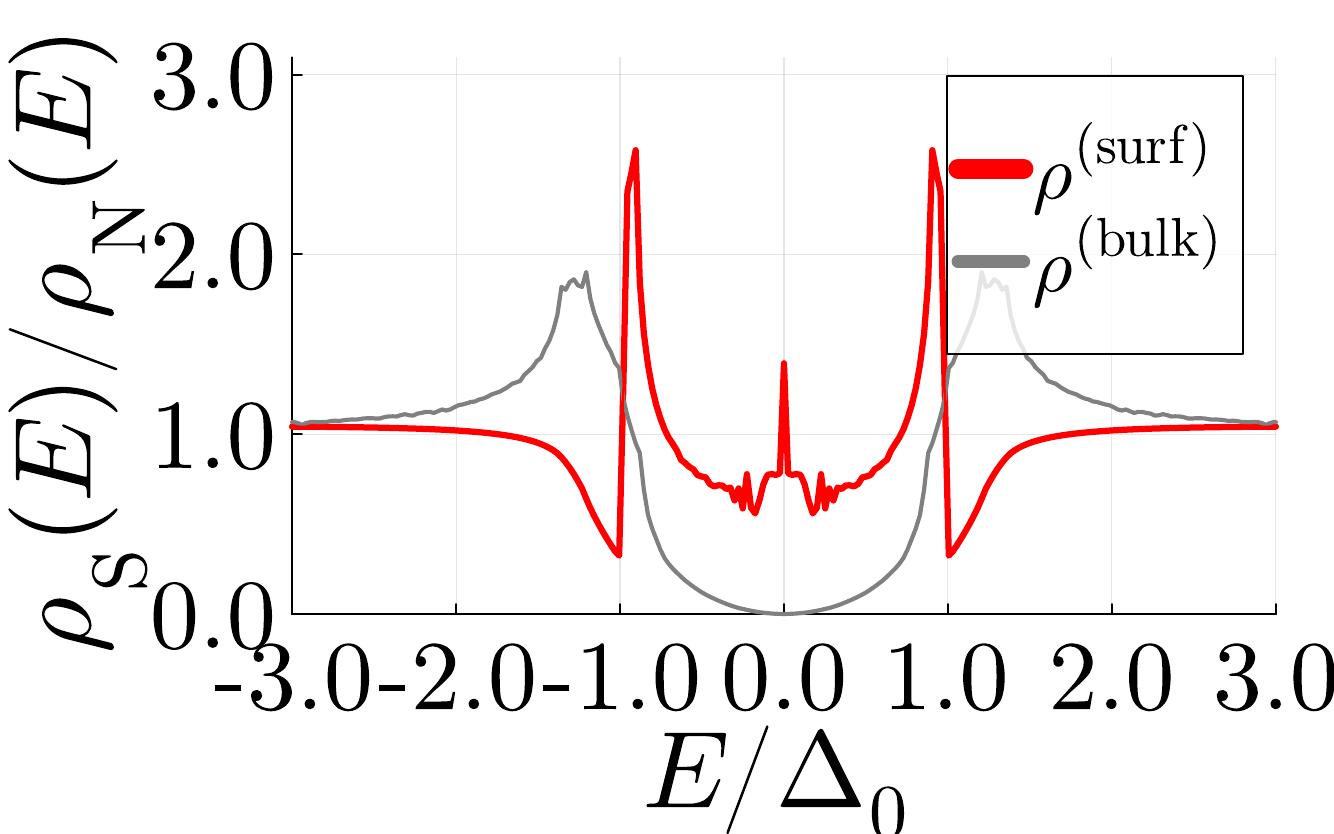}
    \end{minipage}\\
    \rotatebox[origin=c]{0}{(001)}&
    \begin{minipage}[c]{0.35\linewidth}
        \subcaption{}\vspace{-1mm}
        \label{sfig: sdos_B3u_c_dos3d_1}
          \includegraphics[trim={0 20px 0 150px},clip,width=\linewidth]{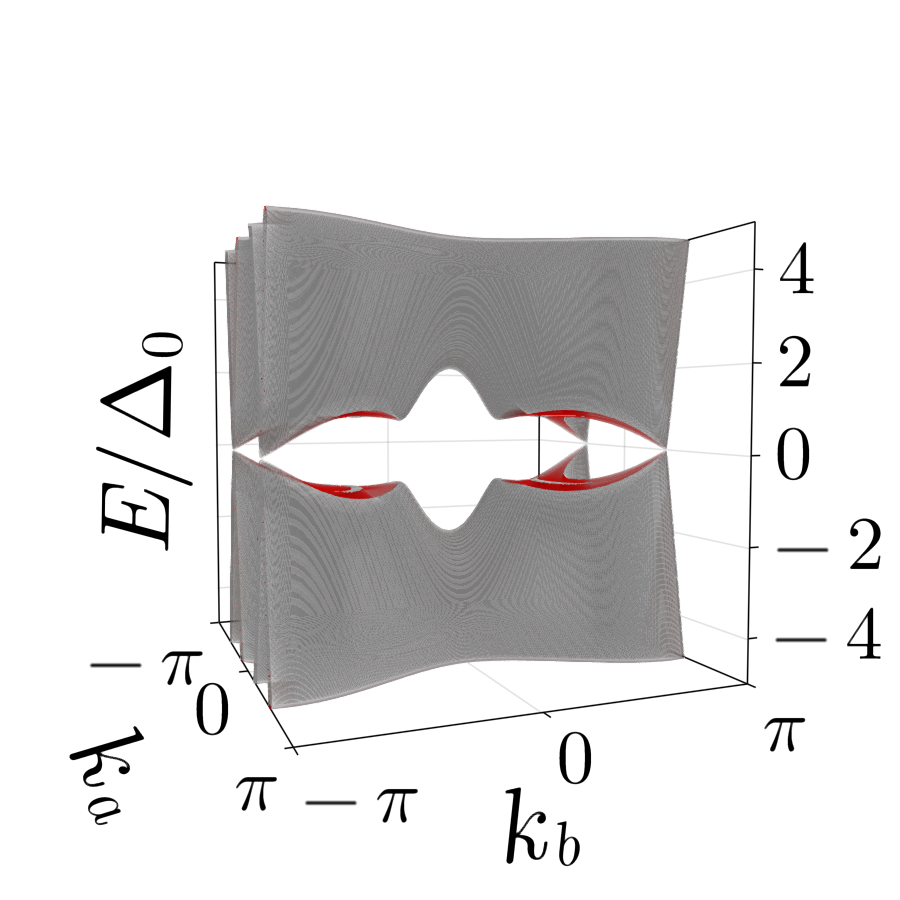}
    \end{minipage}&
    \begin{minipage}[c]{0.45\linewidth}
        \subcaption{}\vspace{-1mm}
        \label{sfig: sdos_B3u_c_sdos}
          \includegraphics[width=\linewidth]{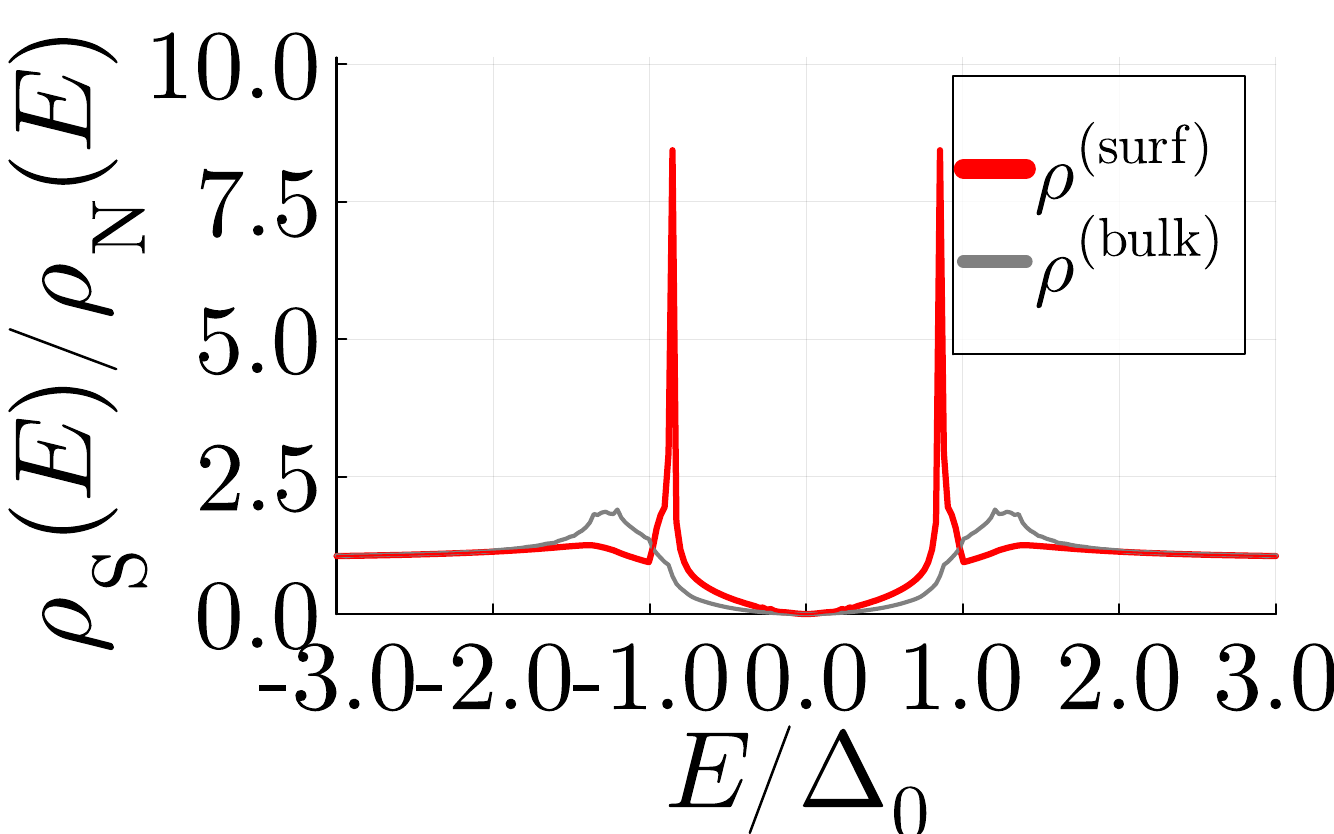}
    \end{minipage}
  \end{tabular}
  \caption{
    Local density of states for the $B_{3u}$ pairing state.
    In all of these figures, the red (grey) color shows the surface (bulk) LDOS. 
    Each row represents the results on the $(100)$, $(010)$ and $(001)$ surface, respectively. 
    The first column represents $\vb*{k}_{∥}$-resolved LDOS by coloring in points with finite $ρ_\mathrm{S}(\vb*{k}_{∥})$ values.
    The bulk $B_{3u}$ state has point nodes, while the ZEP structure appears due to the contribution of the surface flat Fermi arcs in the MSS to the LDOS as shown in (a) and (b).
  }
  \label{fig: sdos_B3u}
\end{figure}

Lastly, we consider the $B_{3u}$ state, where the $\vb*{d}$-vector is given by \cref{eq:d-vector_IR_B3u}.
In this state, there are flat Fermi arc states protected by crystalline symmetry in the (010) and (001) planes~\cite{tei_2023}.
The results of the LDOS on the surfaces are summarized in \cref{fig: sdos_B3u}. 
As shown in \cref{sfig: sdos_B3u_a_dos3d_1,sfig: sdos_B3u_a_sdos}, no surface-localized state appears on the (100) plane.
Therefore, a nodal structure is seen for the surface LDOS and the bulk DOS.

\Cref{sfig: sdos_B3u_b_dos3d_1}, which are results on the (010) planes, show the existence of the surface-localized state and the formation of the flat Fermi arcs on $k_{c} = 0 \mathrm{\; and \;}±π$. 
The flat Fermi arcs are protected by the topological invariant defined by $Γ_{\mathcal{M}_{ab}}$ (see \cref{tb: Majorana_states_and_chiral_operator}). 
The bulk DOS exhibits the nodal structure, whereas the surface LDOS has a ZEP structure in the in-gap region due to the flat Fermi arcs contribution, as shown in \cref{sfig: sdos_B3u_b_sdos}.

\Cref{sfig: sdos_B3u_c_dos3d_1} shows that the surface-localized states exist on the (001) plane. 
However, they are not topologically protected and have no zero energy state.
Therefore, a nodal structure is seen for the surface LDOS and the bulk DOS in \cref{sfig: sdos_B3u_c_sdos}.

In summary, for the $B_{3u}$ pairing state, in the (010) plane, there are flat Fermi arcs, which cause the surface LDOS to have a ZEP structure. 
The height of the ZEP is smaller compared to the $B_{1u}$ case because the flat Fermi arcs are split and partially present.

\subsection{Summary on surface LDOSs}
\label{ssec:summary_on_surface_LDOSs}
In summary, we find that in all cases where MSS is present, the dispersion of the MSSs produces distinct LDOS structures from the bulk, as summarized in \cref{tb: Majorana_states_and_chiral_operator} and shown in \cref{fig: sdos_Au,fig: sdos_B1u,fig: sdos_B2u,fig: sdos_B3u}.
There exist surface Majorana cones on the (100) and (010) planes of the $A_{u}$ pairing state, which cause the surface LDOS to have a V-shaped structure.
For the $B_{1u}$ pairing state, in the (100) and (010) surfaces, there is a flat Fermi arc, which causes the surface LDOS to have a ZEP structure.
The flat Fermi arcs also appear on the (100) and (001) surfaces in the $B_{2u}$ pairing state and (010) surface in the $B_{3u}$ pairing state. 
The intensity of the ZEP is characterized by the length occupied by the flat Fermi arc in the surface Brillouin zone.
However, the height of the V-shaped structure or ZEP is small and the observation of ZBCP in tunneling spectroscopy may not be easy.
Therefore, in the following section, we focus on the local spin susceptibility in order to look for a more qualitative evaluation.

\section{Local spin susceptibility}
\label{sec:local_spin_susceptibility}

In this section, we discuss the correspondence between the anomalous surface magnetic response and the MSS in \ce{UTe2}.
In general, in spin-triplet superconductors and superfluids, there are spin degrees of freedom linked to the direction of the $\vb*{d}$-vector and associated with the spin susceptibility~\cite{sigrist_2005_Introduction}. 
Additionally, the MSS exhibit Ising anisotropy~\cite{chung_2009,nagato_2009,Shindou2010,mizushima_2012}. 
In contrast to the bulk $\vb*{d}$-vector, the MSS gives rise to the paramagnetic response, resulting in the anomalous enhancement of the spin susceptibility on the surface.

\subsection{Majorana Ising spin and paramagnetic response}
\label{ssec:Majorana_Ising_spin_and_paramagnetic_response}

We first show that the Ising spin character of the MSS is a generic consequence of the chiral symmetry~\cref{eq:gamma}~\cite{mizushima_2012,mizushima_2012a,uen13,tsu13,sat14,shi14,miz16}. 
Let $\bm{ψ}≡\qty[ψ_{↑},ψ_{↓},ψ^{†}_{↑},\psi^{†}_{↓}]^\mathrm{t}$ be the electron operator in the particle-hole space, obeying $C\bm{ψ}=\bm{ψ}$. 
Ignoring finite energy eigenstates, the operator is expanded as $\bm{ψ}(\bm{r})=∑_{a}\bm{φ}^{(a)}_0(\bm{r})γ^{(a)}$ in terms of the real operators $γ^{(a)}$, where $a$ labels the zero energy states and $\bm{φ}_0(\bm{r})$ is the zero-energy eigenfunctions of $\mathcal{H}_\mathrm{BdG}$ in Eq.~\eqref{eq:Hbdg}.
When the chiral symmetry is maintained, the zero energy states are simultaneous eigenstates of the chiral operator $Γ_U$, i.e., $Γ_U\bm{φ}_0=λ_{Γ}\bm{φ}^{(a)}_0$ with the eigenvalue $λ_{Γ}=± 1$ 
\cite{Sato2011}. 
The index theorem implies that the winding number $w_\mathrm{1d}=n_{-}-n_{+}$ counts the number of the zero energy states in the sector of $λ_{Γ} = ±1$. 
From the constraint of the chiral symmetry on the field operator, $Γ_U\bm{ψ}=ie^{iϕ_U}\hat{U}Θ C\bm{ψ}=λ_{Γ}\bm{ψ}$, one obtains the following relation, 
\begin{align}
χ(\bm{r})≡ 
\mqty(ψ_{↑} \\ ψ_{↓})
= i e^{iϕ_U}λ_{Γ}{U}
\mqty(ψ^{†}_{↓} \\ -ψ^{†}_{↑}),
\label{eq:majo}
\end{align}
implying that the electron creation directly relates to its annihilation with the unitary matrix ${U}$. 
This is a general consequence of the chiral symmetry and particle-hole symmetry.

Let us now show that the relation in Eq.~\eqref{eq:majo} yields the Ising character of the MSSs. 
The local spin operators, $\bm{S}=\qty(S_x,S_y,S_z)$, are defined in the particle-hole space as $\bm{S}=[ψ_{σ}^{†}(\bm{σ})_{σσ^{\prime}}ψ_{σ^{\prime}}-ψ_{σ}(\bm{σ}^\mathrm{t})_{σσ^{\prime}}ψ^{†}_{σ^{\prime}}]/4$. 
Using the relation in Eq.~\eqref{eq:majo}, one obtains
\begin{align}
S_{μ}(\bm{r})
=\frac{1}{4}e^{-iϕ_U}λ_{Γ}
χ^\mathrm{t}(\bm{r})σ_y
\qty( {U} σ_{μ}-σ_{μ}{U}^{†})
χ(\bm{r}).
\label{eq:ising}
\end{align}
This implies that the spin operator of the MSS becomes identically zero ($S_{μ}=0$) when ${U} σ_{μ}=σ_{μ}{U}^{†}$ is satisfied. 
The unitary operator introduced in Sec.~\ref{sec:symmetry} is expressed as $U=i\bm{σ}⋅\bm{n}$, where $\bm{n}$ is the unit vector pointing to the rotation axis for the $π$ rotation symmetry ($U=C_{μ}$) or the perpendicular direction to the mirror reflection plane for the mirror symmetry ($U=\mathcal{M}$). 
Then, Eq.~\eqref{eq:ising} reduces to $S_{μ}∝ \acomm{σ_{μ}}{\bm{σ}⋅\bm{n}}$, implying the Ising spin of the MSS, ${\bm S} ∥ \hat{\bm{n}}$. 
The spin operator of the MSS points to the rotation axis for the chiral symmetry with $U=\mathcal{C}_{μ}$, while it is locked to the direction perpendicular to the mirror reflection plane for $U=\mathcal{M}_{μν}$. 
Hence, the MSSs can exhibit magnetic response when the magnetic field is applied along the orientation of the Ising spin.

In addition to the Ising spin, the MSS exhibits a negative contribution to the superfluid density and paramagnetic response \cite{Higashitani1997}. 
This is because Majorana zero modes, or more generally zero-energy ABSs are equivalent to odd-frequency pairs~\cite{odd3,odd3b,Tanaka12,
dai12,hig12,asa13,mizushima_2014}. 
In stark contrast to the magnetic response in bulk, the MSSs and the surface ABS exhibit the paramagnetic response to the applied magnetic field \cite{Higashitani1997,Proximityp3,Asano2011}  resulting in an anomalous increase in the surface LSS. 
Therefore, the anomalous surface magnetic response in \ce{UTe2} reflects the Ising-like anisotropy and the paramagnetic response of the MSSs. 
We note that although only the zero energy states are topologically protected, the entire dispersion of the MSSs can hold approximately the Ising-like anisotropic magnetic response.

\subsection{Numerical results}
\label{ssec:LSS_numerical_results}

To calculate the LSS, a lattice Hamiltonian is initially constructed as follows:
\begin{multline}
  H^\mathrm{lattice}_\mathrm{BdG}(\vb*{k}_∥)
  = ∑_{x_{⟂,i},σ,σ^{\prime}}\bigg( \hat{c}^{†}_{\vb*{k}_∥,x_{⟂},σ} H^{∥}(\vb*{k}_∥)\hat{c}_{\vb*{k}_∥,x_{⟂},σ^{\prime}} \\
  +\hat{c}^{†}_{\vb*{k}_∥,x_{⟂},σ} T^\mathrm{NN}(\vb*{k}_∥)\hat{c}_{\vb*{k}_∥,x_{⟂}+1,σ^{\prime}} \\
  +\hat{c}^{†}_{\vb*{k}_∥,x_{⟂},σ} T^\mathrm{NNN}(\vb*{k}_∥)\hat{c}_{\vb*{k}_∥,x_{⟂}+2,σ^{\prime}} 
  +\mathrm{H.c.} \bigg),
  \label{eq:lattice_hamiltonian}
\end{multline}
where $\hat{c}_{\vb*{k}_∥,x_{⟂},σ} = \qty(c_{\vb*{k}_∥,x_{⟂},σ}, c_{-\vb*{k}_∥,x_{⟂},σ}^{†})^\mathrm{t}$, $H^{∥}$ is the on-site BdG Hamiltonian, $T^\mathrm{(N)NN}$ is the (next-)nearest-neighbor hopping term, and $α$ is a label of each eigenstate.
Let $E^{α}_{\bm{k}_{∥}}$ and $E^{β}_{\bm{k}_{∥}}$  be the eigenenergy of $H^\mathrm{lattice}_\mathrm{BdG}(\vb*{k}_∥)$ and $\ket{α_{\bm{k}_{∥}}}$ and $\ket{β_{\bm{k}_{∥}}}$ be their eigenvectors. 
In the linear response theory, the LSS is calculated using the formula~\cite{mah_00,hiranuma_2021}:
\begin{multline}
  χ_{μ\mu}^{(x_{⟂})}(T)
  = g² μ_\mathrm{B}² ∑_{\vb*{k}_∥} ∑_{α,β} \frac{n_\mathrm{F}(E^{β}_{\vb*{k}_{∥}+δ\vb*{q}})-n_\mathrm{F}(E^{α}_{\vb*{k}_∥})}{E^{β}_{\vb*{k}_∥+δ\vb*{q}}-E^{α}_{\vb*{k}_∥} + iδ_{ϵ}}\\
    × \Tr\Bigg[
  \ket{β_{\vb*{k}_∥+δ\vb*{q}}}\bra{β_{\vb*{k}_∥+δ\vb*{q}}}S_{μ}\ket{α_{\vb*{k}_∥}}\bra{α_{\vb*{k}_∥}}S_{μ, x_{⟂}}
  \Bigg],
\end{multline}
where $n_\mathrm{F}(E)$ is the Fermi distribution function, $S_{μ,x_{⟂}}$ is the site-dependent total angular momentum operator of the \textit{f}-orbital $1/2$-electron in \ce{U}, $S_{μ}$ is the full site total angular momentum operator.  
The temperature dependence of the superconducting pair potential is assumed for the BCS case: $Δ_0(T)=Δ_0\tanh(1.74\sqrt{T_\mathrm{c}/T-1})$ where $T_\mathrm{c}$ is the transition temperature. 
The system size $L$ along $x_{⟂}$ is set as $L = 63$ for the calculations, with $x_{⟂}=1,63$ for the surface and $x_{⟂}=32$ for the bulk.

\begin{figure*}[htbp]
  \begin{tabular}{ccccc}
     & \scalebox{1.5}{(100)} & \scalebox{1.5}{(010)} & \scalebox{1.5}{(001)} & \scalebox{1.5}{bulk} \\
    \scalebox{1.5}{\fbox{$A_{u}$}} &
    \begin{minipage}[c]{0.22\linewidth}
      \subcaption{}\vspace{-1mm}
      \label{sfig: LSS_total_Au_a}
      \includegraphics[width=\linewidth]{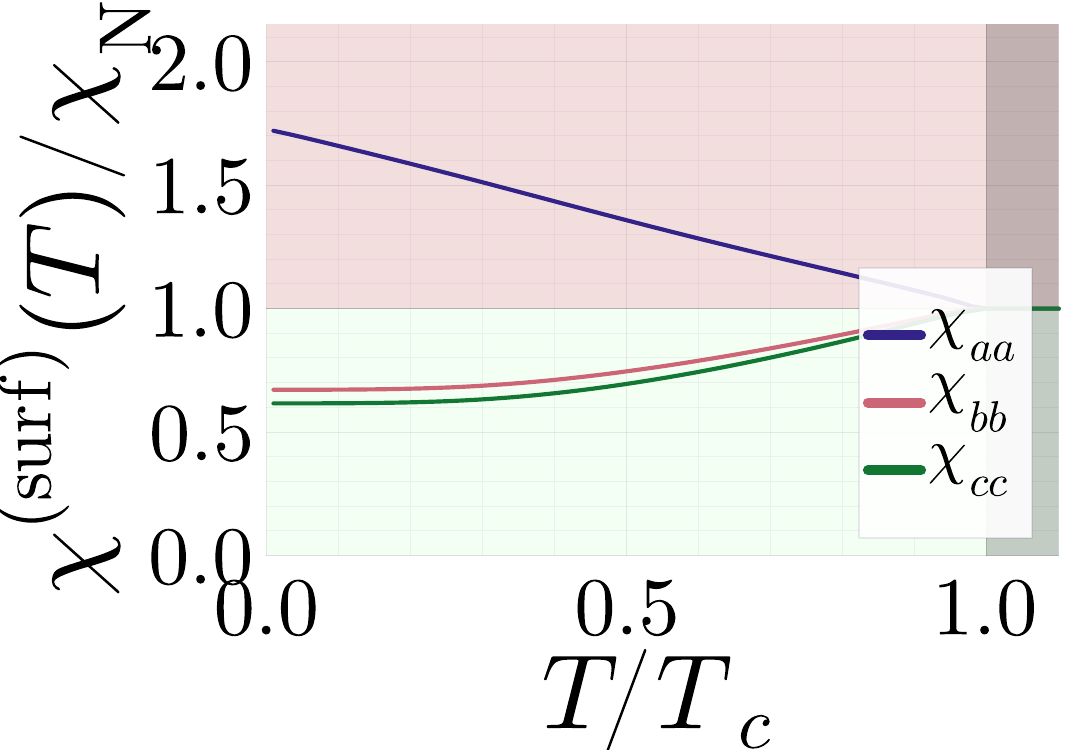}
    \end{minipage}&
    \begin{minipage}[c]{0.22\linewidth}
      \subcaption{}\vspace{-1mm}
      \includegraphics[width=\linewidth]{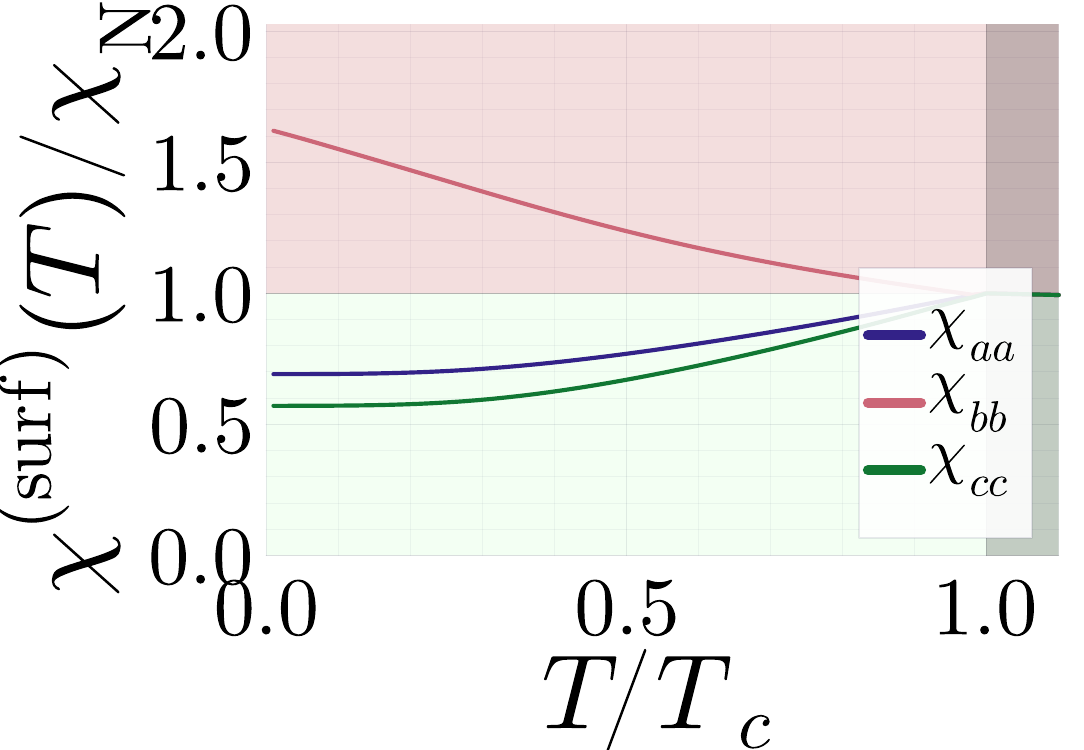}
      \label{sfig: LSS_total_Au_b}
    \end{minipage}&
    \begin{minipage}[c]{0.22\linewidth}
      \subcaption{}\vspace{-1mm}
      \label{sfig: LSS_total_Au_c}
      \includegraphics[width=\linewidth]{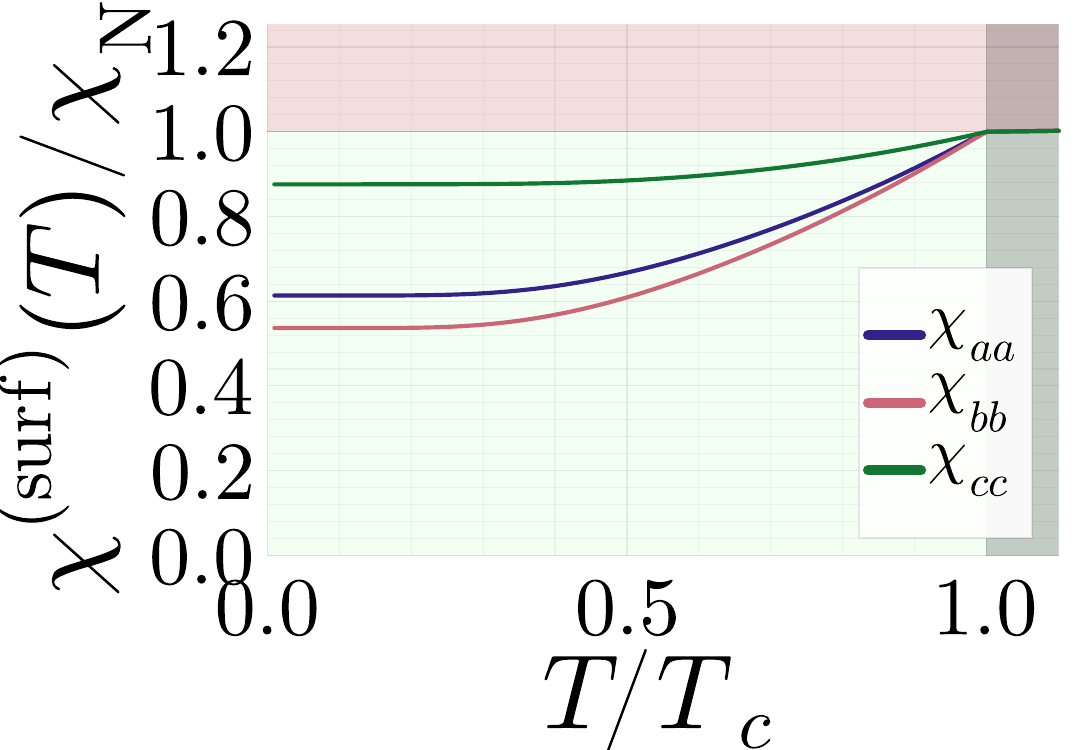}
    \end{minipage}& 
    \begin{minipage}[c]{0.22\linewidth}
      \subcaption{}\vspace{-1mm}
      \label{sfig: LSS_total_Au_bulk}
      \includegraphics[width=\linewidth]{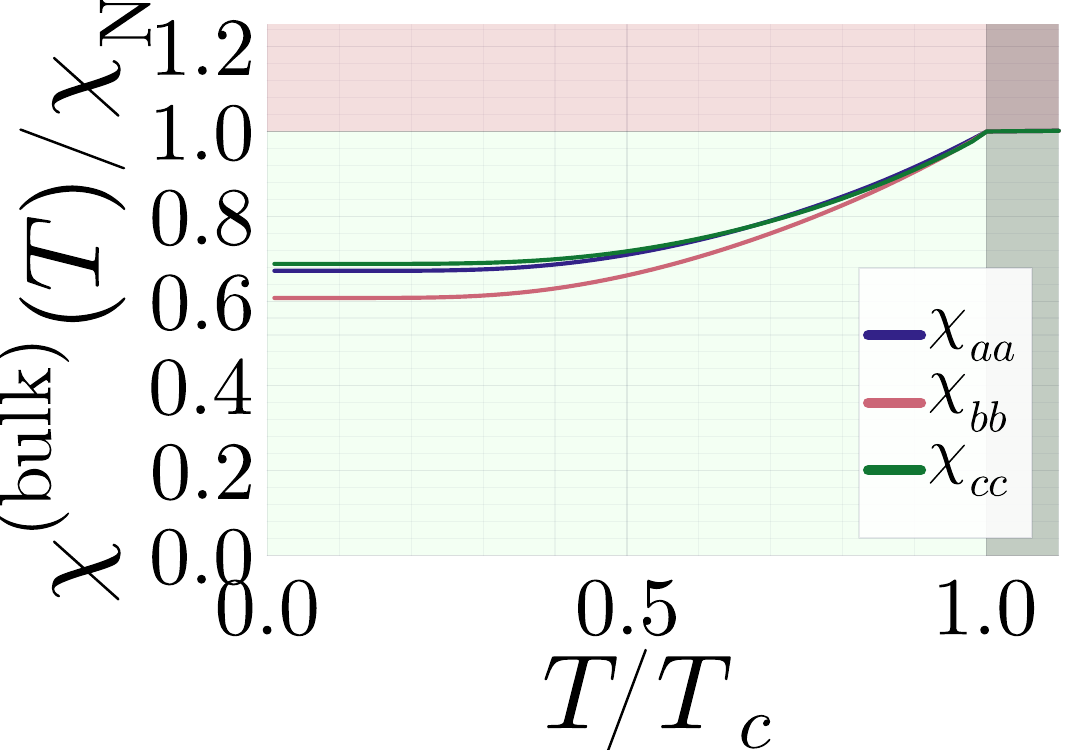}
    \end{minipage} \\

    \scalebox{1.5}{\fbox{$B_{1u}$}} &
    \begin{minipage}[c]{0.22\linewidth}
      \subcaption{}\vspace{-1mm}
      \label{sfig: LSS_total_B1u_a}
      \includegraphics[width=\linewidth]{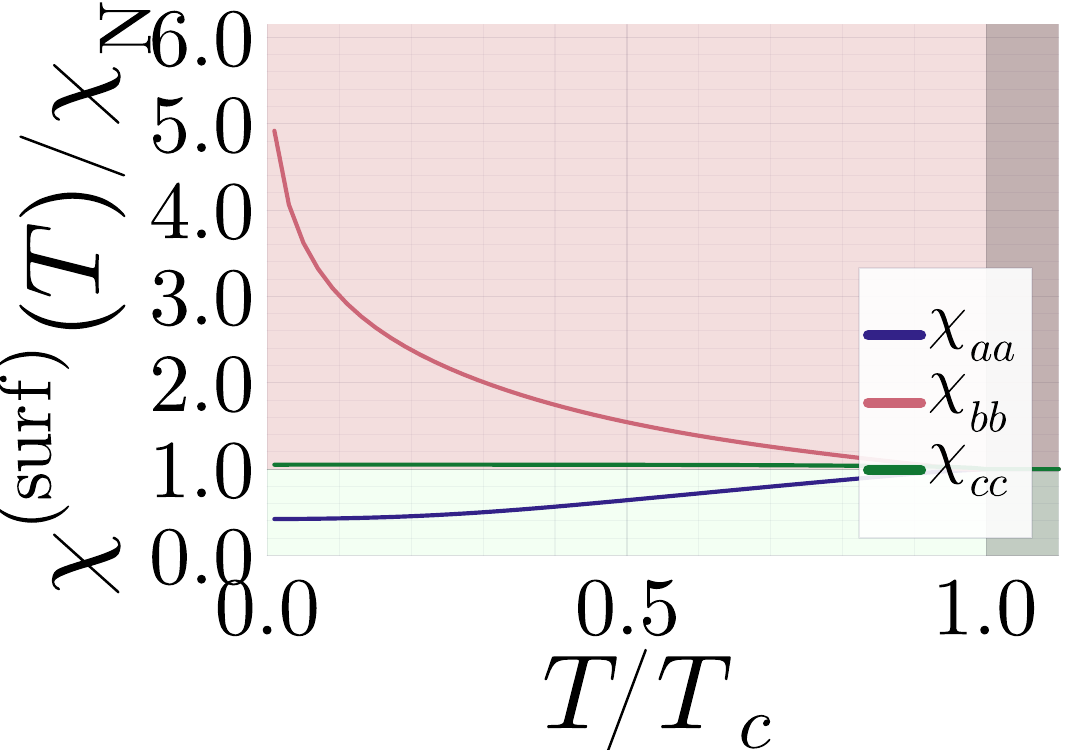}
    \end{minipage}&
    \begin{minipage}[c]{0.22\linewidth}
      \subcaption{}\vspace{-1mm}
      \label{sfig: LSS_total_B1u_b}
      \includegraphics[width=\linewidth]{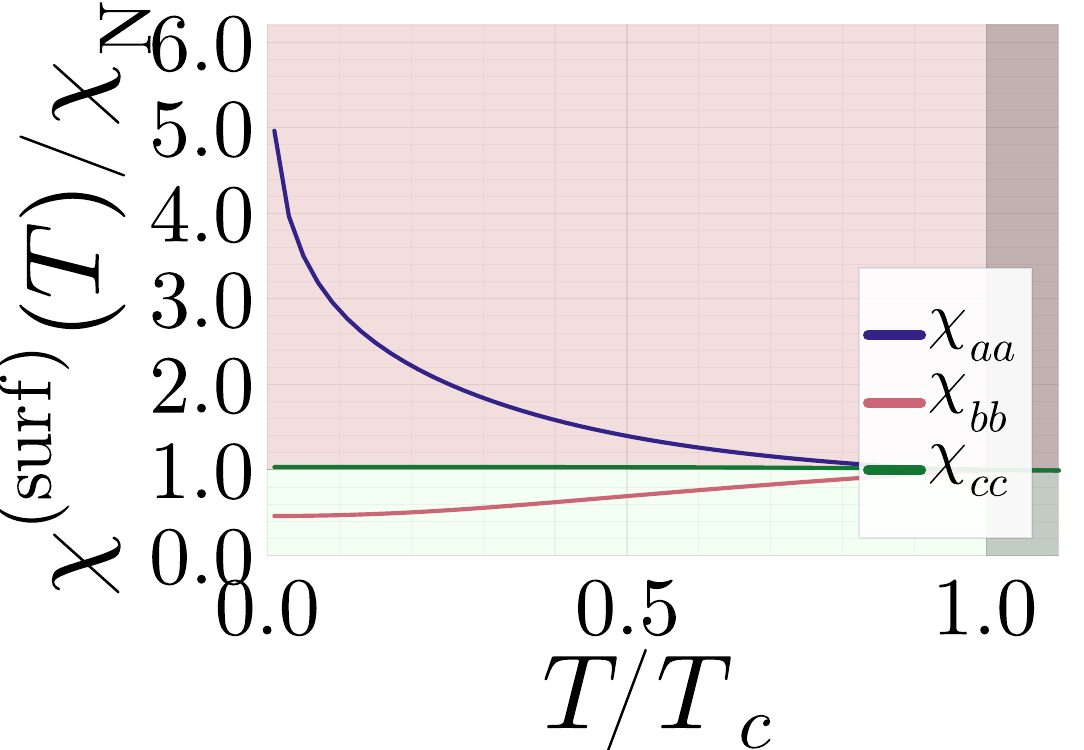}
    \end{minipage}&
    \begin{minipage}[c]{0.22\linewidth}
      \subcaption{}\vspace{-1mm}
      \label{sfig: LSS_total_B1u_c}
      \includegraphics[width=\linewidth]{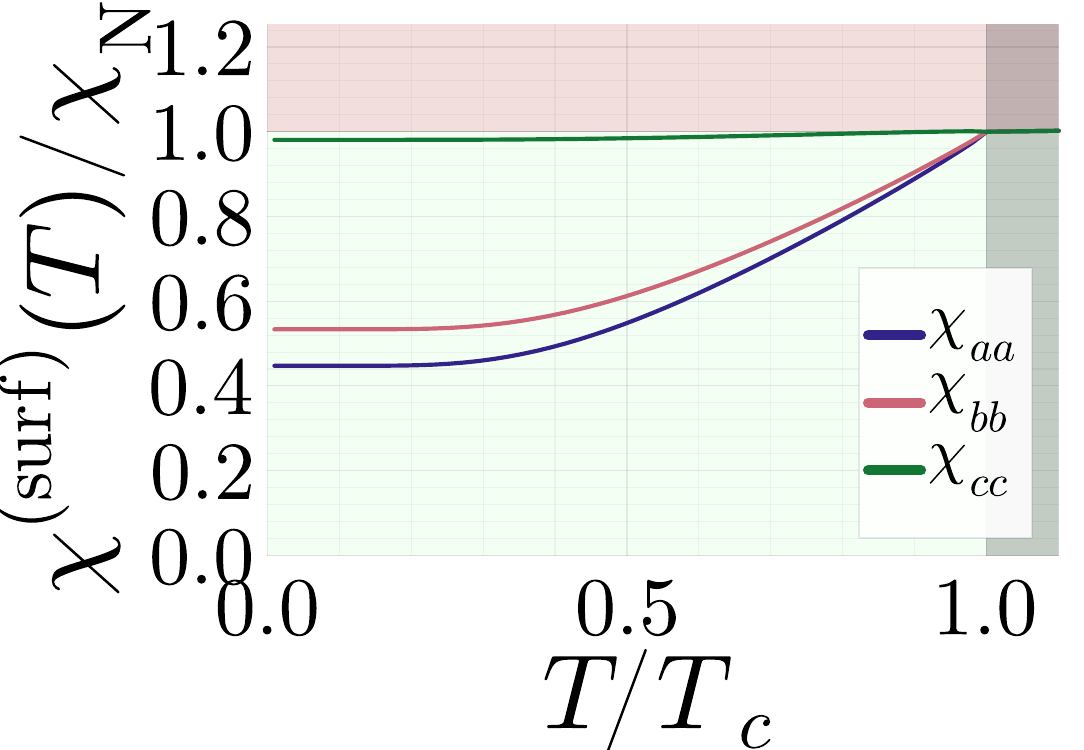}
    \end{minipage}& 
    \begin{minipage}[c]{0.22\linewidth}
      \subcaption{}\vspace{-1mm}
      \label{sfig: LSS_total_B1u_bulk}
      \includegraphics[width=\linewidth]{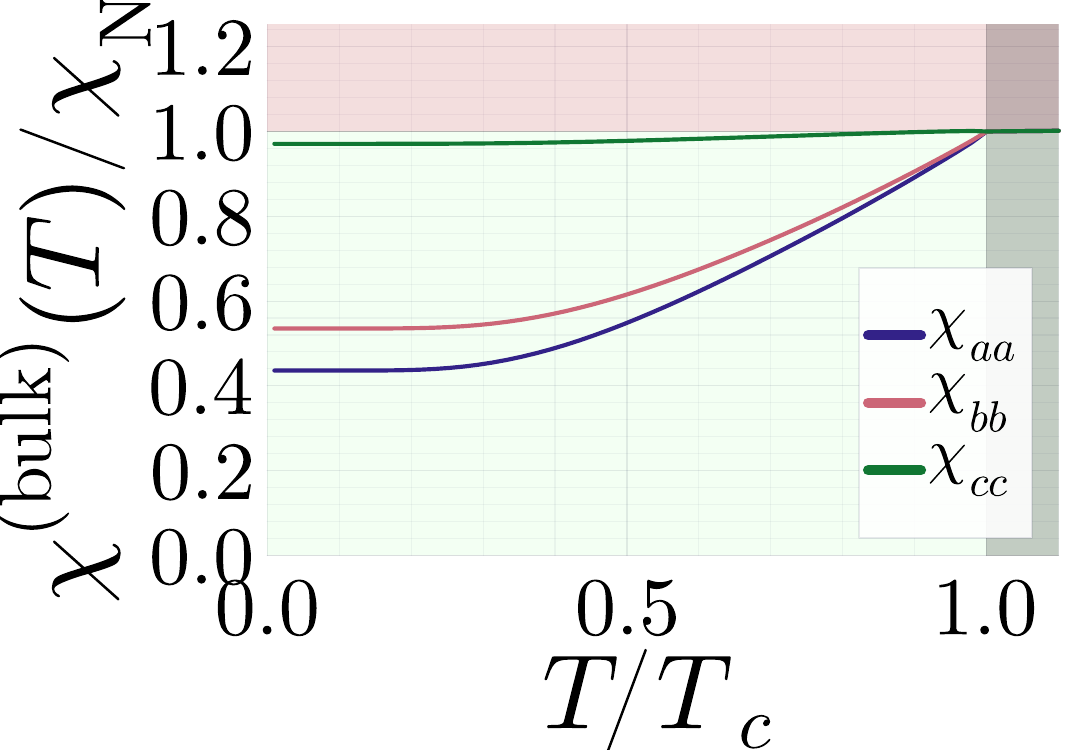}
    \end{minipage} \\

    \scalebox{1.5}{\fbox{$B_{2u}$}} &
    \begin{minipage}[c]{0.22\linewidth}
      \subcaption{}\vspace{-1mm}
      \label{sfig: LSS_total_B2u_a}
      \includegraphics[width=\linewidth]{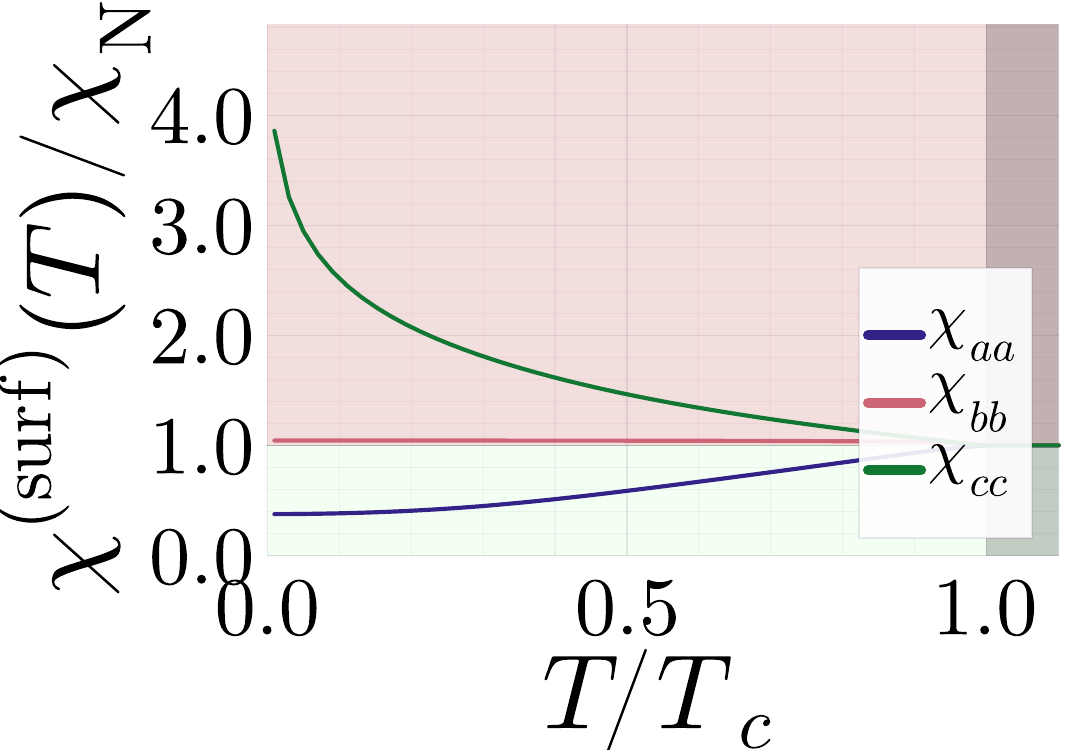}
    \end{minipage}&
    \begin{minipage}[c]{0.22\linewidth}
      \subcaption{}\vspace{-1mm}
      \label{sfig: LSS_total_B2u_b}
      \includegraphics[width=\linewidth]{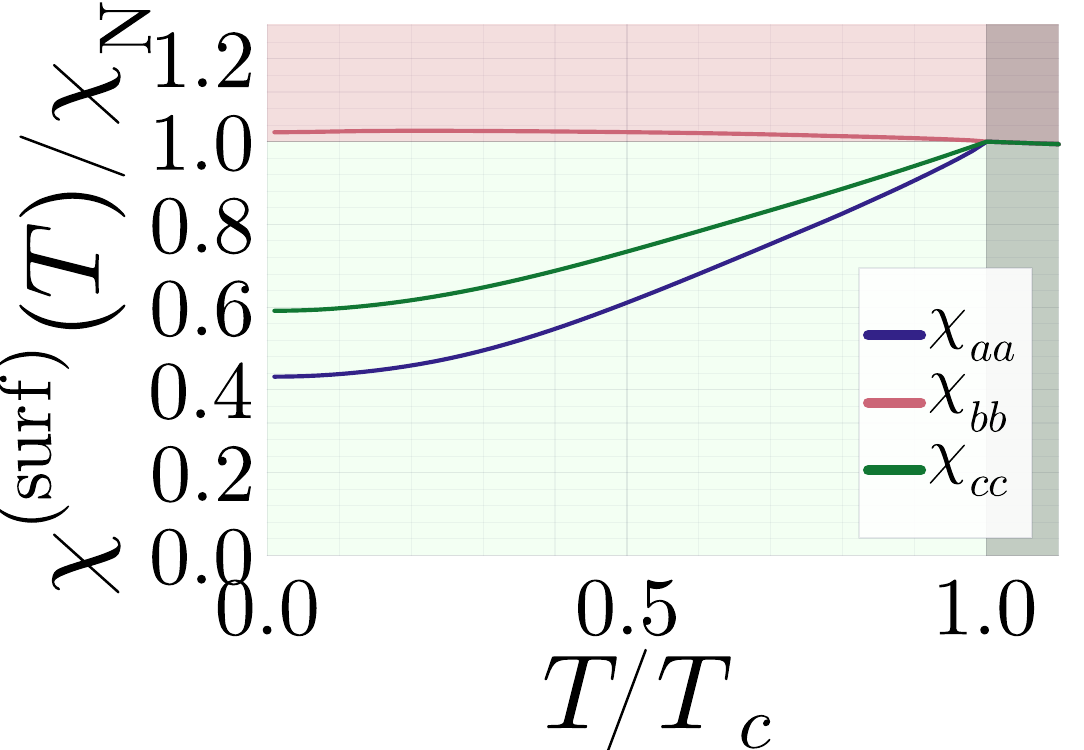}
    \end{minipage}&
    \begin{minipage}[c]{0.22\linewidth}
      \subcaption{}\vspace{-1mm}
      \label{sfig: LSS_total_B2u_c}
      \includegraphics[width=\linewidth]{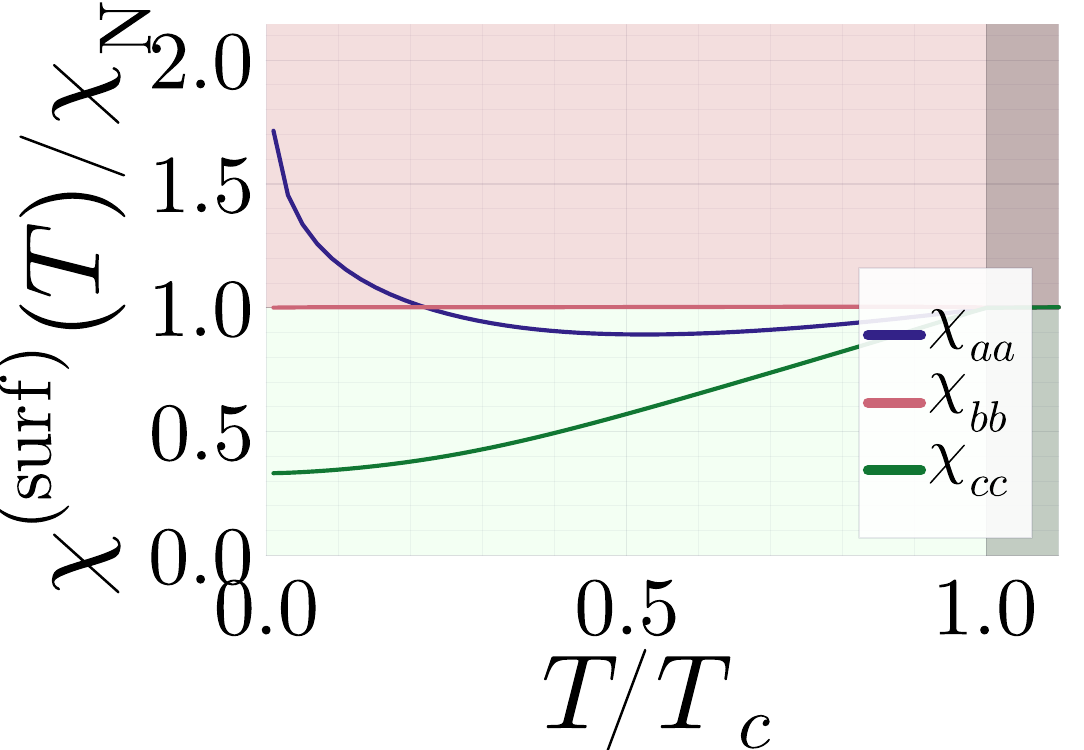}
    \end{minipage}& 
    \begin{minipage}[c]{0.22\linewidth}
      \subcaption{}\vspace{-1mm}
      \label{sfig: LSS_total_B2u_bulk}
      \includegraphics[width=\linewidth]{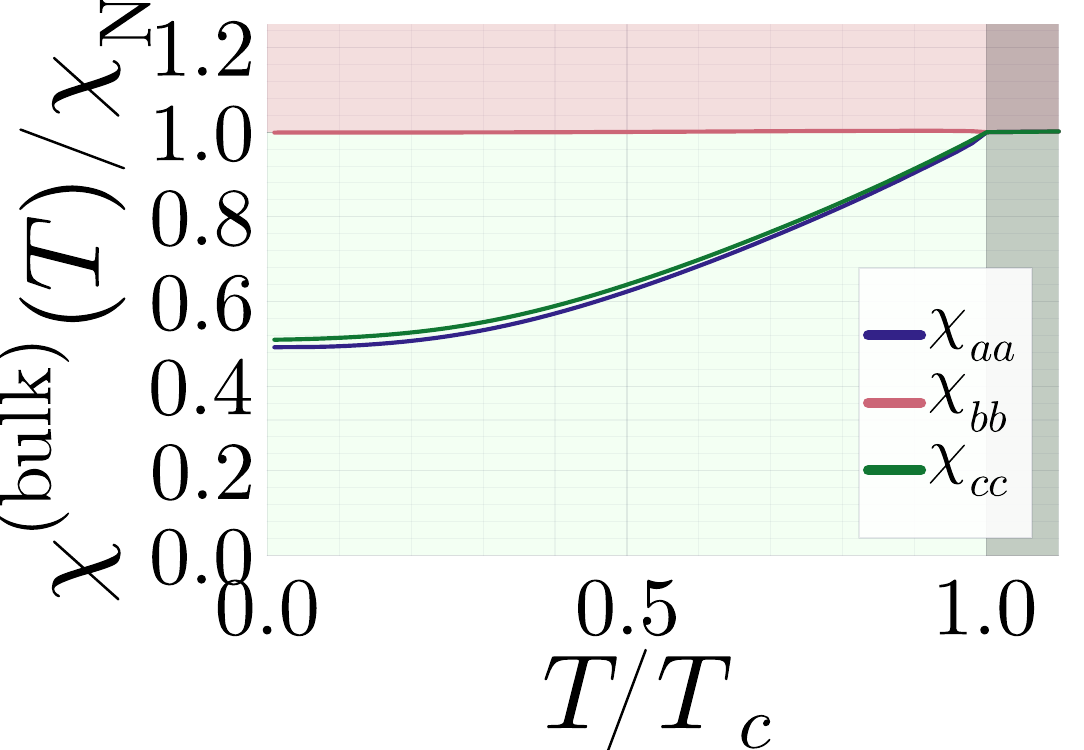}
    \end{minipage} \\

    \scalebox{1.5}{\fbox{$B_{3u}$}} &
    \begin{minipage}[c]{0.22\linewidth}
      \subcaption{}\vspace{-1mm}
      \label{sfig: LSS_total_B3u_a}
      \includegraphics[width=\linewidth]{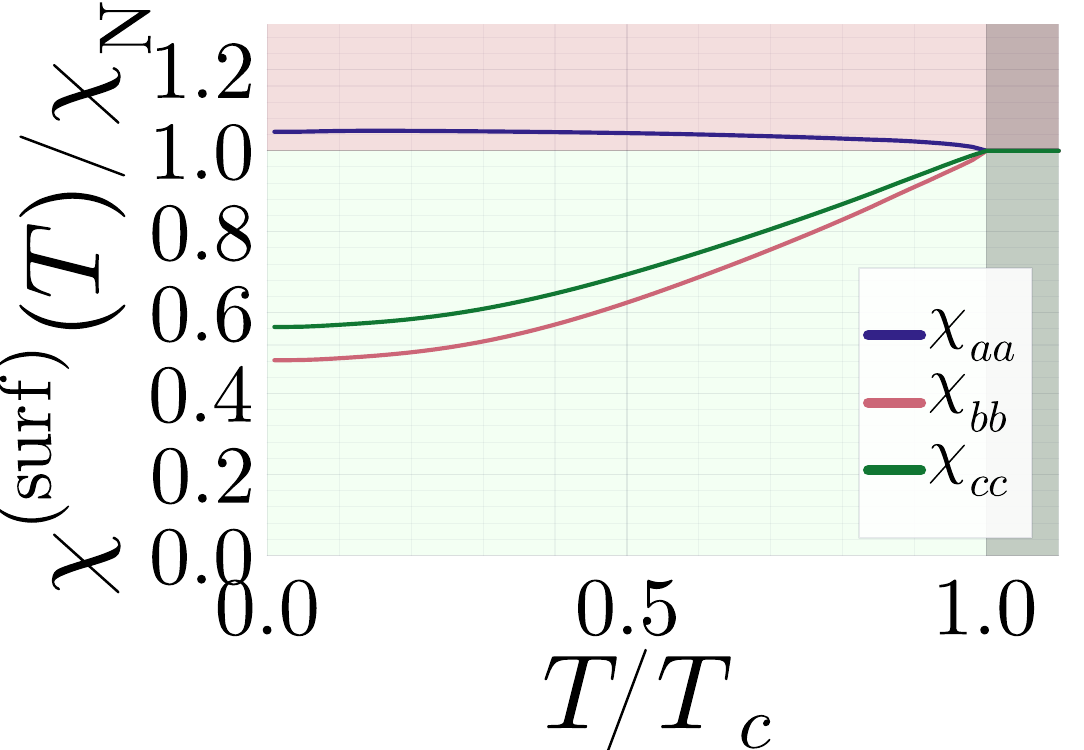}
    \end{minipage}&
    \begin{minipage}[c]{0.22\linewidth}
      \subcaption{}\vspace{-1mm}
      \label{sfig: LSS_total_B3u_b}
      \includegraphics[width=\linewidth]{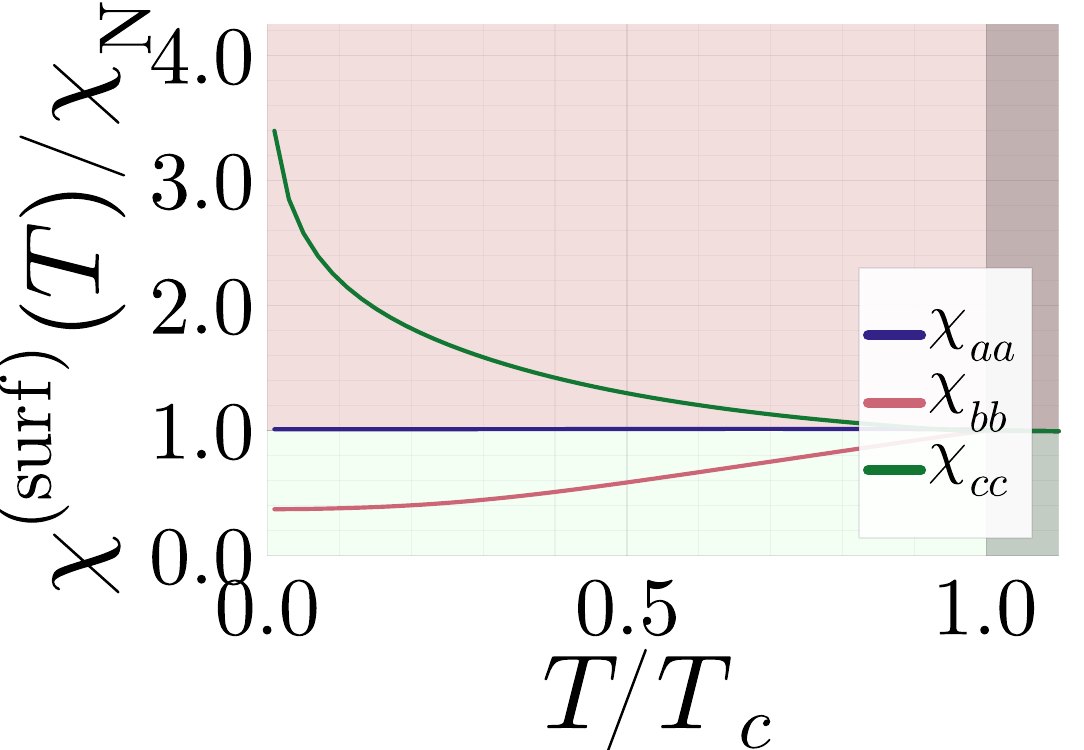}
    \end{minipage}&
    \begin{minipage}[c]{0.22\linewidth}
      \subcaption{}\vspace{-1mm}
      \label{sfig: LSS_total_B3u_c}
      \includegraphics[width=\linewidth]{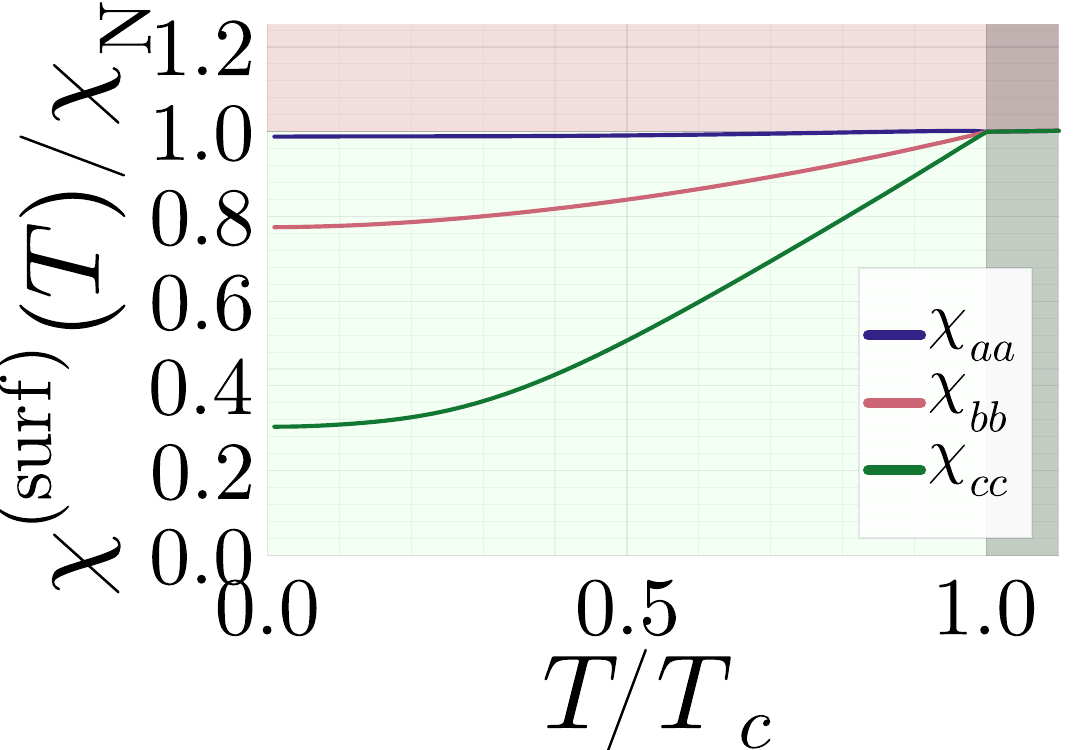}
    \end{minipage}& 
    \begin{minipage}[c]{0.22\linewidth}
      \subcaption{}\vspace{-1mm}
      \label{sfig: LSS_total_B3u_bulk}
      \includegraphics[width=\linewidth]{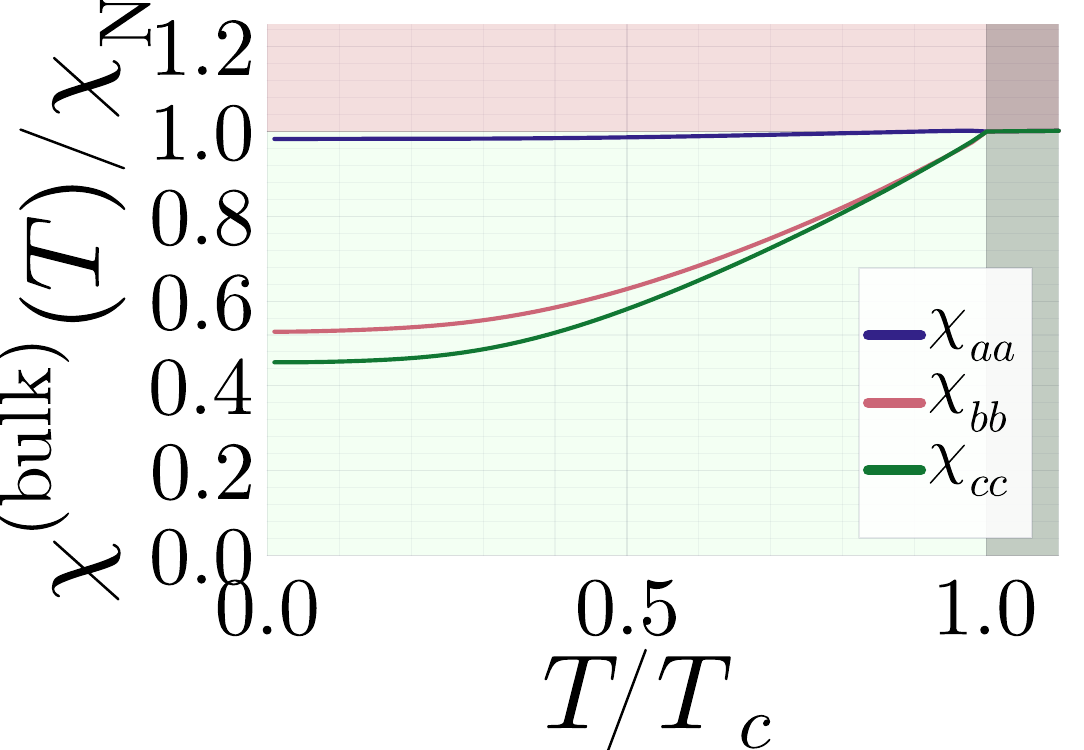}
    \end{minipage}
  \end{tabular}
  \caption{
    Summary of the LSS $χ_{μ\mu}$.
    (a-d), (e-h), (i-l), and (m-p) show the $A_{u}$, $B_{1u}$, $B_{2u}$ and $B_{3u}$ IR cases, respectively.
    The first, second, and third columns show the$(100)$, $(010)$ and $(001)$ surfaces, respectively. 
    The LSSs in bulk are shown in the fourth column.
    The red-shaded region corresponds to the enhancement of the LSS due to the Majorana states.
    }
  \label{fig: LSS_total}
\end{figure*}

\textit{Overview}:
\Cref{fig: LSS_total} and \cref{tb:surface_SS} summarize the LSS of \ce{UTe2}.
For each IR, we clarify the relations between anomalous enhancement of the surface LSS and surface ABS.
In the following subsections, we will evaluate the results in detail for each of the IRs.

For the $A_u$ pairing state, the surface LDOS exhibits a V-shaped structure due to the presence of surface Majorana cones. 
This Majorana state exhibits Ising-like anisotropy in the magnetic response. 
We observe a strong enhancement of the surface LSS in the direction where the applied magnetic field breaks the chiral symmetry protecting the Majorana state.
In the $B_{1u}$ pairing state, the surface LDOS shows a ZEP structure due to the flat Fermi arc surface states. 
Similar to the $A_u$ case, the flat Fermi arc state also exhibits Ising-like anisotropy in the magnetic response. 
An anomalous enhancement of the surface LSS occurs in the direction where the protecting symmetry of the flat Fermi arc is broken.
For the $B_{2u}$ and $B_{3u}$ pairing states, the bulk has nodal structures, but the surface LDOS still exhibits ZEPs due to the presence of flat Fermi arcs. 
Again, we observe an anisotropic enhancement of the surface LSS correlated with the directions where the Fermi arc states are protected.

In the following subsections, we will provide a detailed evaluation of the results on the LSS for each IR, focusing on the connections between the surface ABS, the surface anomalous magnetic response, and the underlying topological properties.

\begin{table}[htbp]
  \centering
  \caption{
    Summary of correspondence between surface LSS, chiral operator, and anomalous LSS enhancement of \ce{UTe2}.
    }
  \label{tb:surface_SS}
\begin{tabular}{ccccc}
  \toprule
  IR & \makecell{surface}  & surface ABS  & \makecell{chiral\\ operator} & \makecell{enhanced\\ surface LSS}  \\
  \midrule\midrule
  \multirow{3}{*}{$A_{u}$} & (100) & Majorana cones & $Γ_{\mathcal{C}_{a}}$ & $χ_{aa}$  \\
   & (010) & Majorana cones & $Γ_{\mathcal{C}_{b}}$ & $χ_{bb}$ \\
   & (001) & -- & -- & -- \\
  \midrule
  \multirow{3}{*}{$B_{1u}$} &(100) & Flat Fermi arc & $Γ_{\mathcal{M}_{ca}}$ & $χ_{bb}$  \\
   & (010) & Flat Fermi arc & $Γ_{\mathcal{M}_{bc}}$ & $χ_{aa}$ \\
   & (001) & -- & -- & -- \\
  \midrule
  \multirow{3}{*}{$B_{2u}$} & (100) & Flat Fermi arcs & $Γ_{\mathcal{M}_{ab}}$ & $χ_{cc}$ \\
   & (010) & -- & -- & -- \\
   & (001) & Flat Fermi arcs & $Γ_{\mathcal{M}_{bc}}$ & $χ_{aa}$ \\
  \midrule
  \multirow{3}{*}{$B_{3u}$} & (100) & -- & -- & -- \\
   & (010) & Flat Fermi arcs  & $Γ_{\mathcal{M}_{ab}}$ & $χ_{cc}$  \\
   & (001) & -- & -- & -- \\
  \bottomrule
\end{tabular}
\end{table}

\subsubsection{\texorpdfstring{$A_{u}$}{Au} pairing state}
\label{ssec:SS_Au}

\begin{figure}[htbp]
  \begin{tabular}{ccc}
    \multicolumn{3}{l}{\scalebox{1.0}{\fbox{$A_{u}$}}} \\
    (100) & (010) & (001) \\
    \begin{minipage}[c]{0.3\linewidth}
        \subcaption{}\vspace{-1mm}
        \label{sfig: LSS_site_Au_a_aa}
          \includegraphics[width=\linewidth]{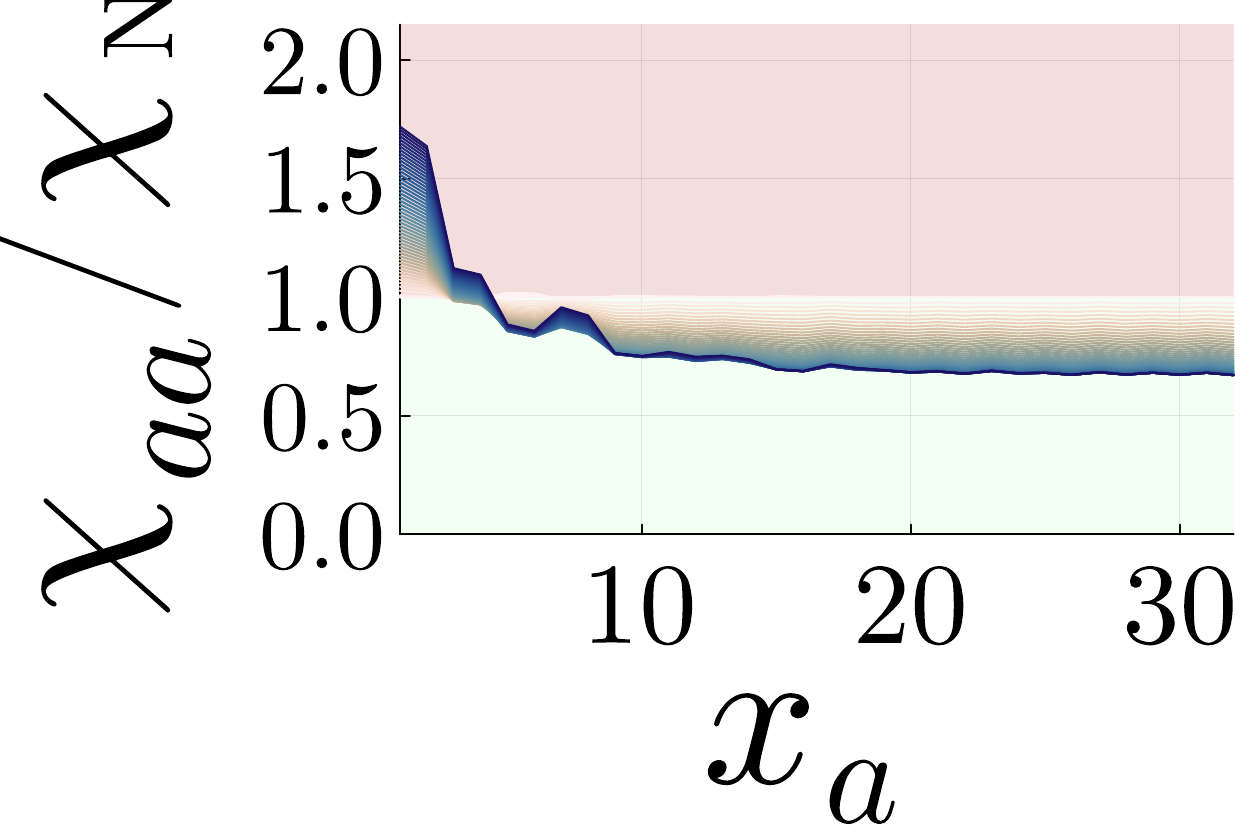}
    \end{minipage}&
    \begin{minipage}[c]{0.30\linewidth}
        \subcaption{}\vspace{-1mm}
        \label{sfig: LSS_site_Au_b_aa}
          \includegraphics[width=\linewidth]{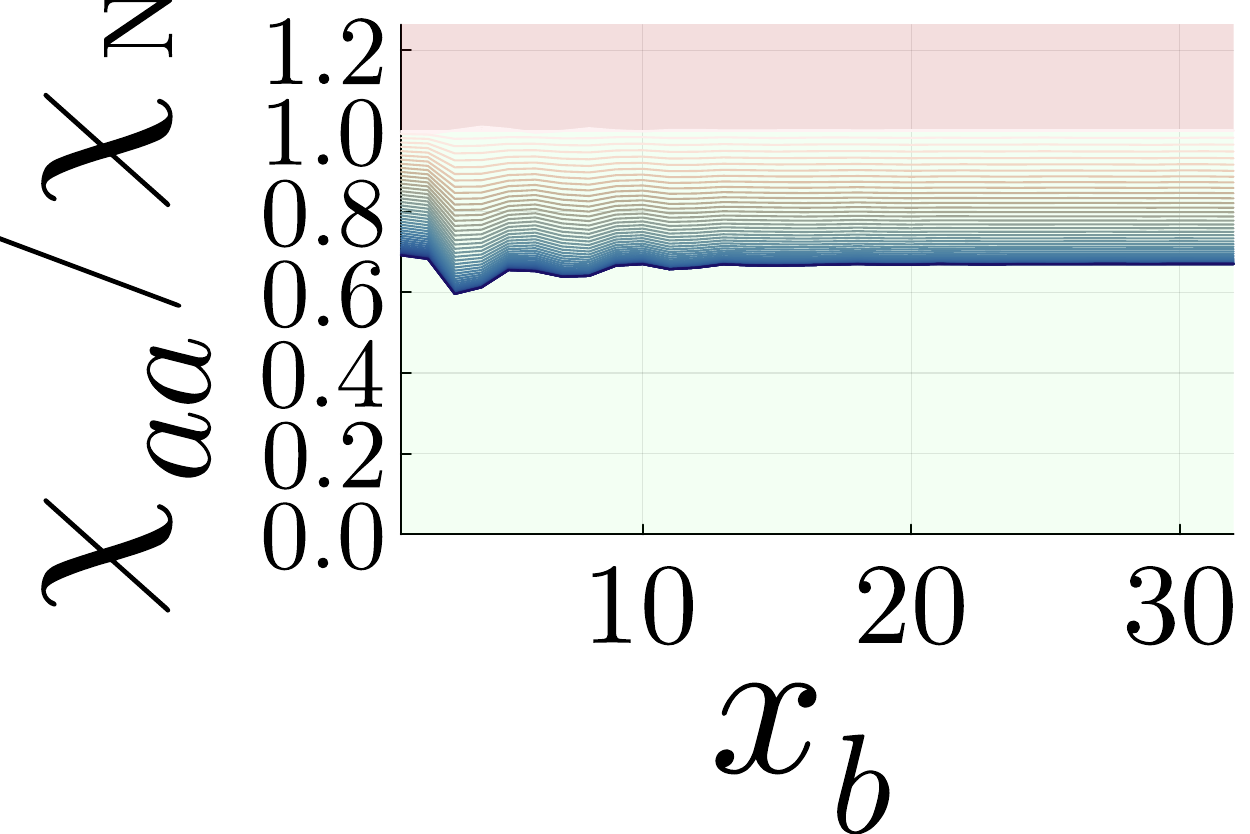}
    \end{minipage}&
    \begin{minipage}[c]{0.30\linewidth}
        \subcaption{}\vspace{-1mm}
        \label{sfig: LSS_site_Au_c_aa}
          \includegraphics[width=\linewidth]{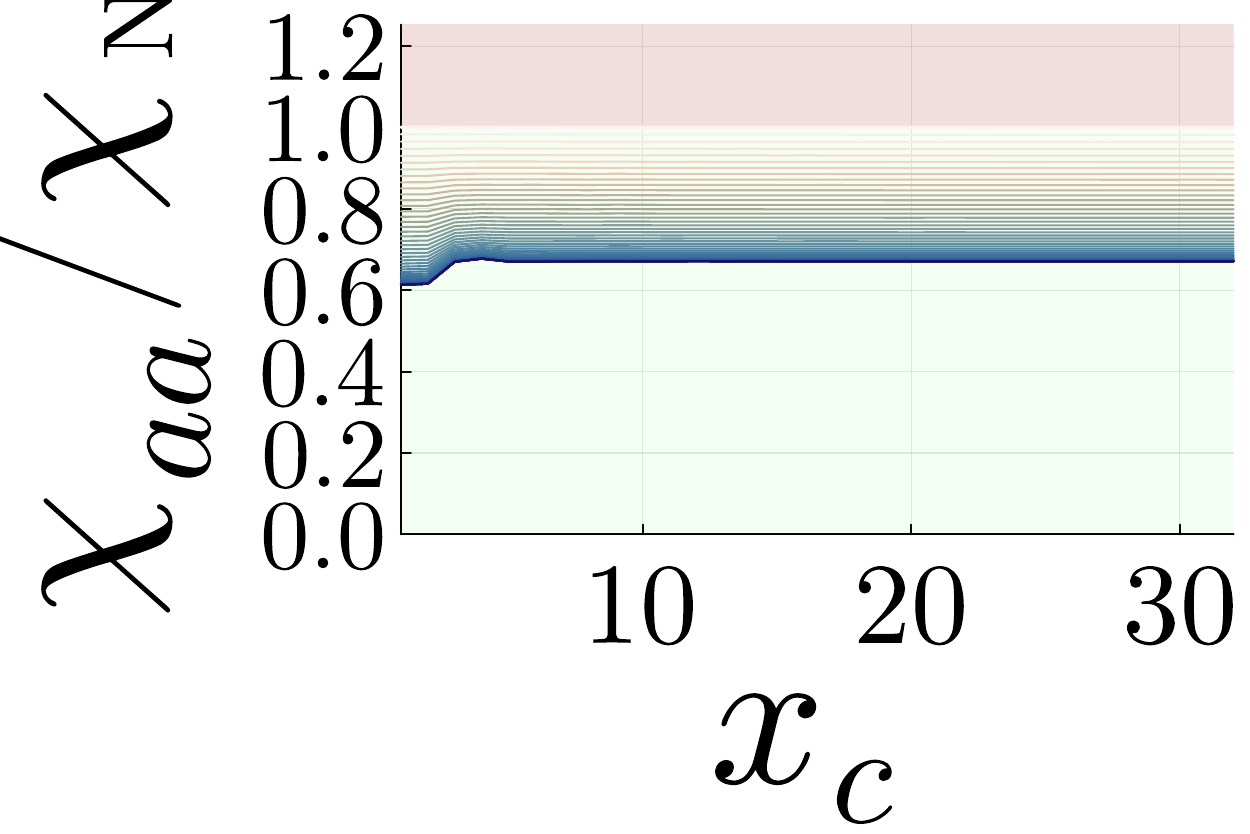}
    \end{minipage} \\ 
    \begin{minipage}[c]{0.30\linewidth}
        \subcaption{}\vspace{-1mm}
        \label{sfig: LSS_site_Au_a_bb}
          \includegraphics[width=\linewidth]{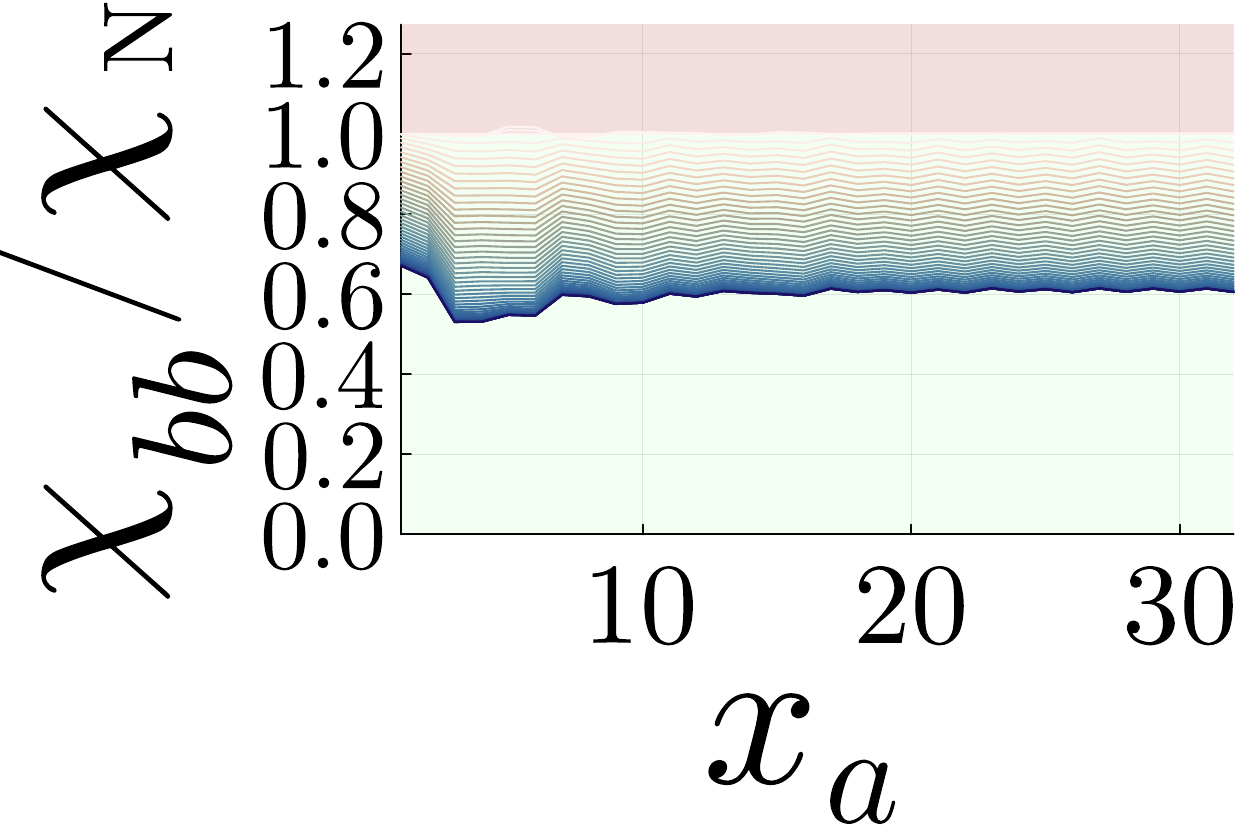}
    \end{minipage}&
    \begin{minipage}[c]{0.30\linewidth}
        \subcaption{}\vspace{-1mm}
        \label{sfig: LSS_site_Au_b_bb}
          \includegraphics[width=\linewidth]{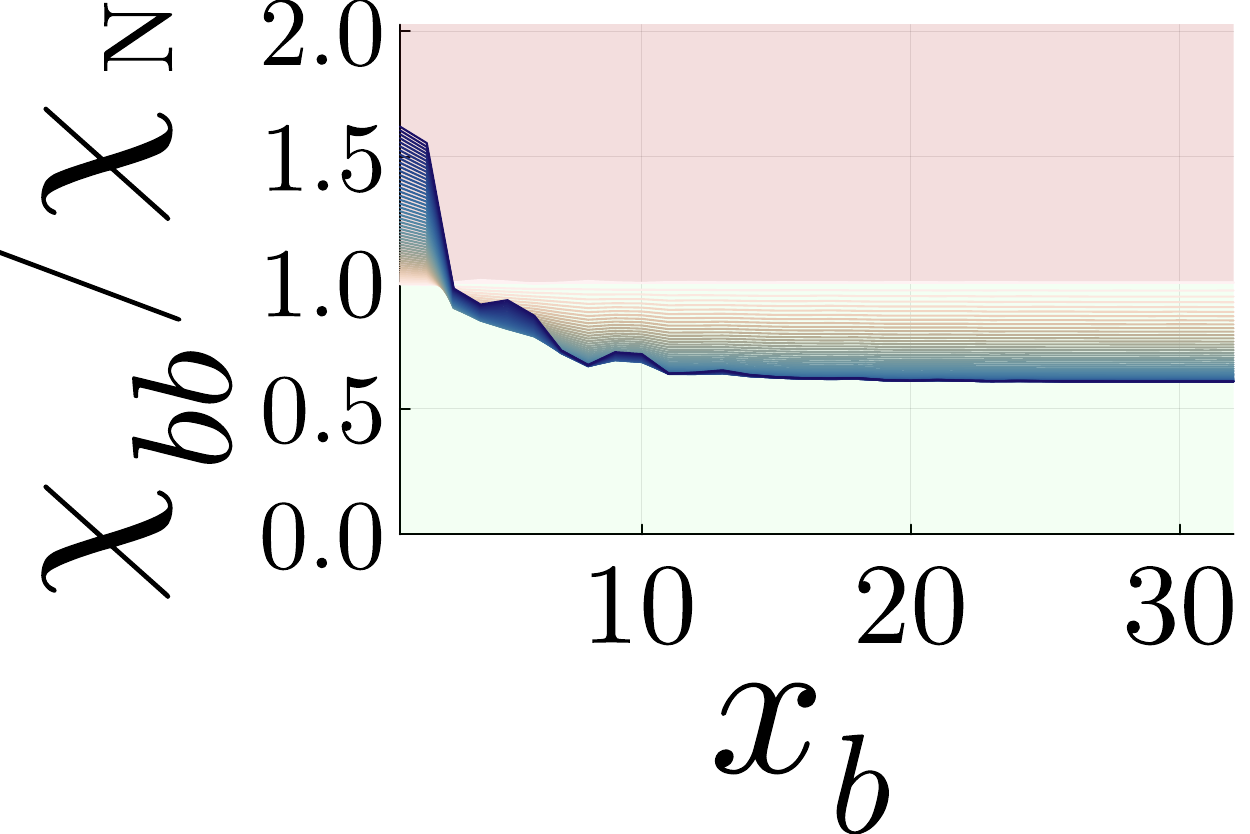}
    \end{minipage}&
    \begin{minipage}[c]{0.30\linewidth}
        \subcaption{}\vspace{-1mm}
        \label{sfig: LSS_site_Au_c_bb}
          \includegraphics[width=\linewidth]{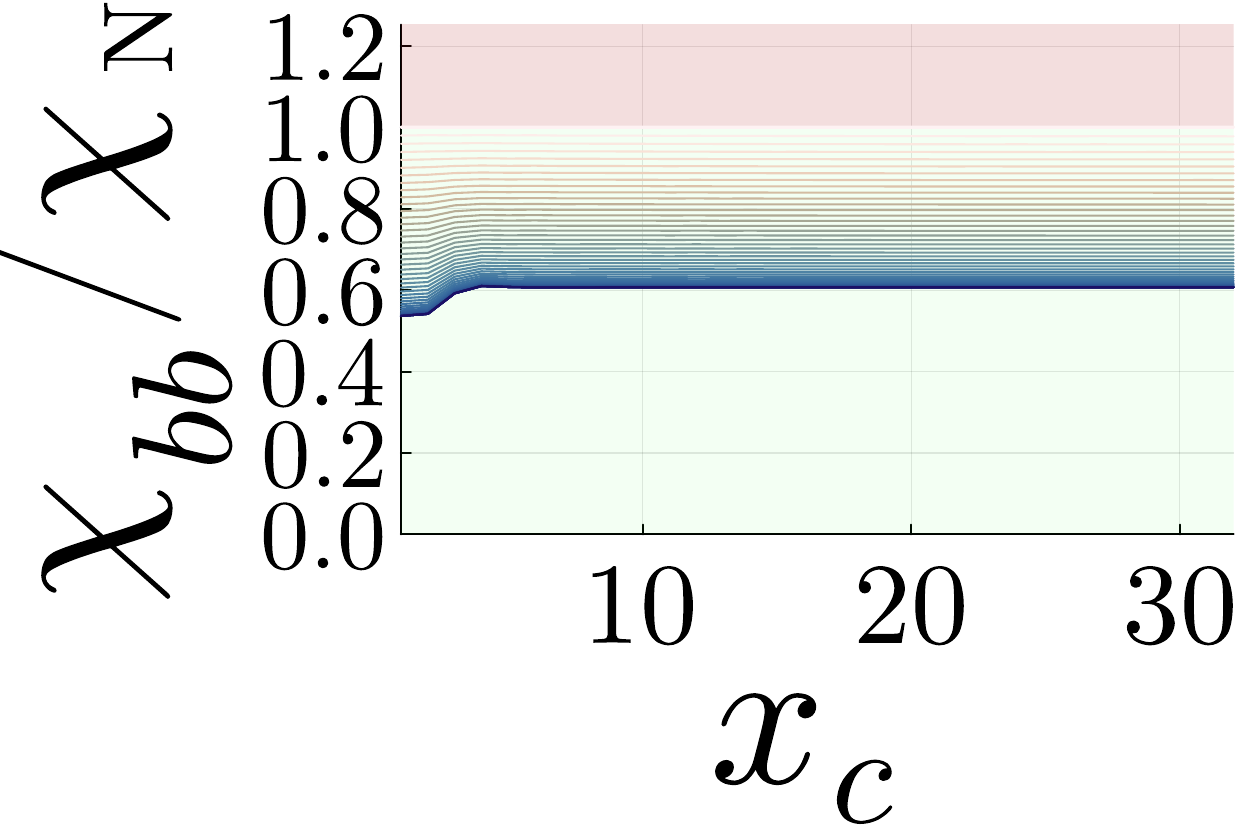}
    \end{minipage} \\
    \begin{minipage}[c]{0.30\linewidth}
        \subcaption{}\vspace{-1mm}
        \label{sfig: LSS_site_Au_a_cc}
          \includegraphics[width=\linewidth]{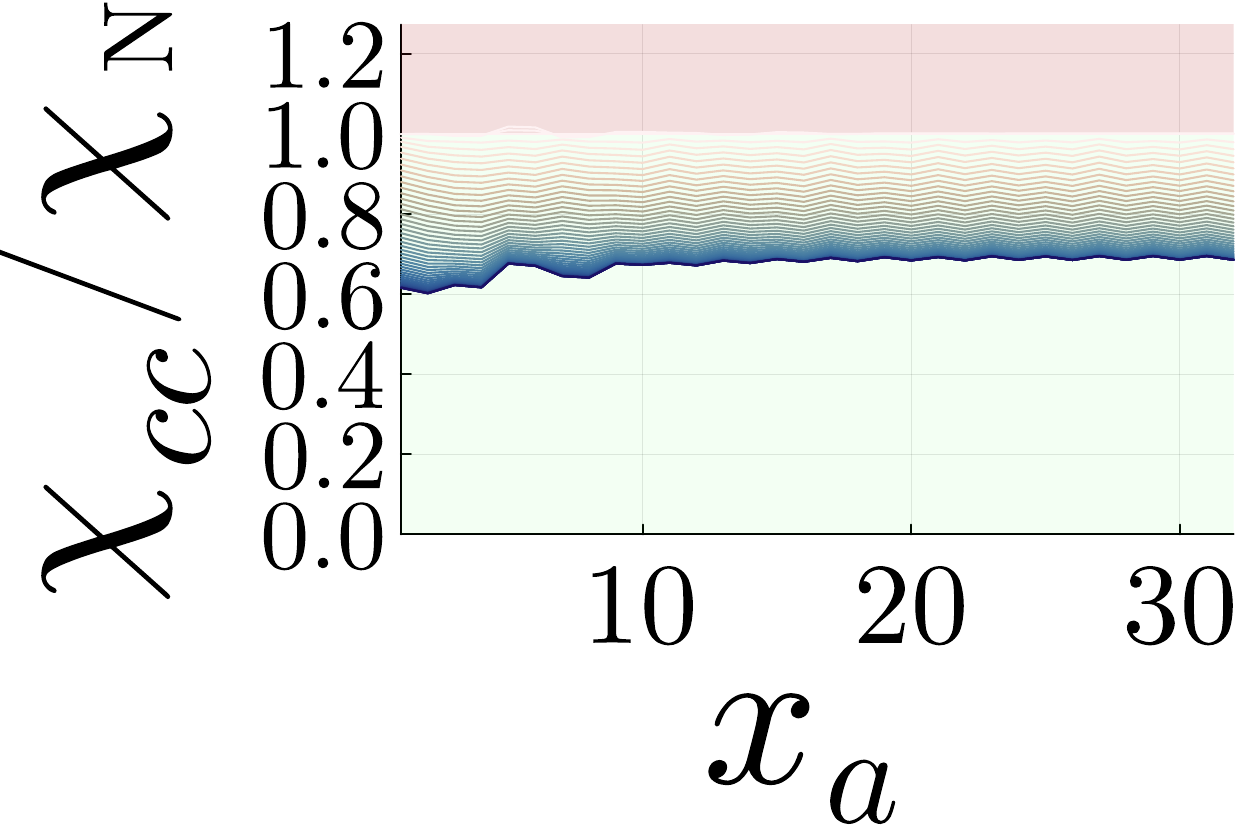}
    \end{minipage}&
    \begin{minipage}[c]{0.30\linewidth}
        \subcaption{}\vspace{-1mm}
        \label{sfig: LSS_site_Au_b_cc}
          \includegraphics[width=\linewidth]{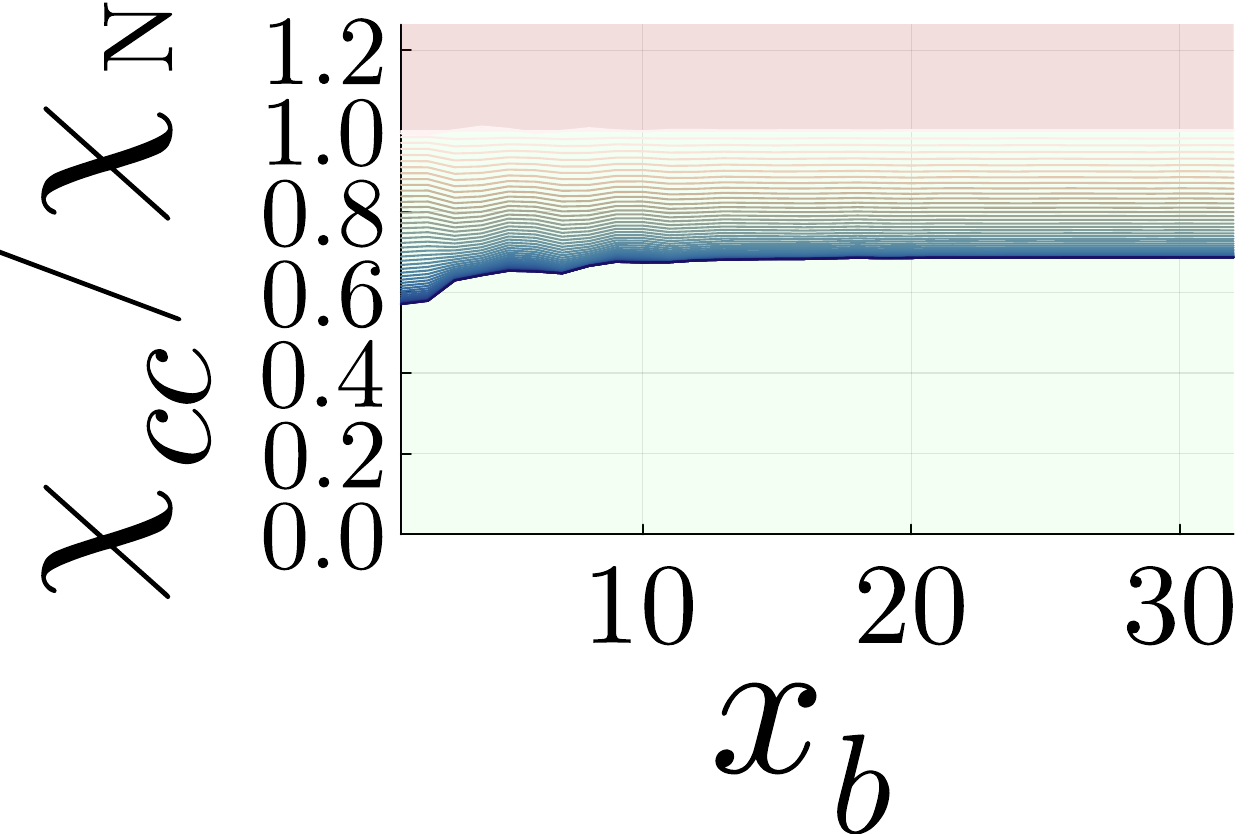}
    \end{minipage}&
    \begin{minipage}[c]{0.30\linewidth}
        \subcaption{}\vspace{-1mm}
        \label{sfig: LSS_site_Au_c_cc}
          \includegraphics[width=\linewidth]{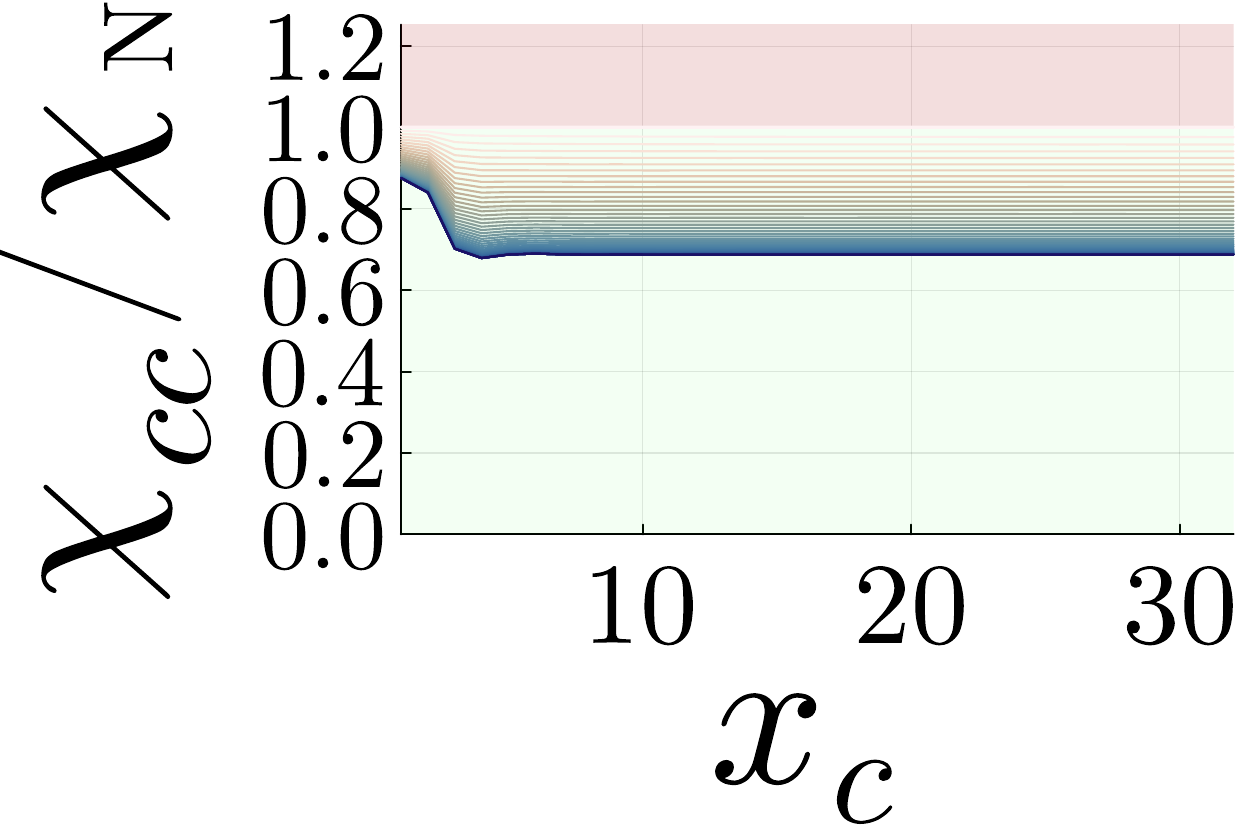}
    \end{minipage} \\
    \multicolumn{3}{c}{
      \includegraphics[width=0.9\linewidth]{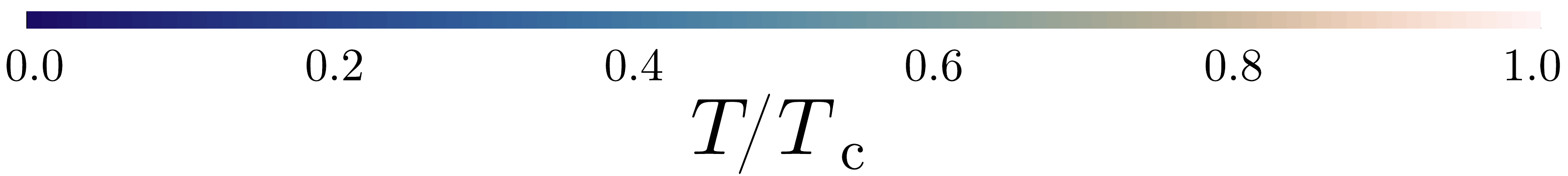}
    }\\
  \end{tabular}
  \caption{
    Site dependence of the LSS at each temperature for $A_{u}$ state.
    $x_{⟂}=1$ for the surface and $x_{⟂}=32$ for the bulk are concerned, respectively.
    \subref{sfig: LSS_site_Au_a_aa} and \subref{sfig: LSS_site_Au_b_bb} show the anomalous enhancement of the surface LSS due to MSS.
  }
  \label{fig: LSS_site_Au}
\end{figure}
First, we consider the $A_{u}$ pairing state discussed in \cref{ssec:sdos_Au}.
The temperature dependences of surface and bulk LSS calculations are summarized in the first row in \cref{fig: LSS_total}. 
\Cref{fig: LSS_site_Au} also plots the LSS in the $a, b$ and $c$ directions from the surface ($x_{⟂}=1$) to the bulk ($x_⟂=32$) for each surface.
The LSS of the bulk, as shown in \cref{sfig: LSS_total_Au_bulk}, decreases with decreasing temperature in all directions, as is known for the \ce{^3He}-B phase~\cite{vollhardt}.

However, when ABS is present on the surface of \ce{UTe2}, the behavior of the surface LSS dramatically changes from that of the bulk.
As shown in \cref{sfig: LSS_total_Au_a} and the first column of \cref{fig: LSS_site_Au}, $χ_{aa}$ on the (100) plane enhances with decreasing temperature.
Moreover, this enhancement appears only in the LSS near the surface, and its effect diminishes as one goes to the bulk, as shown in \cref{sfig: LSS_site_Au_a_aa}.
Majorana cones appear as the topological protected ABS on the (100) plane, as confirmed in \cref{sfig: sdos_Au_a_dos3d_1,sfig: sdos_Au_a_sdos}, giving rise to the enhancement of the LSS on the surface as mentioned in Eq.~\eqref{eq:ising}.
In the case of the (100) plane, the enhancement appears only in $χ_{aa}$, implying the strong Ising anisotropy of the MSS.
This Ising anisotropy can be understood in terms of the crystalline symmetry that protects the Majorana state as follows:
The Majorana cones are protected by the chiral operator $Γ_{\mathcal{C}_{a}}$~\cite{tei_2023}.
This symmetry is broken when a magnetic field is applied to the $a$ direction. 
In other words, the zero energy states can be gapped out when the magnetic field is applied along the direction that coincides with the Ising spin of the MSS. 
We discuss how Zeeman magnetic fields affect the dispersion of the MSS in \cref{app:sdos_zeeman}.

As shown in \cref{sfig: LSS_total_Au_b} and the second column of \cref{fig: LSS_site_Au}, $χ_{bb}$ on the (010) plane also enhances with decreasing temperature and this enhancement appears only in the LSS near the surface, as shown in \cref{sfig: LSS_site_Au_b_bb}.
In the case of the (010) plane, the Majorana cones are protected by the chiral operator $Γ_{\mathcal{C}_{b}}$~\cite{tei_2023} which is broken when a magnetic field is applied in the $b$ direction.
Thus, the enhancement of the LSS occurs due to the coupling of the Ising spin of the MSS with the magnetic field in the $b$ direction.

For the (001) plane, on the other hand, no zero-energy state exists on the surface, and no magnetic anisotropy in the LSS is observed as shown in the third column of \cref{fig: LSS_site_Au}.
Comparing the LSS between the bulk and the surface, $χ_{cc}$ in \cref{sfig: LSS_site_Au_c_cc} slightly increases in the surface region. 
This enhancement is attributed to the topologically trivial surface ABS with finite energies (see the third row of \cref{fig: sdos_Au}) and not protected by chiral symmetry.

\subsubsection{\texorpdfstring{$B_{1u}$}{B1u} pairing state}
\label{ssec:SS_B1u}

\begin{figure}[htbp]
  \begin{tabular}{ccc}
    \multicolumn{3}{l}{\scalebox{1.0}{\fbox{$B_{1u}$}}} \\
    (100) & (010) & (001) \\
    \begin{minipage}[c]{0.30\linewidth}
        \subcaption{}\vspace{-1mm}
        \label{sfig: LSS_site_B1u_a_aa}
          \includegraphics[width=\linewidth]{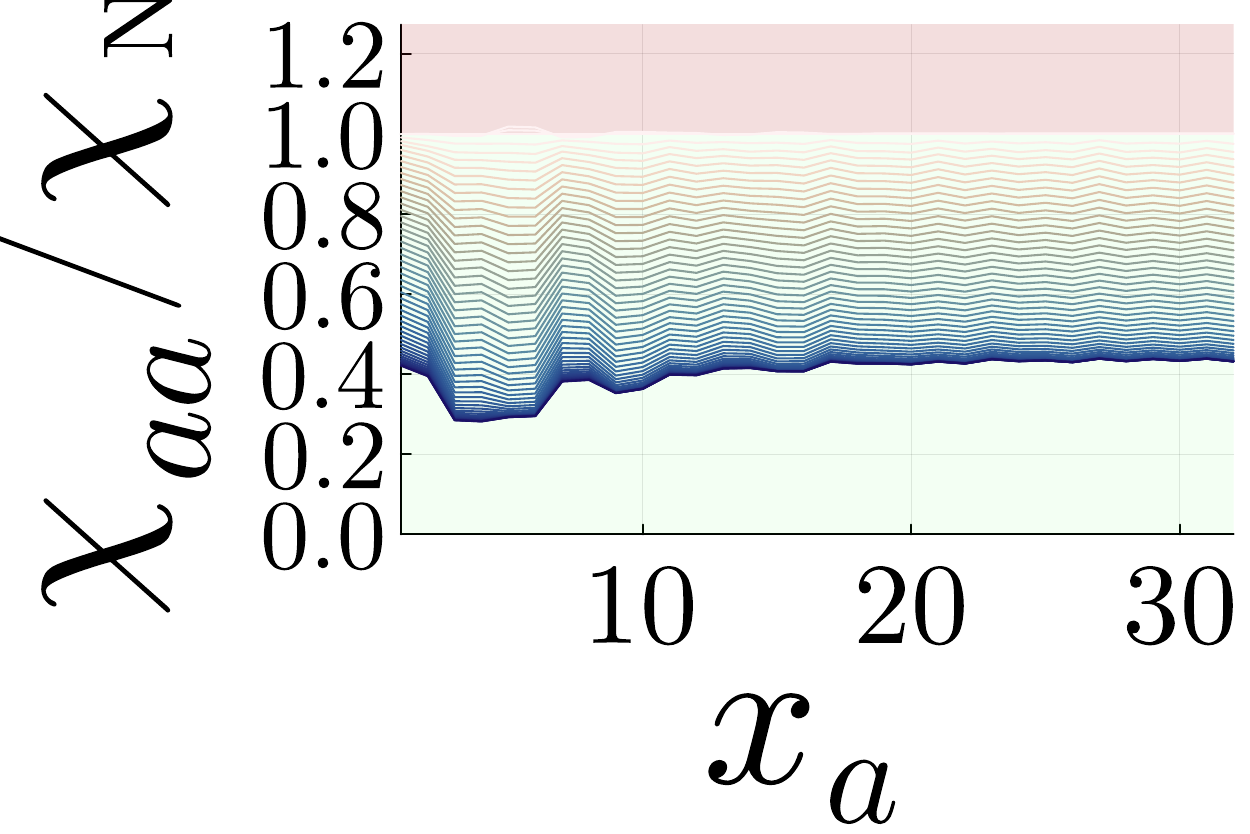}
    \end{minipage}&
    \begin{minipage}[c]{0.30\linewidth}
        \subcaption{}\vspace{-1mm}
        \label{sfig: LSS_site_B1u_b_aa}
          \includegraphics[width=\linewidth]{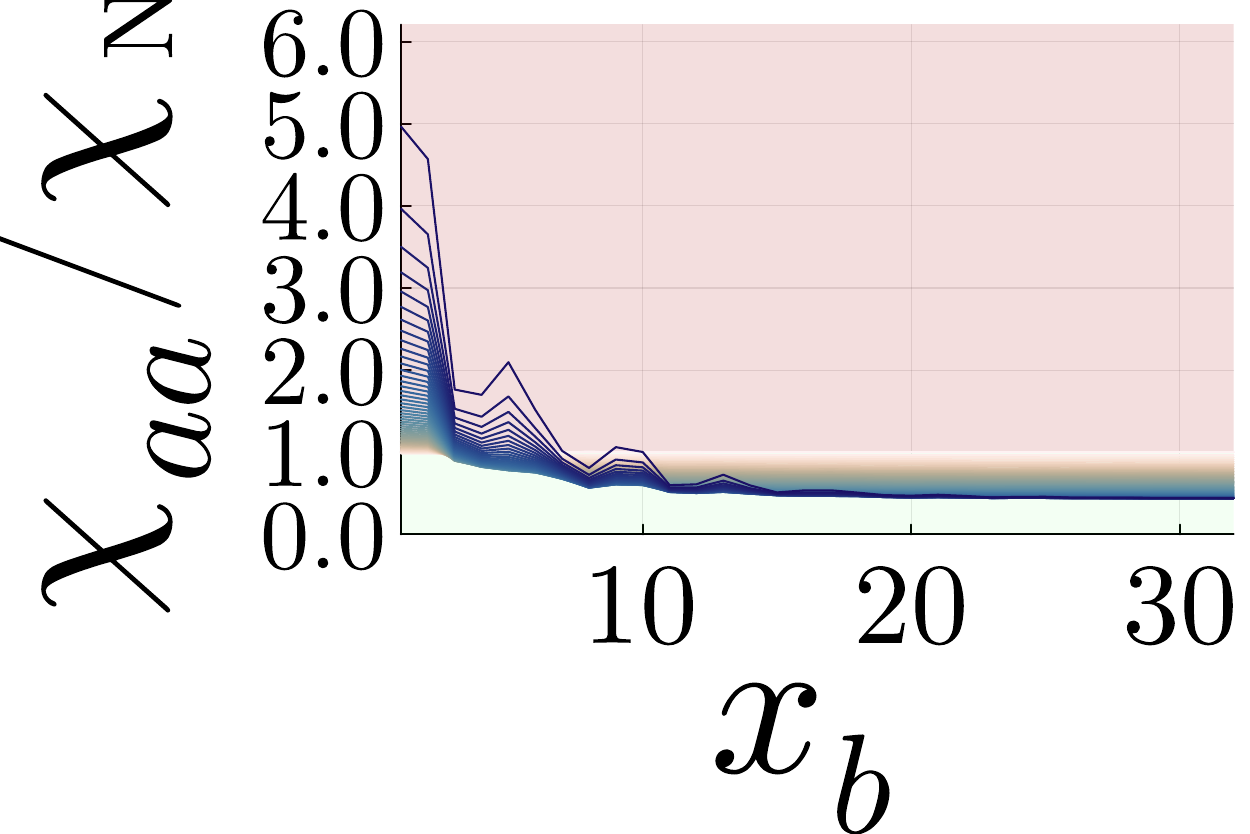}
    \end{minipage}&
    \begin{minipage}[c]{0.30\linewidth}
        \subcaption{}\vspace{-1mm}
        \label{sfig: LSS_site_B1u_c_aa}
          \includegraphics[width=\linewidth]{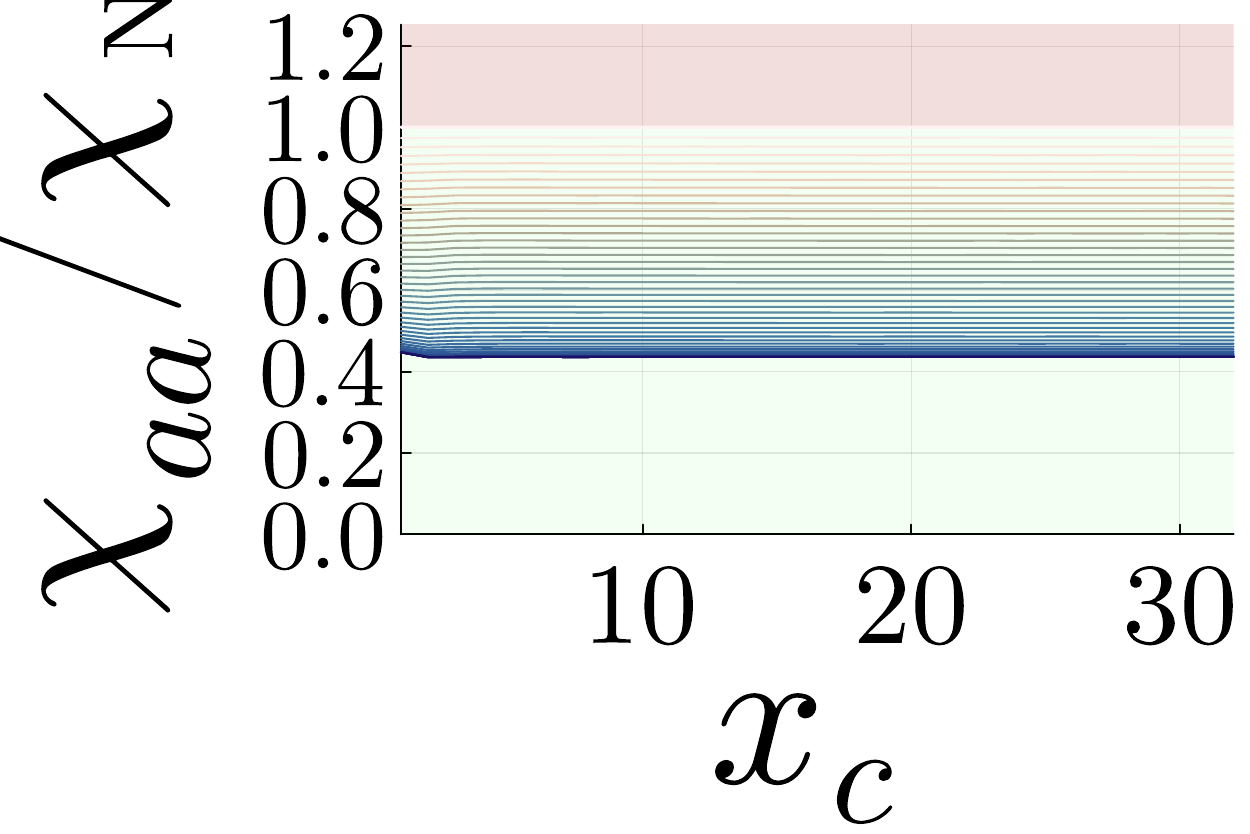}
    \end{minipage} \\
    \begin{minipage}[c]{0.30\linewidth}
        \subcaption{}\vspace{-1mm}
        \label{sfig: LSS_site_B1u_a_bb}
          \includegraphics[width=\linewidth]{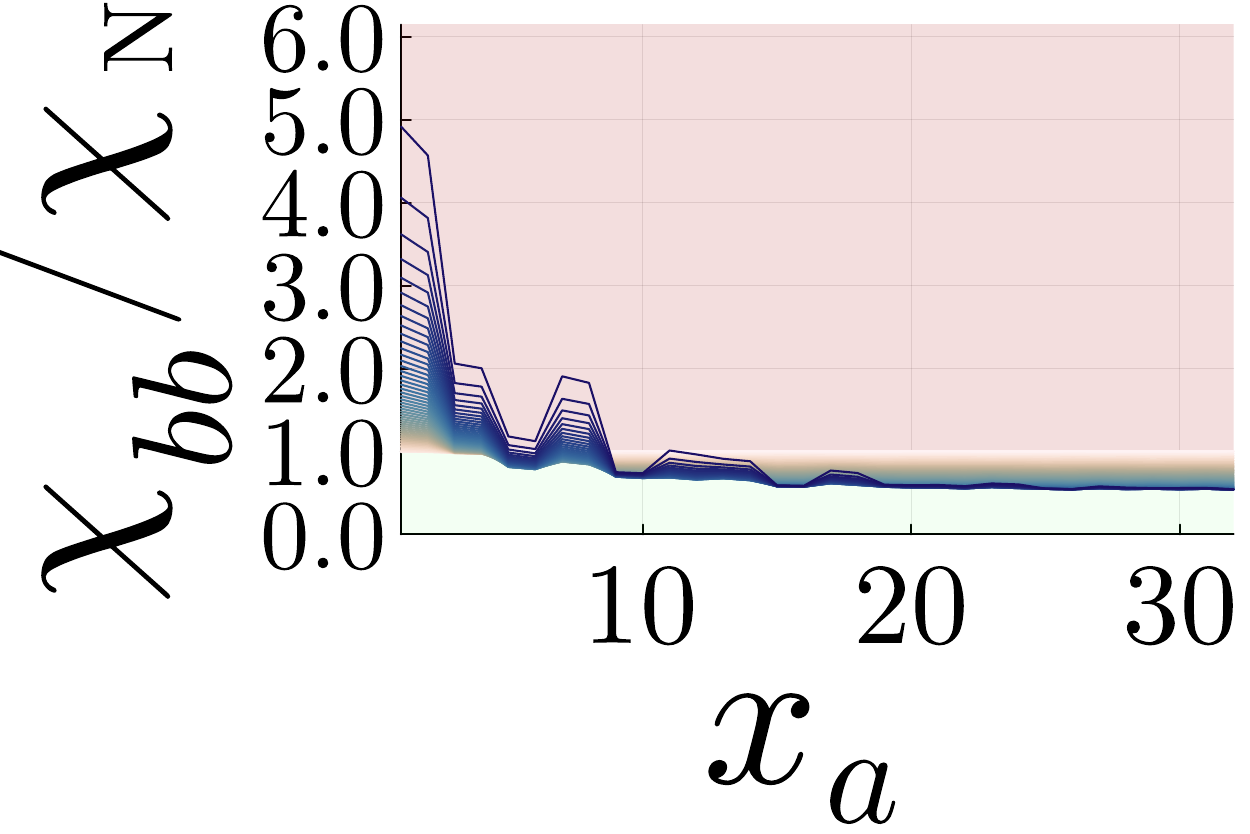}
    \end{minipage}&
    \begin{minipage}[c]{0.30\linewidth}
        \subcaption{}\vspace{-1mm}
        \label{sfig: LSS_site_B1u_b_bb}
          \includegraphics[width=\linewidth]{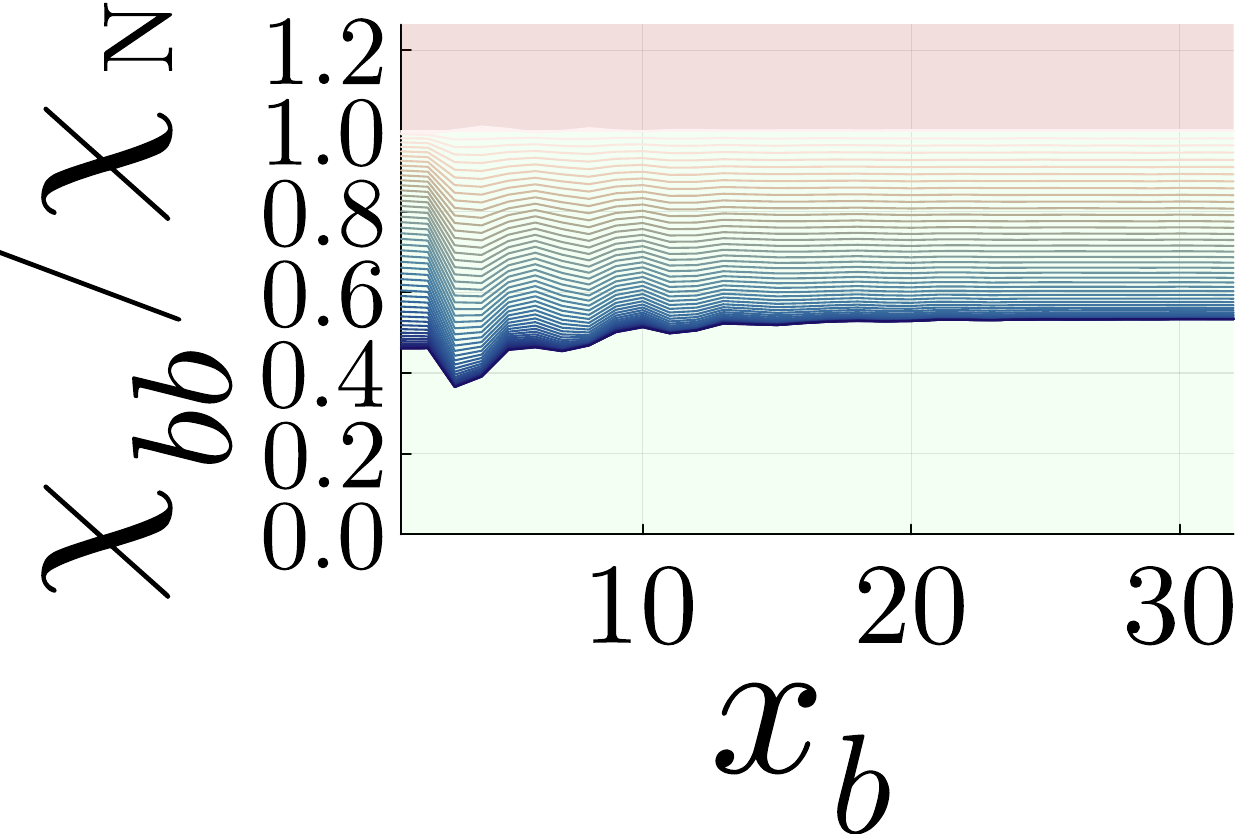}
    \end{minipage}&
    \begin{minipage}[c]{0.30\linewidth}
        \subcaption{}\vspace{-1mm}
        \label{sfig: LSS_site_B1u_c_bb}
          \includegraphics[width=\linewidth]{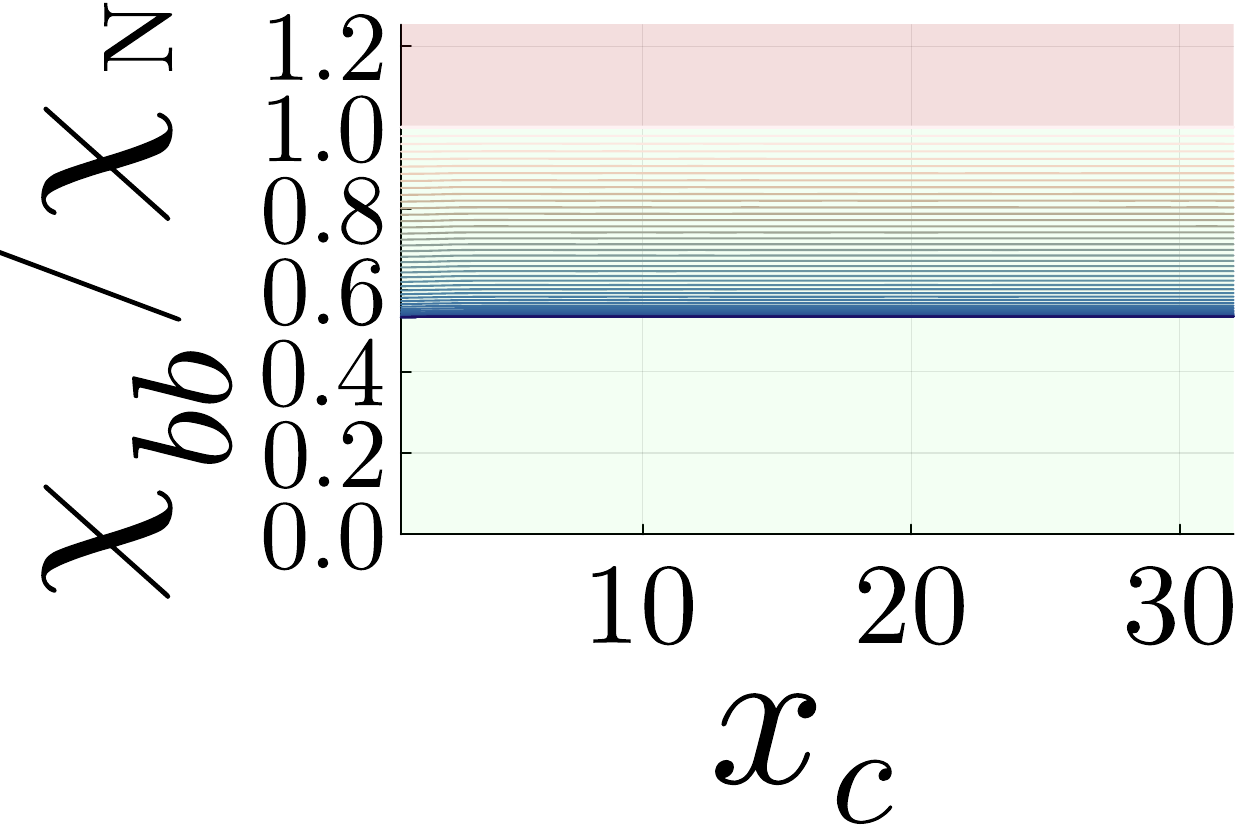}
    \end{minipage}\\
    \begin{minipage}[c]{0.30\linewidth}
        \subcaption{}\vspace{-1mm}
        \label{sfig: LSS_site_B1u_a_cc}
          \includegraphics[width=\linewidth]{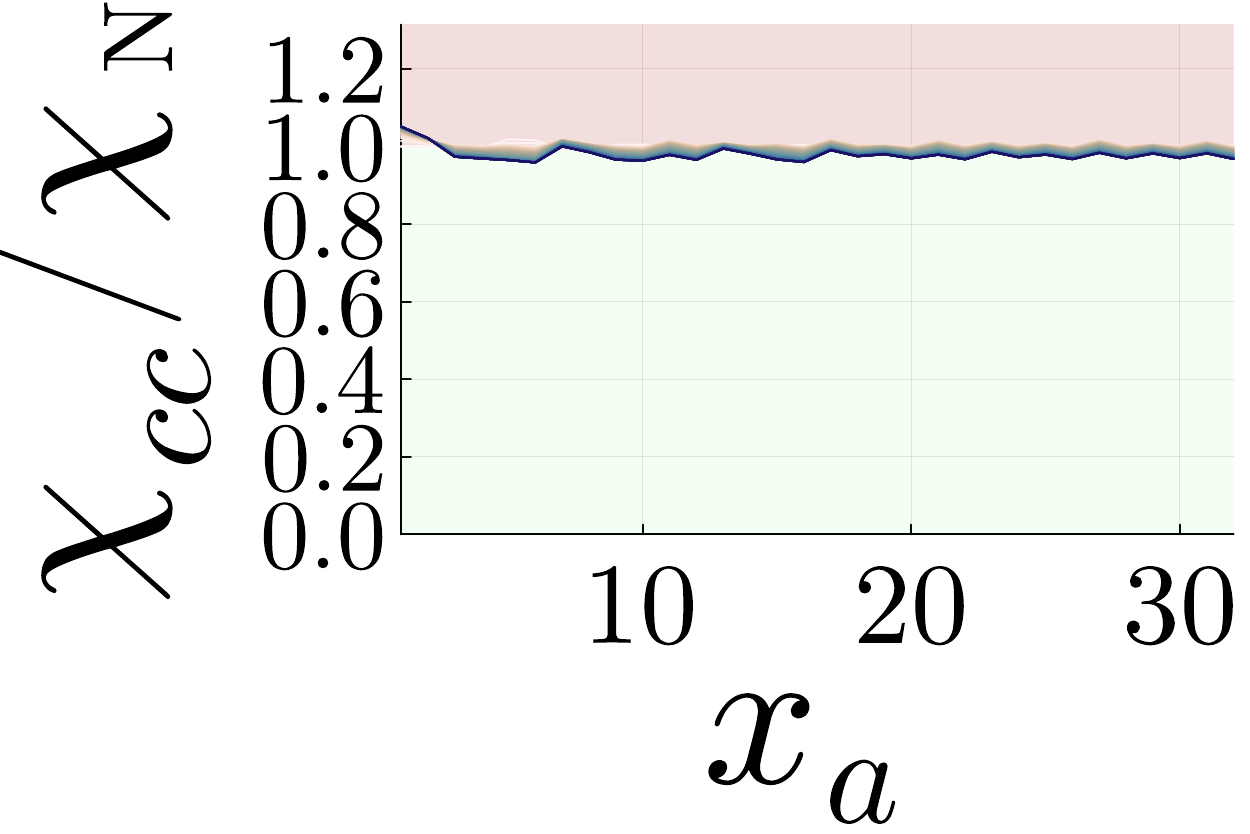}
    \end{minipage}&
    \begin{minipage}[c]{0.30\linewidth}
        \subcaption{}\vspace{-1mm}
        \label{sfig: LSS_site_B1u_b_cc}
          \includegraphics[width=\linewidth]{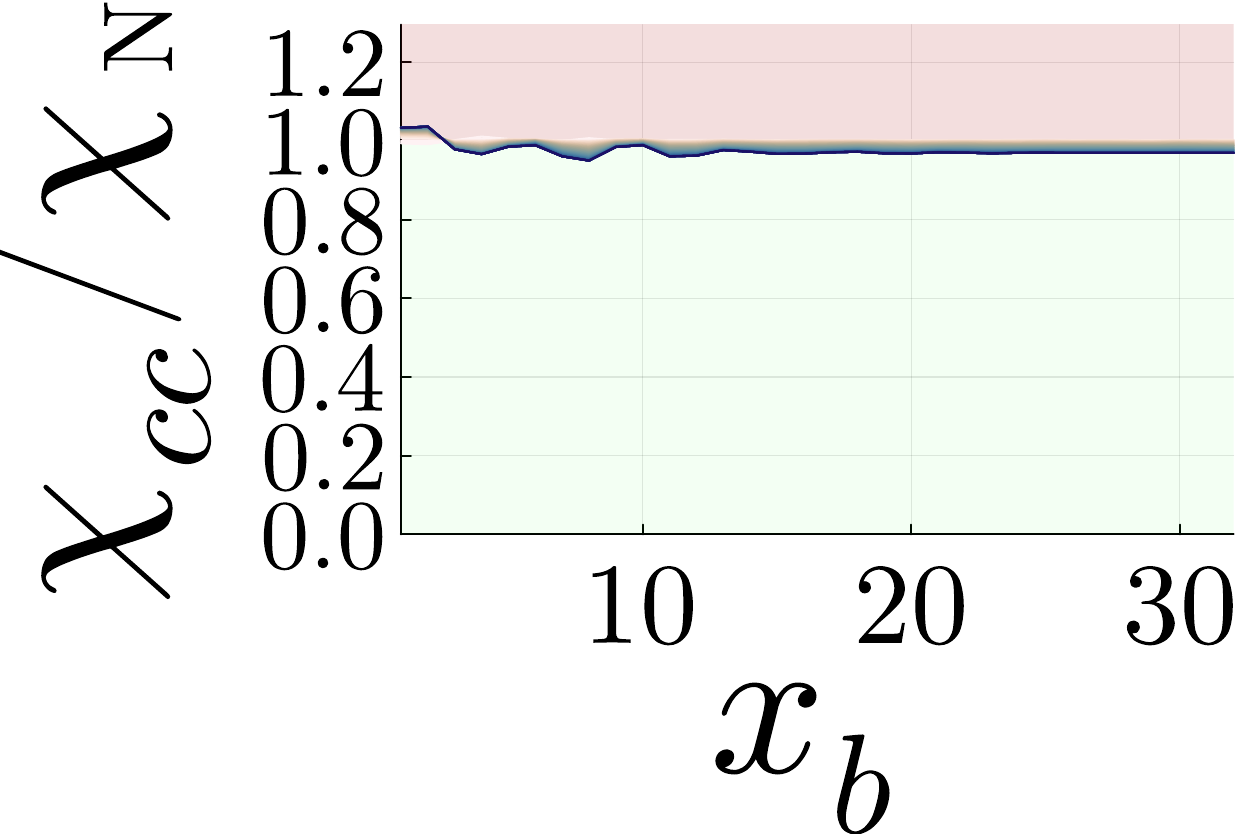}
    \end{minipage}&
    \begin{minipage}[c]{0.30\linewidth}
        \subcaption{}\vspace{-1mm}
        \label{sfig: LSS_site_B1u_c_cc}
          \includegraphics[width=\linewidth]{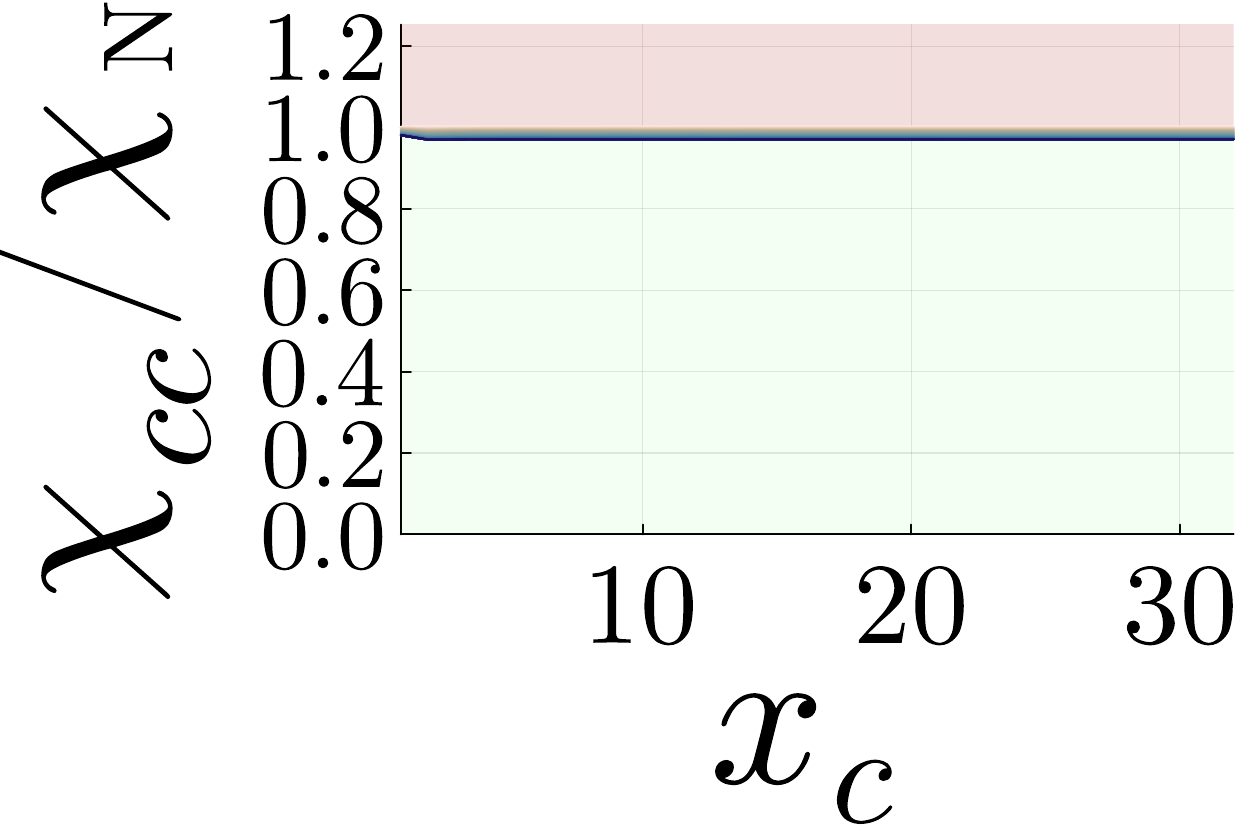}
    \end{minipage}\\
    \multicolumn{3}{c}{
      \includegraphics[width=0.9\linewidth]{cbar-spin-sus-site.pdf}
    }\\
  \end{tabular}
  \caption{
    Site dependence of the LSS at each temperature for $B_{1u}$ state.
    $x_{⟂}=1$ for the surface and $x_{⟂}=32$ for the bulk are concerned, respectively.
    \subref{sfig: LSS_site_B1u_b_aa} and \subref{sfig: LSS_site_B1u_a_bb} show the anomalous enhancement of the surface LSS due to MSS.
    $χ_{cc}$ has a constant value regardless of temperature because the directions of the magnetic field and $\vb*{d}$-vector (\cref{eq:d-vector_IR_B1u}) are orthogonal.
  }
  \label{fig: LSS_site_B1u}
\end{figure}
Next, we consider the $B_{1u}$ state whose surface LDOSs are discussed in \cref{ssec:sdos_B1u}.
The results of LSSs on the surface and bulk are summarized in the second row in \cref{fig: LSS_total}.
\Cref{fig: LSS_site_B1u} also shows the LSS as a function of the distance from the surface for each surface.
The LSS of the bulk, as shown in \cref{sfig: LSS_total_B1u_bulk}, decreases with decreasing temperature in $a$ and $b$ directions, which is consistent with the previous results~\cite{hiranuma_2021}.
$χ_{cc}/χ_\mathrm{N}$ stays constant at $χ_{cc}(T)/χ_\mathrm{N}=1$ for all temperatures because the directions of the magnetic field and $\vb*{d}$-vector [\cref{eq:d-vector_IR_B1u}] are orthogonal.

As shown in \cref{sfig: LSS_total_B1u_a} and the first column of \cref{fig: LSS_site_B1u}, $χ_{bb}$ on the (100) plane enhances with decreasing temperature.
Moreover, this enhancement appears only in the LSS near the surface and its effect diminishes in the region far from the surface, as shown in \cref{sfig: LSS_site_B1u_a_bb}. 
The enhancement appears only in $χ_{bb}$ on the (100) plane, implying the strong Ising anisotropy of the surface states. 
As seen in \cref{sfig: sdos_B1u_a_dos3d_2,sfig: sdos_B1u_a_sdos}, in the $B_{1u}$ state, the Flat Fermi arc on the (100) plane is protected by the chiral operator $Γ_{\mathcal{M}_{ca}}$.
This symmetry is broken when a magnetic field is applied in the $b$ direction.
In addition, $\chi_{bb}$ on the (100) plane of the $B_{1u}$ state significantly enhances, compared with the case of the $A_{u}$ pairing state in \cref{sfig: LSS_total_Au_a} and \cref{sfig: LSS_total_B1u_a}. 
This is because the zero-energy state in the $A_{u}$ state appears in the surface Brillouin zone as a point node, whereas the $B_{1u}$ state has a flat Fermi arc which forms a nodal line in the surface Brillouin zone, resulting in a significant paramagnetic contribution to the spin susceptibility.

As shown in \cref{sfig: LSS_total_B1u_b} and the second column of \cref{fig: LSS_site_B1u}, $χ_{aa}$ on the (010) plane also enhances with decreasing temperature and this enhancement appears only in LSS near the surface, as shown in \cref{sfig: LSS_site_B1u_b_aa}.
The flat Fermi arc appears on the (010) plane, which is protected by the chiral operator $Γ_{\mathcal{M}_{bc}}$~\cite{tei_2023}. 
Therefore, a magnetic field along the $a$ direction breaks the symmetry and the Ising spin of the MSS exhibits the paramagnetic response.

In the case of the (001) plane, the magnetic anisotropy shown in \cref{sfig: LSS_total_B1u_b} and the third column of \cref{fig: LSS_site_B1u} is understandable with the orientation of the $\bm{d}$ vector as the surface ABS is absent in the (001) plane.

\subsubsection{\texorpdfstring{$B_{2u}$}{B2u} pairing state}
\label{ssec:SS_B2u}

\begin{figure}[htbp]
  \begin{tabular}{ccc}
    \multicolumn{3}{l}{\scalebox{1.0}{\fbox{$B_{2u}$}}} \\
    (100) & (010) & (001) \\
    \begin{minipage}[c]{0.30\linewidth}
        \subcaption{}\vspace{-1mm}
        \label{sfig: LSS_site_B2u_a_aa}
          \includegraphics[width=\linewidth]{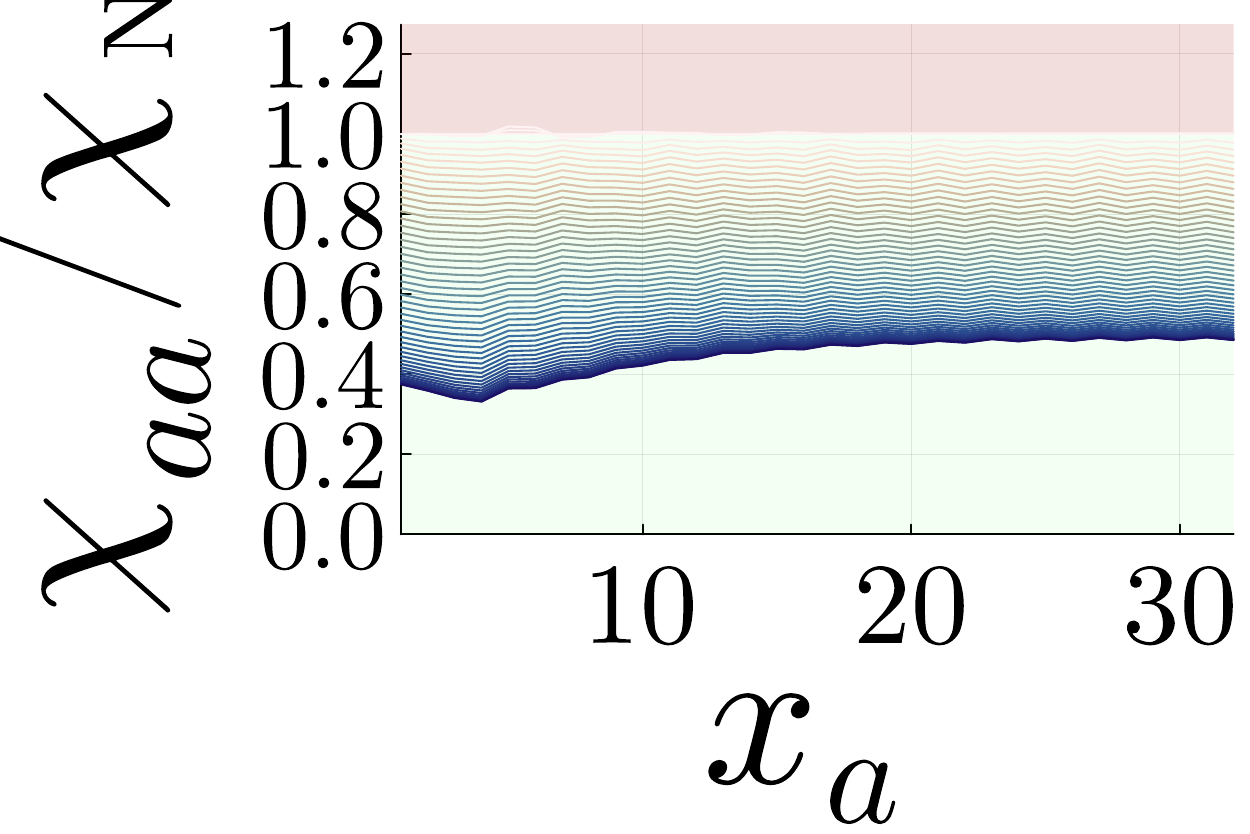}
    \end{minipage}&
    \begin{minipage}[c]{0.30\linewidth}
        \subcaption{}\vspace{-1mm}
        \label{sfig: LSS_site_B2u_b_aa}
          \includegraphics[width=\linewidth]{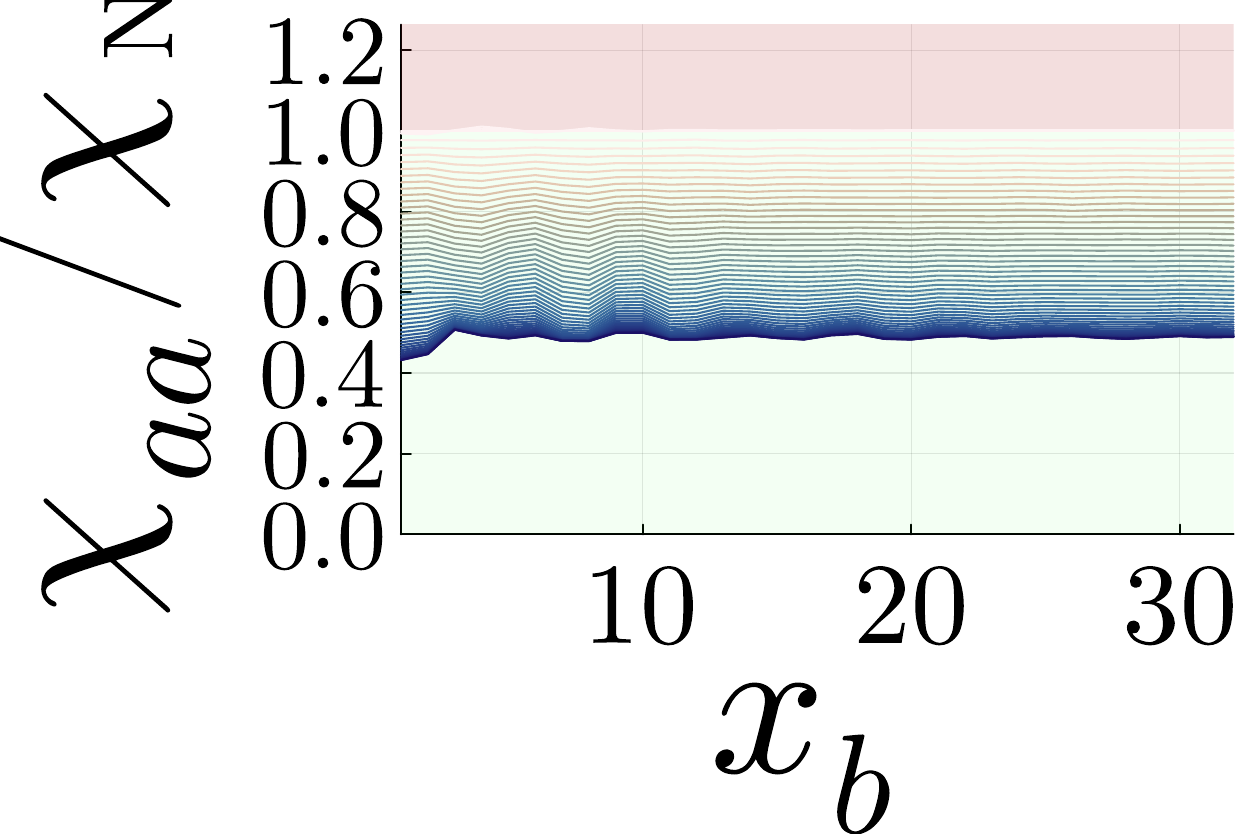}
    \end{minipage}&
    \begin{minipage}[c]{0.30\linewidth}
        \subcaption{}\vspace{-1mm}
        \label{sfig: LSS_site_B2u_c_aa}
          \includegraphics[width=\linewidth]{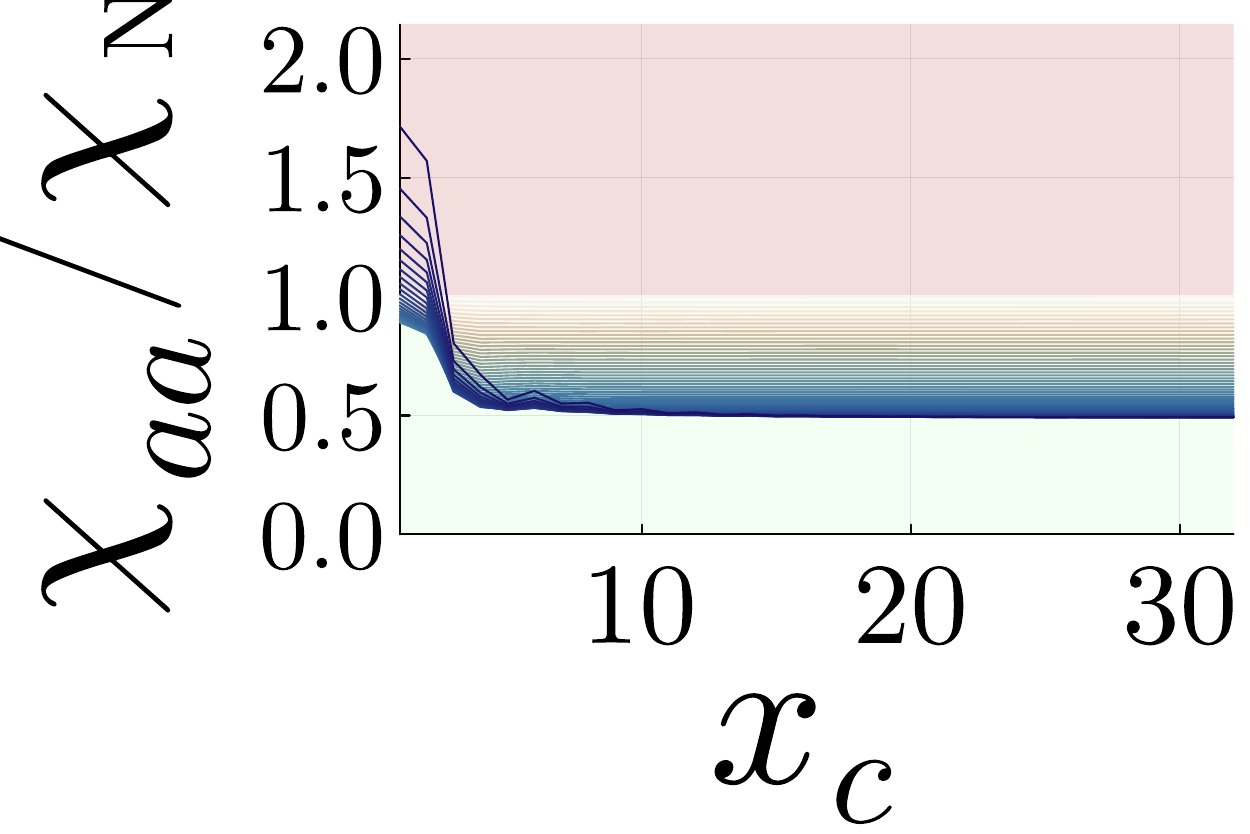}
    \end{minipage}\\
    \begin{minipage}[c]{0.30\linewidth}
        \subcaption{}\vspace{-1mm}
        \label{sfig: LSS_site_B2u_a_bb}
          \includegraphics[width=\linewidth]{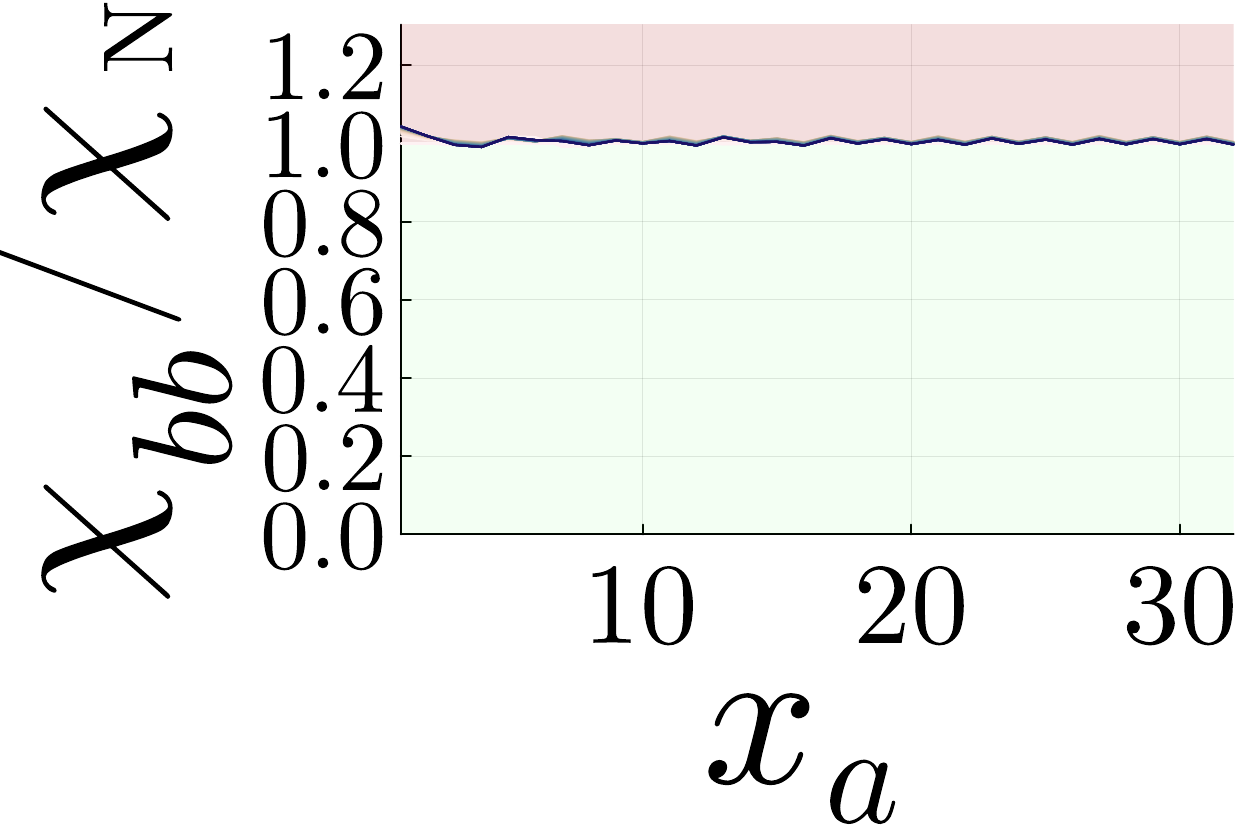}
    \end{minipage}&
    \begin{minipage}[c]{0.30\linewidth}
        \subcaption{}\vspace{-1mm}
        \label{sfig: LSS_site_B2u_b_bb}
          \includegraphics[width=\linewidth]{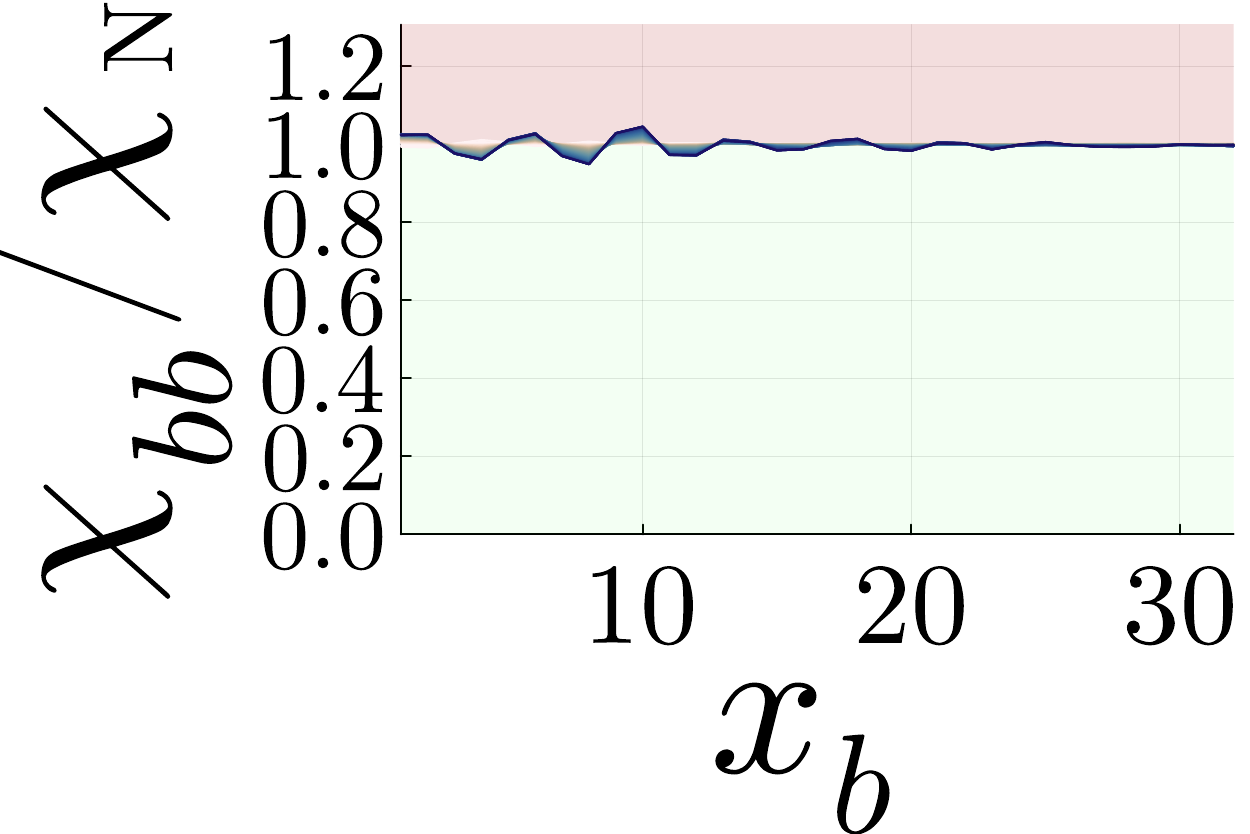}
    \end{minipage}&
    \begin{minipage}[c]{0.30\linewidth}
        \subcaption{}\vspace{-1mm}
        \label{sfig: LSS_site_B2u_c_bb}
          \includegraphics[width=\linewidth]{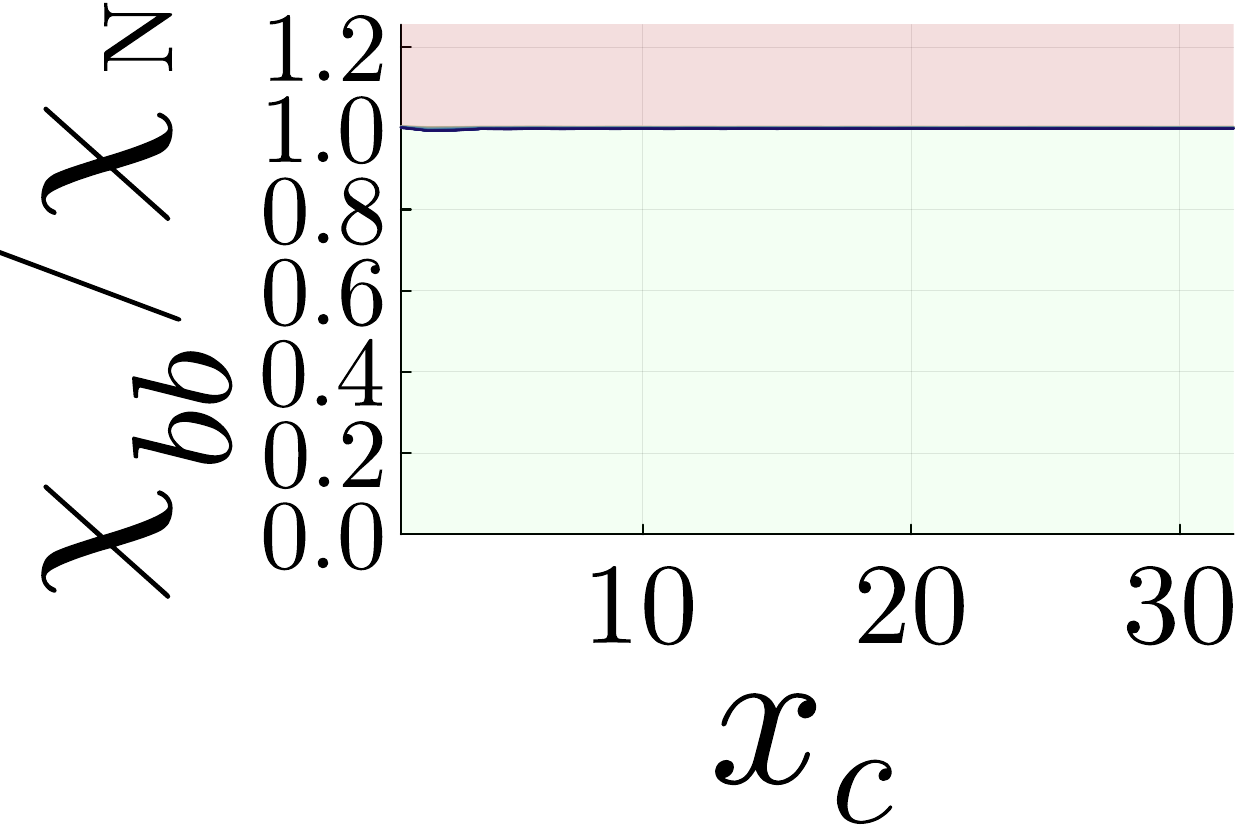}
    \end{minipage}\\
    \begin{minipage}[c]{0.30\linewidth}
        \subcaption{}\vspace{-1mm}
        \label{sfig: LSS_site_B2u_a_cc}
          \includegraphics[width=\linewidth]{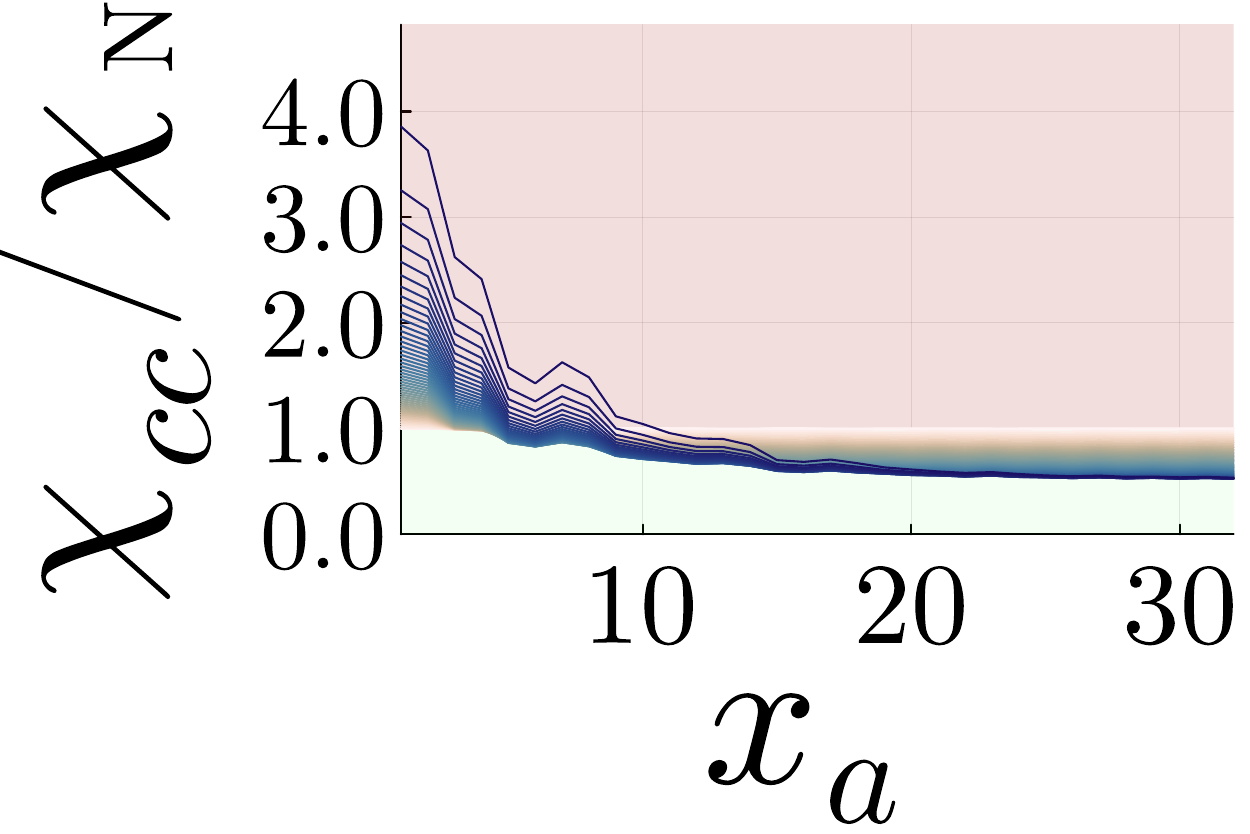}
    \end{minipage}&
    \begin{minipage}[c]{0.30\linewidth}
        \subcaption{}\vspace{-1mm}
        \label{sfig: LSS_site_B2u_b_cc}
          \includegraphics[width=\linewidth]{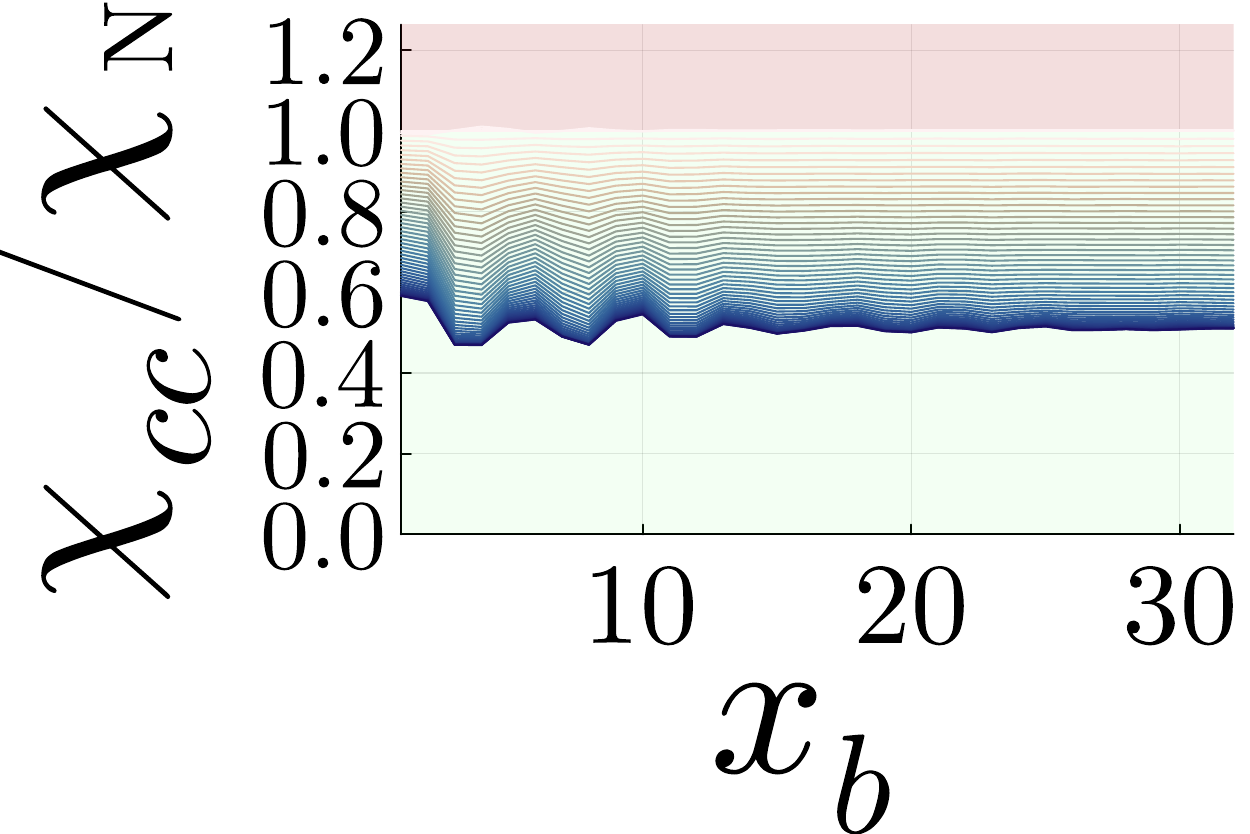}
    \end{minipage}&
    \begin{minipage}[c]{0.30\linewidth}
        \subcaption{}\vspace{-1mm}
        \label{sfig: LSS_site_B2u_c_cc}
          \includegraphics[width=\linewidth]{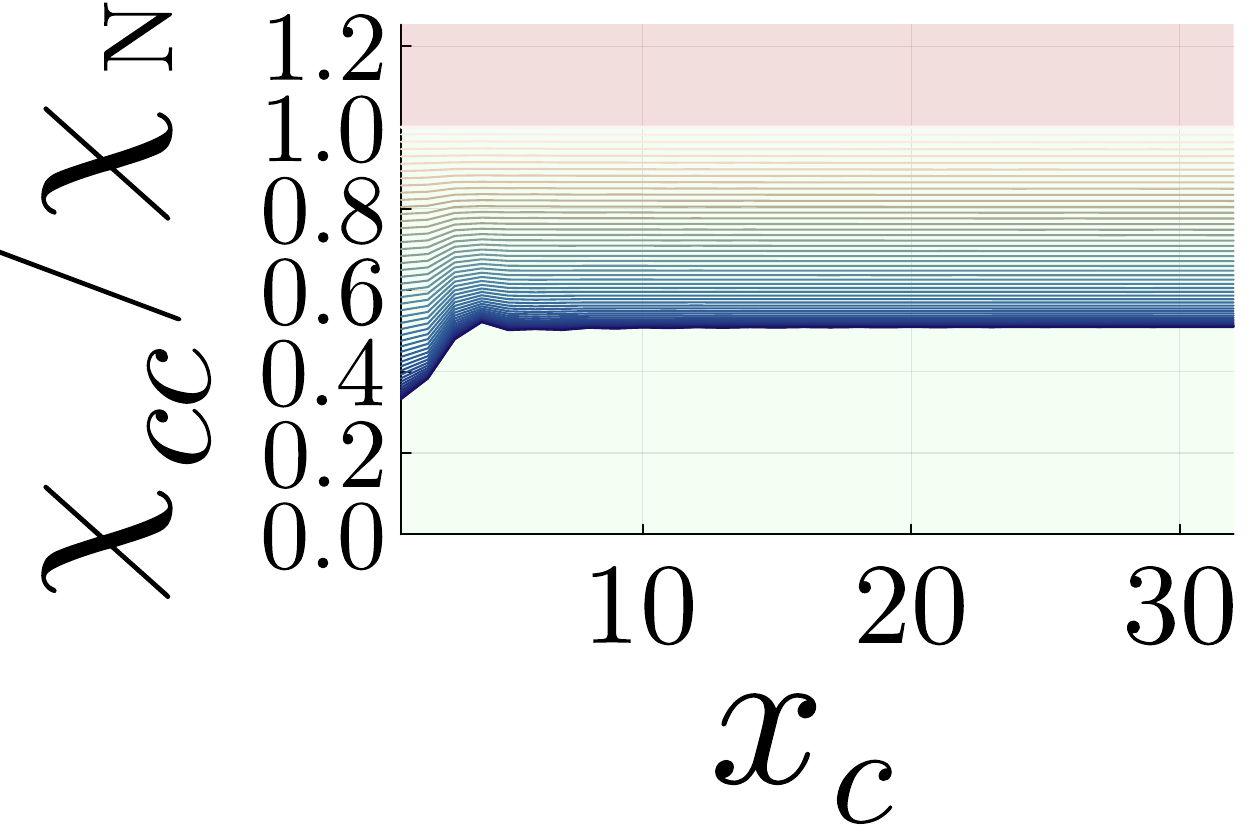}
    \end{minipage}\\
    \multicolumn{3}{c}{
      \includegraphics[width=0.9\linewidth]{cbar-spin-sus-site.pdf}
    }\\
  \end{tabular}
  \caption{
    Site dependence of the LSS at each temperature for $B_{2u}$ state.
    $x_{⟂}=1$ for the surface and $x_{⟂}=32$ for the bulk are concerned, respectively.
    \cref{sfig: LSS_site_B2u_c_aa} and \cref{sfig: LSS_site_B2u_a_cc} show the anomalous enhancement of the surface LSS due to MSS.
    $χ_{bb}$ has a constant value regardless of temperature because the directions of the magnetic field and $\vb*{d}$-vector (\cref{eq:d-vector_IR_B2u}) are orthogonal.
  }
  \label{fig: LSS_site_B2u}
\end{figure}
We also consider the $B_{2u}$ state discussed in \cref{ssec:sdos_B2u}.
The results of surface and bulk LSS calculations are summarized in the third row in \cref{fig: LSS_total}. 
\Cref{fig: LSS_site_B2u} plots the LSS as a function of the distance from the surface for each surface orientation.
The LSS of the bulk, as shown in \cref{sfig: LSS_total_B2u_bulk}, decreases with decreasing temperature in $a$ and $c$ directions, which is consistent with the previous work~\cite{hiranuma_2021}.
For all temperatures, $χ_{bb}$ has a value of $χ_{bb}(T)/χ_\mathrm{N}=1$ as the applied magnetic field is perpendicular to the $\vb*{d}$-vector [\cref{eq:d-vector_IR_B2u}].

As shown in \cref{sfig: LSS_total_B2u_a} and the first column of \cref{fig: LSS_site_B2u}, $χ_{cc}$ on the (100) plane enhances with decreasing temperature and this enhancement appears only in the LSS near the surface, as shown in \cref{sfig: LSS_site_B2u_a_cc}.
The enhancement appears only in the $χ_{cc}$, reflecting the Ising anisotropic character of the Fermi arcs in \cref{sfig: sdos_B2u_a_dos3d_1,sfig: sdos_B2u_a_sdos}. 
The flat Fermi arcs are protected by the chiral operator $Γ_{\mathcal{M}_{ab}}$~\cite{tei_2023}. 
When a magnetic field is applied in the $c$ direction, the chiral symmetry is broken, and the Fermi arcs exhibit paramagnetic contribution to the spin susceptibility.

As shown in \cref{sfig: LSS_total_B2u_c} and the third column of \cref{fig: LSS_site_B2u}, $χ_{aa}$ on the (001) plane also enhances with decreasing temperature. 
This enhancement appears only in LSS near the surface, as shown in \cref{sfig: LSS_site_B2u_c_aa}.
In (001) plane case, the flat Fermi arcs are protected by the chiral operator $\Gamma_{\mathcal{M}_{ab}}$~\cite{tei_2023}, which is broken when a magnetic field is applied in the $a$ direction.

In the case of the (010) plane, the magnetic anisotropy shown in \cref{sfig: LSS_total_B2u_b} and the second column of \cref{fig: LSS_site_B2u} is understandable with the orientation of the $\bm{d}$ vector as the surface ABS is absent in the (010) plane.

\subsubsection{\texorpdfstring{$B_{3u}$}{B3u} pairing state}
\label{ssec:SS_B3u}

\begin{figure}[htbp]
  \begin{tabular}{ccc}
    \multicolumn{3}{l}{\scalebox{1.0}{\fbox{$B_{3u}$}}} \\
    (100) & (010) & (001) \\
    \begin{minipage}[c]{0.30\linewidth}
        \subcaption{}\vspace{-1mm}
        \label{sfig: LSS_site_B3u_a_aa}
          \includegraphics[width=\linewidth]{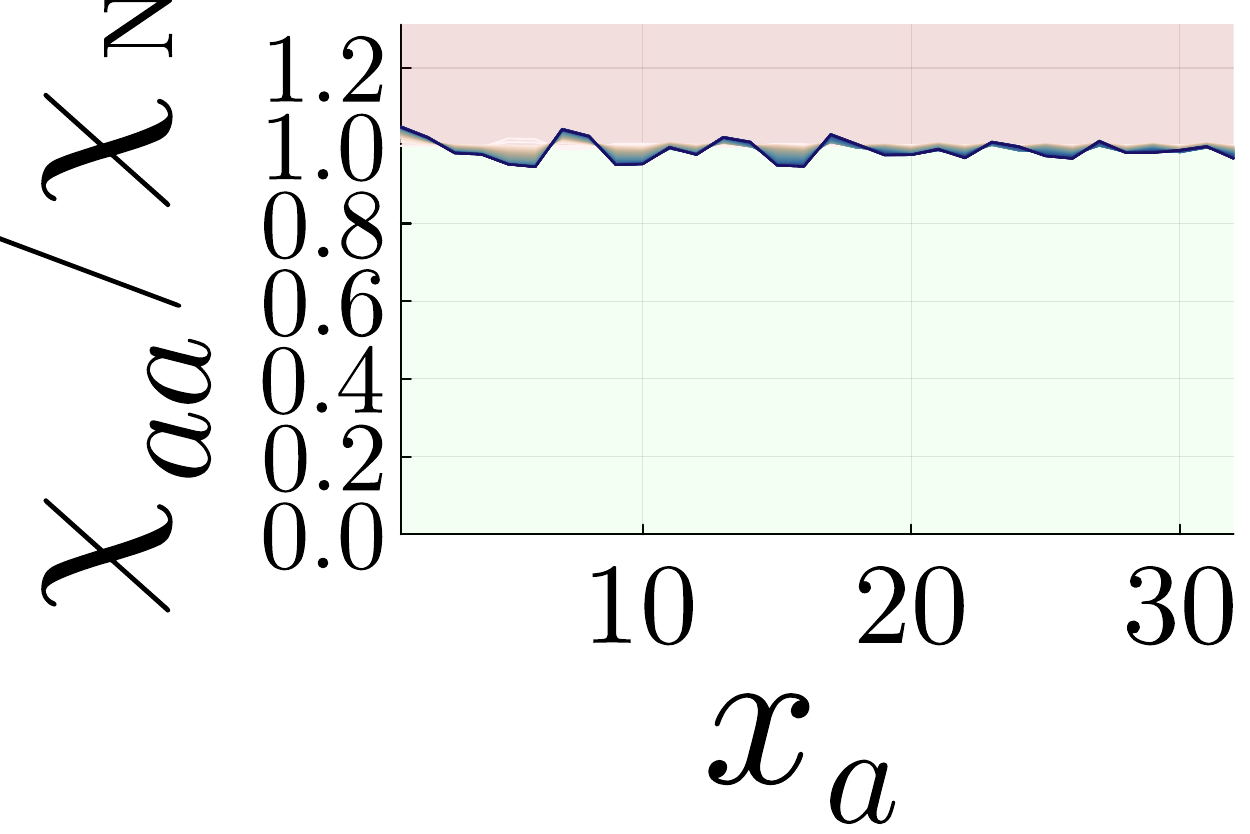}
    \end{minipage}&
    \begin{minipage}[c]{0.30\linewidth}
        \subcaption{}\vspace{-1mm}
        \label{sfig: LSS_site_B3u_b_aa}
          \includegraphics[width=\linewidth]{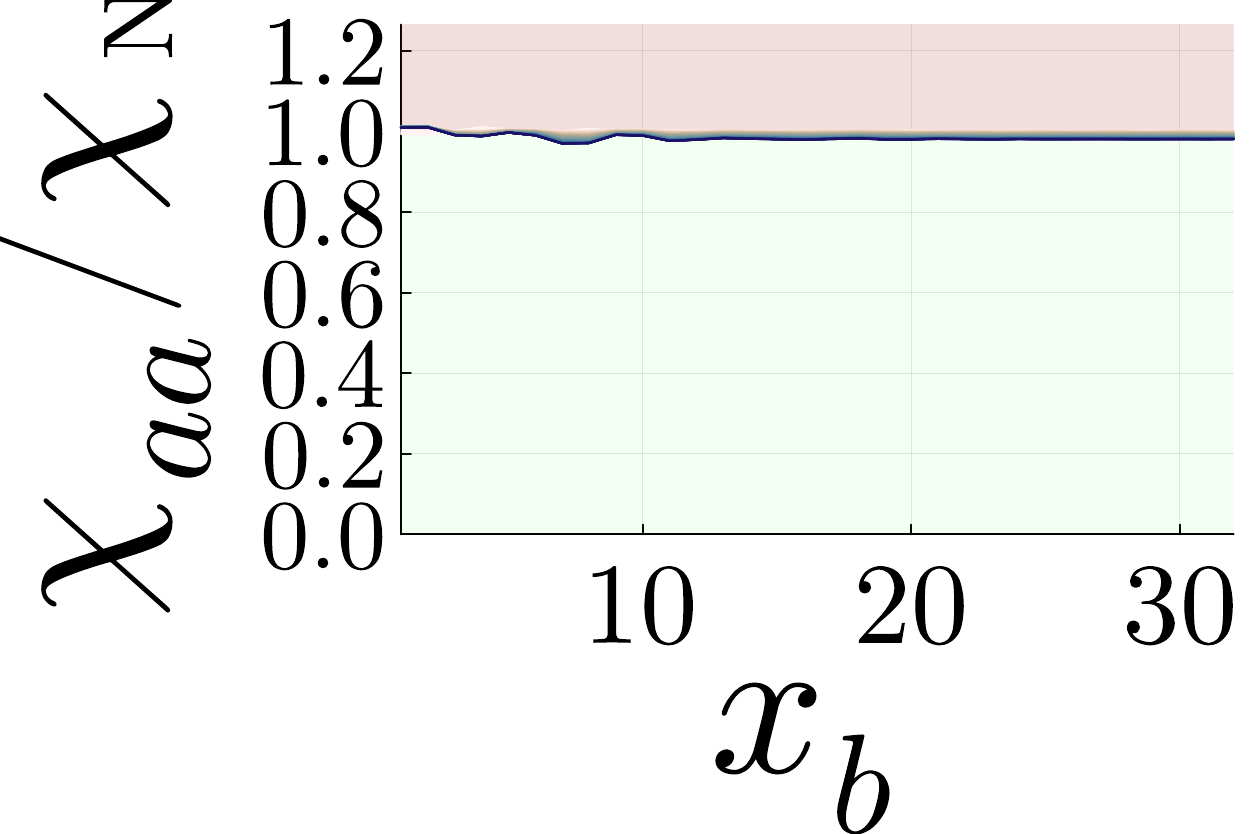}
    \end{minipage}&
    \begin{minipage}[c]{0.30\linewidth}
        \subcaption{}\vspace{-1mm}
        \label{sfig: LSS_site_B3u_c_aa}
          \includegraphics[width=\linewidth]{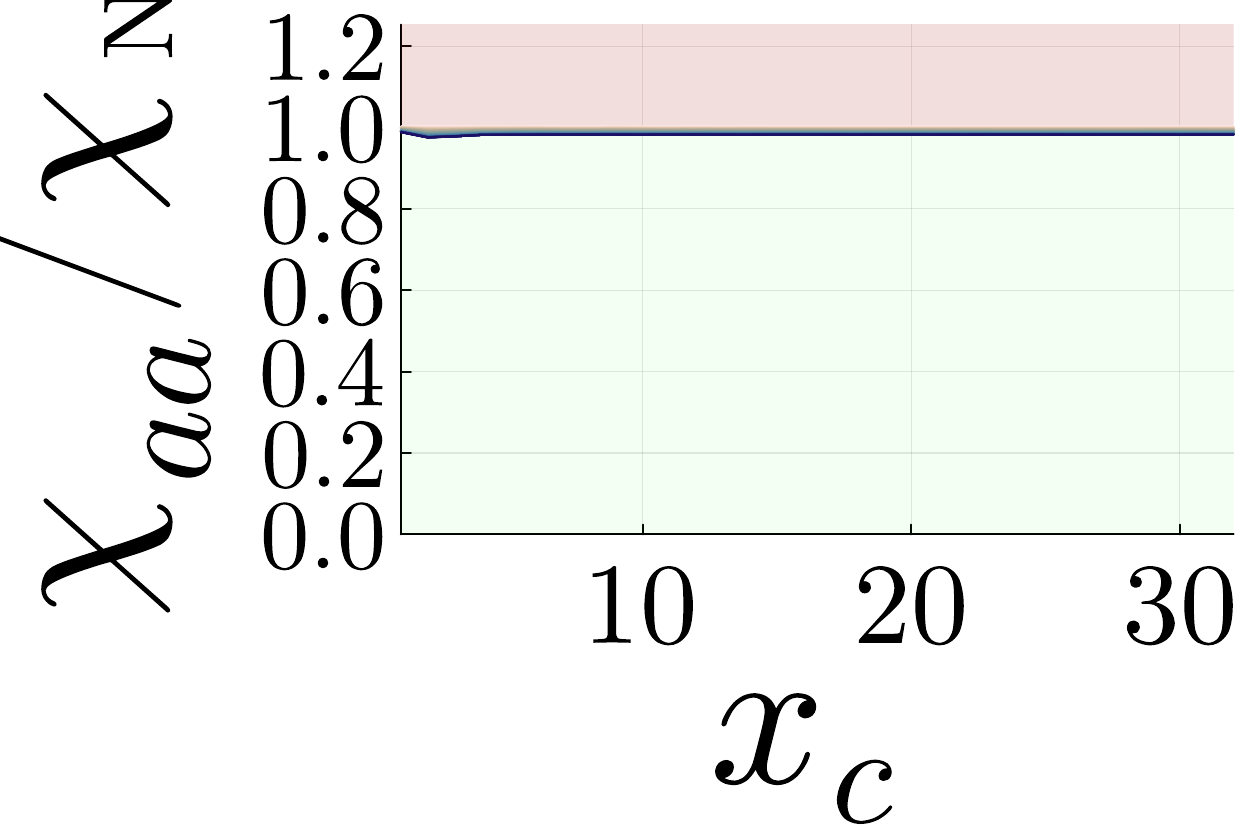}
    \end{minipage}\\
    \begin{minipage}[c]{0.30\linewidth}
        \subcaption{}\vspace{-1mm}
        \label{sfig: LSS_site_B3u_a_bb}
          \includegraphics[width=\linewidth]{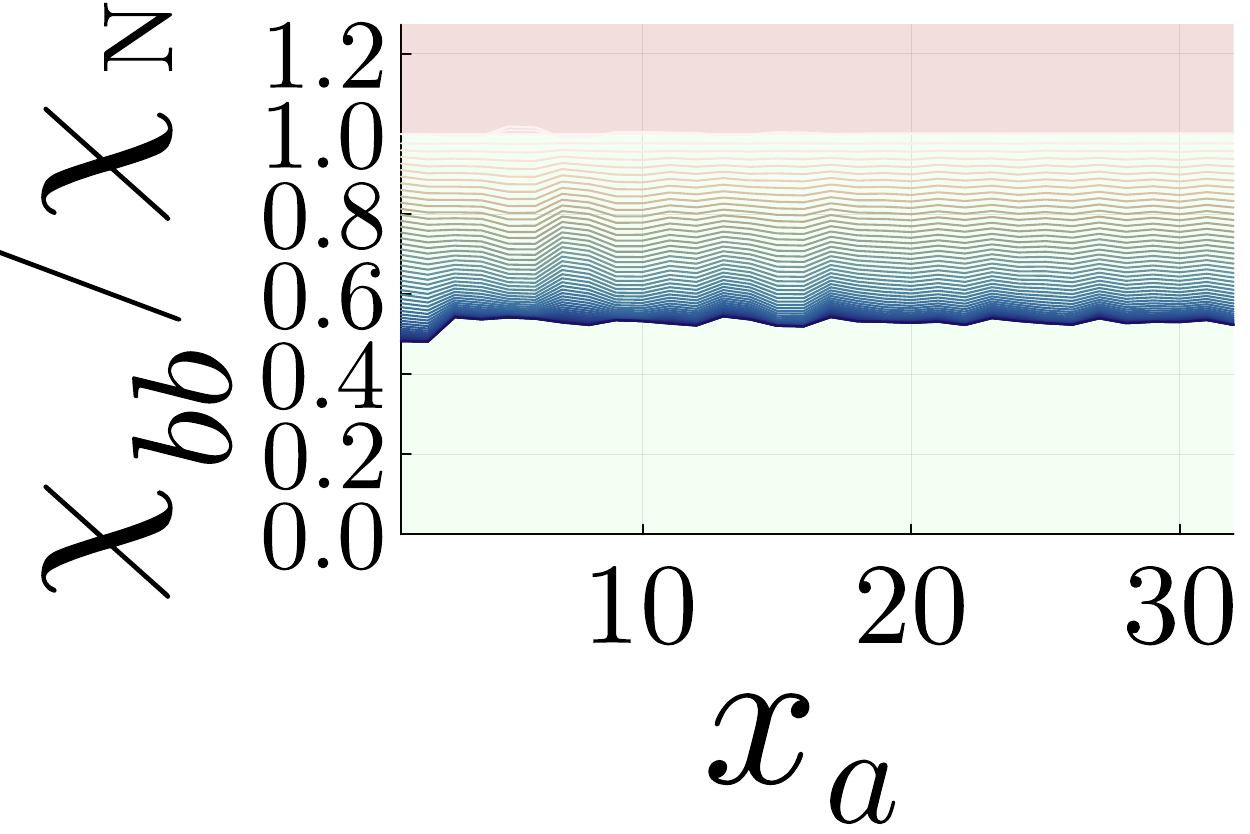}
    \end{minipage}&
    \begin{minipage}[c]{0.30\linewidth}
        \subcaption{}\vspace{-1mm}
        \label{sfig: LSS_site_B3u_b_bb}
          \includegraphics[width=\linewidth]{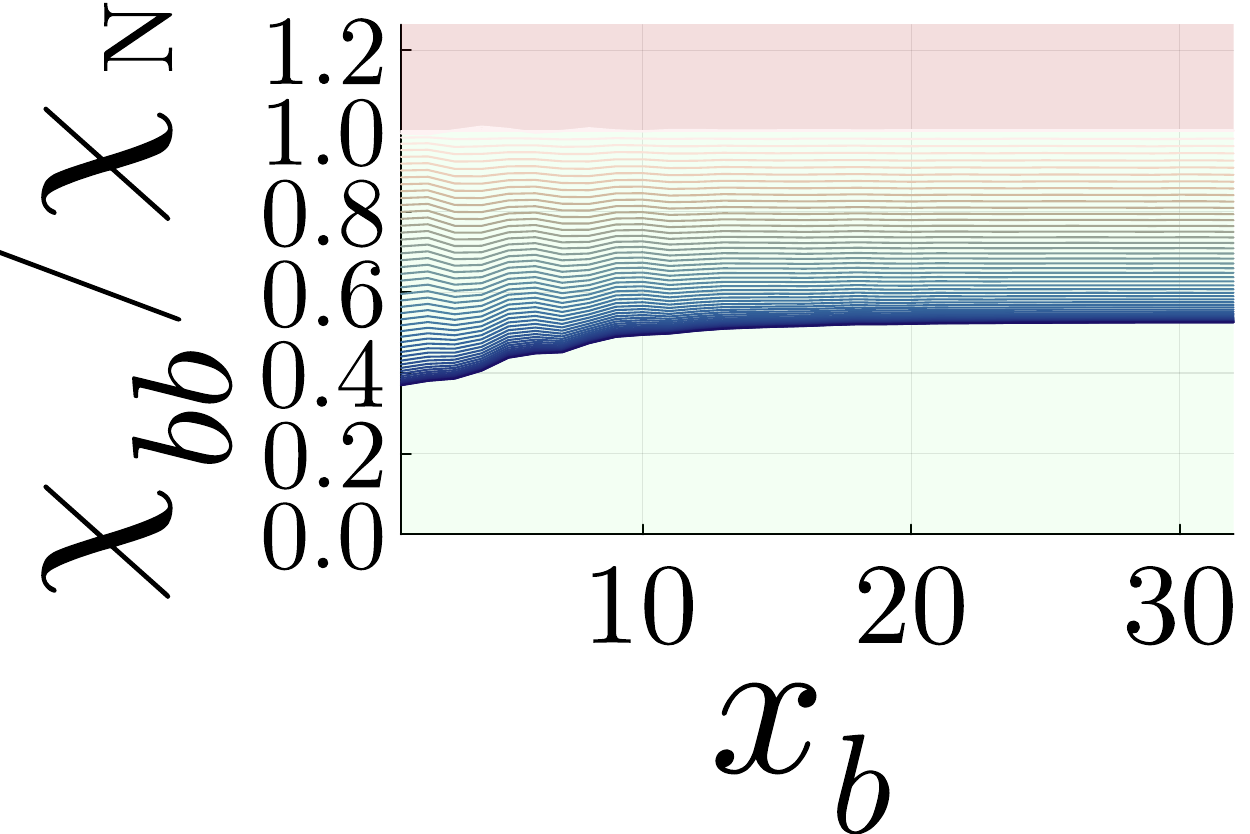}
    \end{minipage}&
    \begin{minipage}[c]{0.30\linewidth}
        \subcaption{}\vspace{-1mm}
        \label{sfig: LSS_site_B3u_c_bb}
          \includegraphics[width=\linewidth]{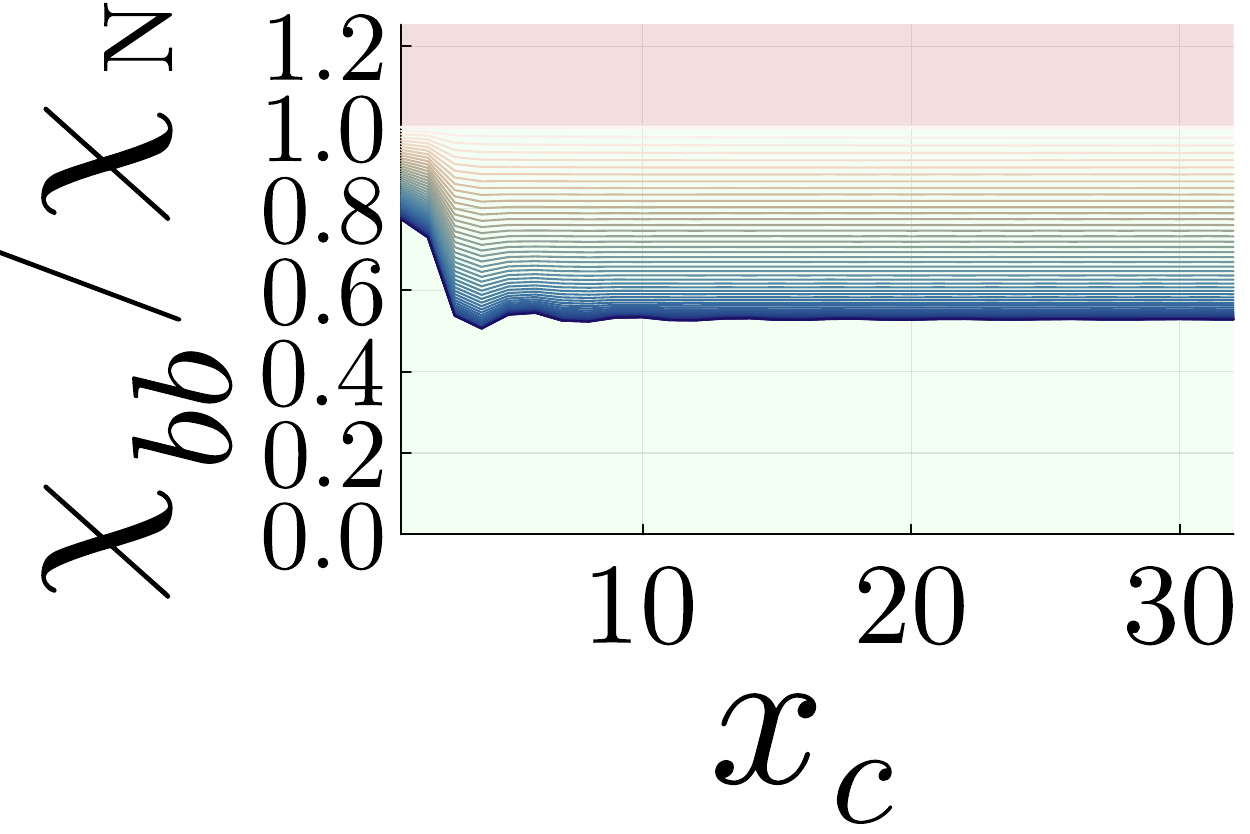}
    \end{minipage}\\
    \begin{minipage}[c]{0.30\linewidth}
        \subcaption{}\vspace{-1mm}
        \label{sfig: LSS_site_B3u_a_cc}
          \includegraphics[width=\linewidth]{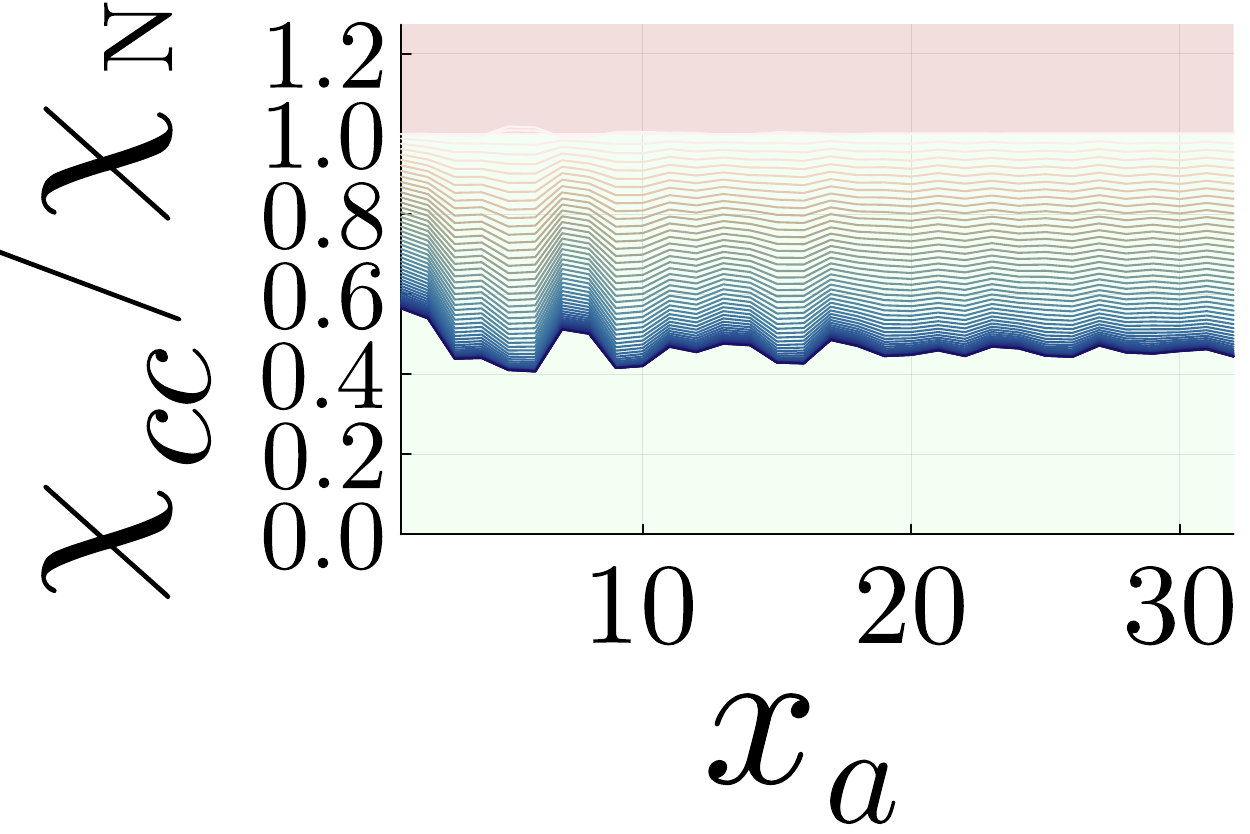}
    \end{minipage}&
    \begin{minipage}[c]{0.30\linewidth}
        \subcaption{}\vspace{-1mm}
        \label{sfig: LSS_site_B3u_b_cc}
          \includegraphics[width=\linewidth]{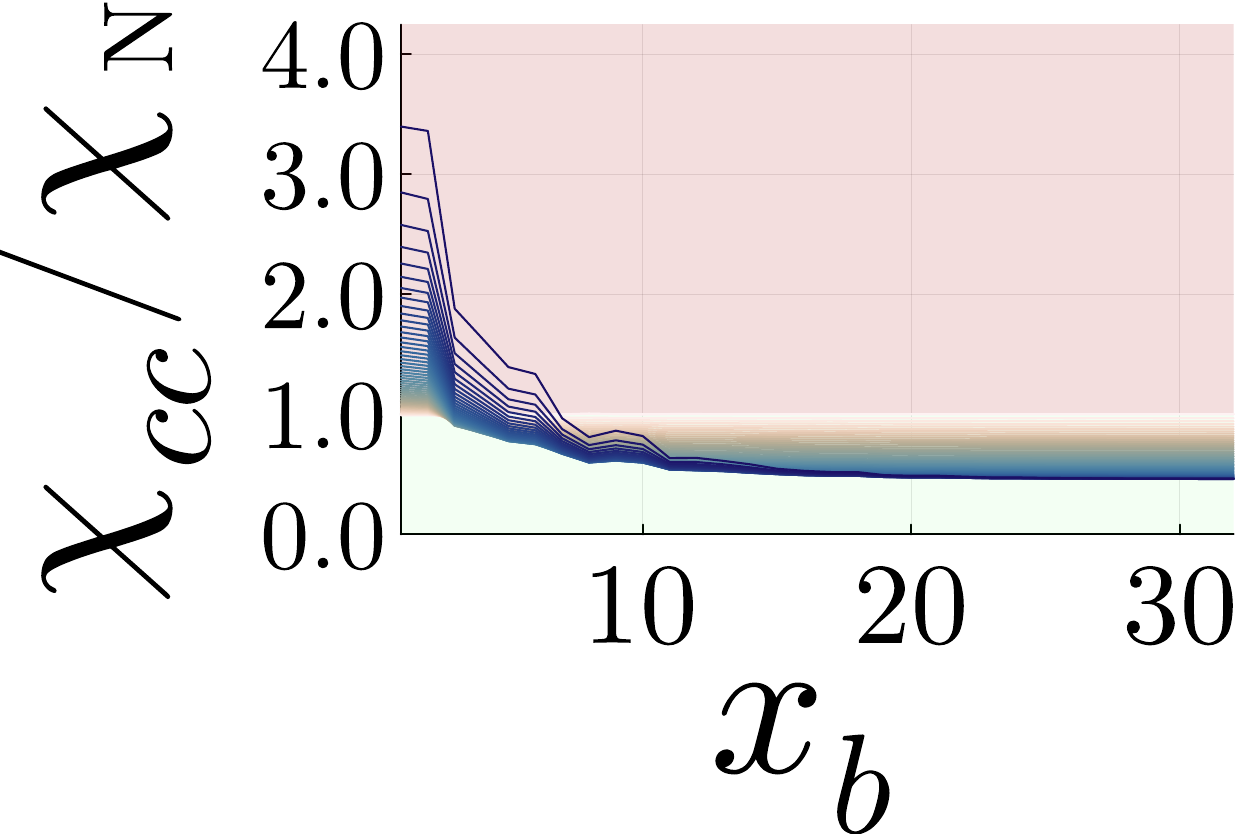}
    \end{minipage}&
    \begin{minipage}[c]{0.30\linewidth}
        \subcaption{}\vspace{-1mm}
        \label{sfig: LSS_site_B3u_c_cc}
          \includegraphics[width=\linewidth]{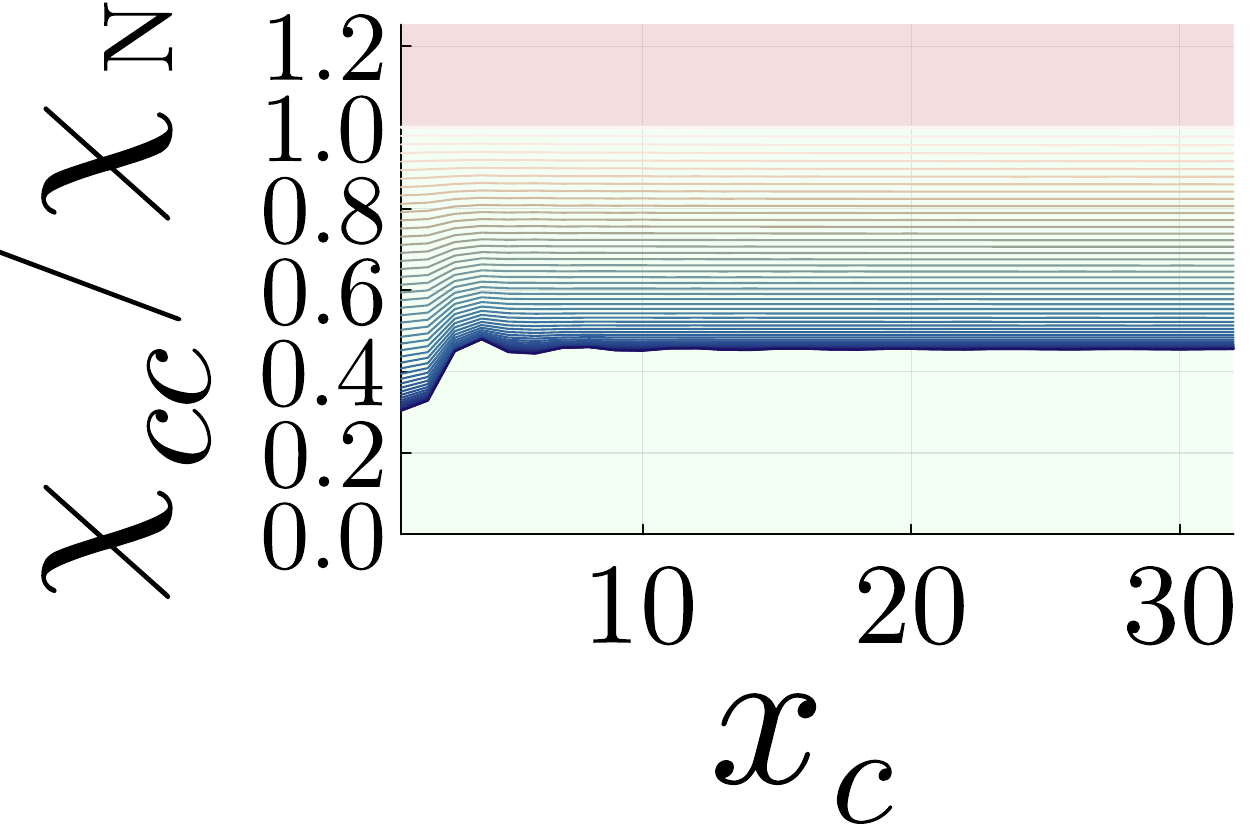}
    \end{minipage}\\
    \multicolumn{3}{c}{
      \includegraphics[width=0.9\linewidth]{cbar-spin-sus-site.pdf}
    }\\
  \end{tabular}
  \caption{
    Site dependence of the LSS at each temperature for $B_{3u}$ state.
    $x_{⟂}=1$ for the surface and $x_{⟂}=32$ for the bulk are concerned, respectively.
    \cref{sfig: LSS_site_B3u_b_cc} shows the anomalous enhancement of the surface LSS due to MSS.
    $χ_{aa}$ has a constant value regardless of temperature because the directions of the magnetic field and $\vb*{d}$-vector (\cref{eq:d-vector_IR_B3u}) are orthogonal.
  }
  \label{fig: LSS_site_B3u}
\end{figure}

Lastly, we consider the $B_{3u}$ state discussed in \cref{ssec:sdos_B3u}.
The results of surface and bulk LSS calculations are summarized in the fourth row of \cref{fig: LSS_total}.
\Cref{fig: LSS_site_B3u} also plots the LSS as a function of the distance from the surface for each surface orientation.
The LSS of the bulk, as shown in \cref{sfig: LSS_total_B3u_bulk}, decreases with decreasing temperature in $b$ and $c$ directions, which is consistent with the previous work~\cite{hiranuma_2021}.
For all results, $χ_{aa}$ has a value of $χ_{aa}(T)/χ_\mathrm{N}=1$ for all temperatures because the magnetic field and $\vb*{d}$-vector [\cref{eq:d-vector_IR_B3u}] are orthogonal.

In the case of $B_{3u}$ pairing state, the enhancement of LSS occurs only in $χ_{cc}$ in the (010) plane, as shown in \cref{sfig: LSS_total_B3u_b} and \cref{sfig: LSS_site_B3u_b_cc}.
As seen from \cref{fig: sdos_B3u}, the flat Fermi arcs emerge only on the (010) plane. 
Therefore, the Ising anisotropy of the surface LSS also occurs only in the case of the (010) plane. 
The spin susceptibility is enhanced when a magnetic field is applied in the $c$ direction and the chiral operator $Γ_{\mathcal{M}_{ab}}$ is broken.

\section{Summary and discussion}
\label{sec: summary_discussion}

We have investigated topologically protected MSSs and their magnetic response in the expected spin-triplet superconducting state of \ce{UTe2}. 
Our purpose is to identify the pairing symmetry of the superconductor \ce{UTe2} and to detect the signature of the MSSs.
To achieve this, we have examined how the Majorana surface state hosted by the spin-triplet superconducting state of \ce{UTe2} contributes to the LDOS and the LSS.

First, we summarize the LDOS results. 
We have demonstrated the existence of the MSSs from the numerical calculation of the surface $\bm{k}_{∥}$-resolved LDOS. 
The Majorana cone appears in the $A_u$ pairing state, while the zero-energy Fermi arcs are formed in the surface Brillouin zone in the certain surface configuration of the other irreducible representation ($B_{1u}$, $B_{2u}$, and $B_{3u}$). 
The difference in the dispersion of the MSS is reflected in the in-gap structure of the LDOS, which is shown in \cref{fig: sdos_Au,fig: sdos_B1u,fig: sdos_B2u,fig: sdos_B3u}.
The V-shaped surface LDOS reflects the cone-like dispersion of the Majorana surface state in the $A_{u}$ pairing state, while the zero-energy peak appears in the $B_{1u}, B_{2u}, B_{3u}$ pairing states as a reflection of the flat Fermi arcs. 
However, the difference in the in-gap structure of the LDOS is not sufficient to detect a strong signal of the MSS and to identify the gap symmetry.

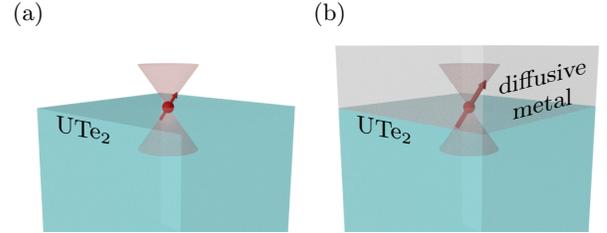
\begin{figure}[htbp]
	\centering
  \begin{minipage}[t]{0.45\linewidth}
    \subcaption{}\vspace{-5mm}
    \label{sfig: majorana_wo_DM}
    \begin{tikzpicture}
      \node[anchor=center,inner sep=0] (image) at (0,0) {\includegraphics[width=\linewidth]{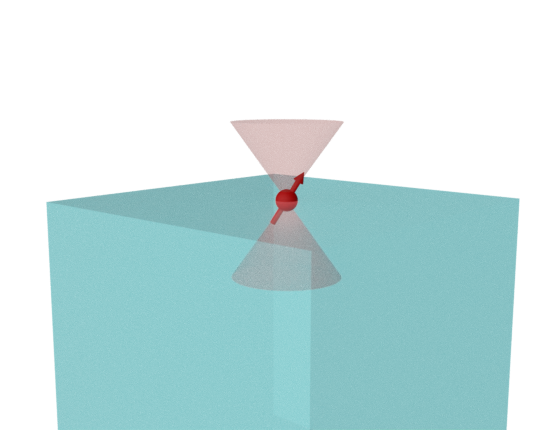}};
      \begin{scope}[every node/.append style={yslant=-0.17}]
        \node [anchor=center] (note) at (-1.0,-0.2) {\ce{UTe2}};
      \end{scope}
    \end{tikzpicture}%
  \end{minipage}
  \begin{minipage}[t]{0.45\linewidth}
    \subcaption{}\vspace{-5mm}
    \label{sfig: majorana_w_DM}
    \begin{tikzpicture}
      \node[anchor=center,inner sep=0] (image) at (0,0) {\includegraphics[width=\linewidth]{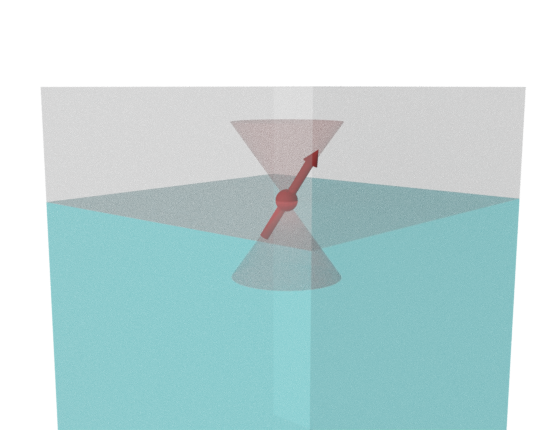}};
      \begin{scope}[every node/.append style={yslant=-0.17}]
        \node [anchor=center] (note) at (-1.0,-0.2) {\ce{UTe2}};
      \end{scope}
      \begin{scope}[every node/.append style={yslant=0.25}]
      \node [anchor=center] (note) at (1.1,0.3) {\begin{tabular}{c}diffusive\\ metal\end{tabular}};
      \end{scope}
    \end{tikzpicture}%
  \end{minipage}
  \caption{
    Schematic illustration of detecting MSS.
    The Ising anisotropic magnetic response caused by the Majorana state is expected to increase when \ce{UTe2} is attached to a diffusive metal junction.
  }
  \label{fig: majorana_DM}
\end{figure}

To capture a clear signature of the gap symmetry and Majorana surface states, we have calculated the LSS at each surface direction and each field direction for four irreducible representations. 
The numerical results are shown in \cref{fig: LSS_total,fig: LSS_site_Au,fig: LSS_site_B1u,fig: LSS_site_B2u,fig: LSS_site_B3u}.
The results show a strong LSS enhancement in the direction that the chiral symmetry protecting the MSS is broken by the applied magnetic field. 
The existence of the MSS and anisotropy of the LSS are summarized in \cref{tb:surface_SS}.
These results indicate that MSSs possess Ising-like anisotropic spin and exhibit paramagnetic response. 
In addition, the anisotropy of the surface LSS reflects the chiral symmetry in each gap state. 
Therefore, measuring the surface LSS allows for identifying the MSS contribution and determining the gap symmetry.

Lastly, we briefly comment on the correspondence between our numerical results and experiments. 
Our results in the LSS suggest the possibility of detecting MSS via NMR measurements. 
However, we have focused on the magnetic response on the surface and it is still unclear whether the signal of the surface states can be detected by the NMR measurement in the bulk superconductor.
A possible approach to amplify the signal of the surface effect is to coat the surface of the superconductor with a diffusive metal as shown in \cref{fig: majorana_DM}. 
It is known that the surface ABS, including the MSS, behaves as odd-frequency spin-triplet $s$-wave pairs that can penetrate and survive in the diffusive metal 
generating ZEP of LDOS
which is called anomalous proximity 
effect~\cite{Proximityp,Proximityp2,tanaka_2007,Higashitani2009,Tanaka12}. 
There are experimental reports in \ce{CoSi2}/\ce{TiSi2} heterostructure which are consistent with ZBCP and ZEP of LDOS induced by odd-frequency pairing~\cite{Lin2021,Lin2023}. 
Since the induced odd-frequency $s$-wave spin-triplet pairs  are robust against impurity scattering~\cite{Proximityp,tanaka_2007,Higashitani2009}, they  
can enhance the magnetic response in the diffusive metal~\cite{nagato_2018}. 
We would like to note that in the superfluid \ce{^3He}-B immersed in aerogels~\cite{Higashitani2009}, the enhancement of the spin susceptibility was observed through NMR measurements~\cite{scott_2023}. 
The gap symmetry of the \ce{^3He}-B is similar to that of $A_{u}$ state and the Majorana cone exists on the surface. 
As the aerogel is regarded as a network of nonmagnetic impurities with a finite diameter, the \ce{^3He}-B immersed in aerogels increases the volume of the surface region and enhances the signal of the surface spin susceptibility. 
Also, it is interesting to study the rough surface effect in spin-triplet $p$-wave superconductor~\cite{nagato_1998}.  
Since the even-frequency $p$-wave pairing which suppresses ZEP of LDOS is suppressed near the surface while the odd-frequency $s$-wave pairing is enhanced,  the zero energy surface LDOS can be enhanced~\cite{Nagai2018,Bakurskiy2014}. 
Hence, it would be worthwhile to study the magnetic response of the superconductor \ce{UTe2} attached to a diffusive metal and that with a rough surface in the future. 
\rocomment{
  Moreover, the analysis of DOS and LSS on the cleavage surface (011) is also an important issue.
  Due to the existence of surface-localized states \cite{tei_2023}, it is highly likely that more enhancement of LSS will occur as in other surfaces.
}

\acknowledgments
This work was supported by 
JST CREST Grant No. JPMJCR19T5, Japan, 
and JSPS KAKENHI (
    Grants No. JP21H01039, 
    No. JP22H01221, 
    No. JP23K17668, 
    No. JP24K00556, 
    No. JP24K00583, 
    and No. JP24KJ1621 
).  

\appendix
\begin{appendix}

\section{Andreev bound state under the magnetic field}
\label{app:sdos_zeeman}

To examine the response of MSS to magnetic fields, we calculate the spectrum of the ABS state when an external magnetic field is applied.
In the calculation, we add a Zeeman term in normal Hamiltonian \cref{eq:normal_hamiltonian}
\begin{align}
  H_\mathrm{N}^\mathrm{Z} = H_\mathrm{N} + \mu_\mathrm{B}\bm{H^\mathrm{ext}}⋅\bm{σ}
\end{align}
where $\bm{H^\mathrm{ext}} = \qty(H^\mathrm{ext}_{a}, H^\mathrm{ext}_{b}, H^\mathrm{ext}_{c})$ is the external magnetic field. 
We use the recursive Green's function method to calculate the surface LDOS and evaluate the behavior of MSS for each magnetic field direction.
In this calculation, we set the $\abs{\mu_\mathrm{B}H^\mathrm{ext}} = 0.2Δ₀$.

\begin{figure}[htbp]
  \begin{tabular}{ccc}
    {\begin{tabular}{c}\fbox{\scalebox{1.5}{$A_{u}$}}\\ (010)\end{tabular}} & $H^\mathrm{ext}_{a}≠ 0$ & $H^\mathrm{ext}_{b}≠ 0$\\
    \rotatebox[origin=c]{90}{$k_{c}=0$}&
    \begin{minipage}[c]{0.40\linewidth}
        \subcaption{}\vspace{-1mm}
          \includegraphics[width=\linewidth]{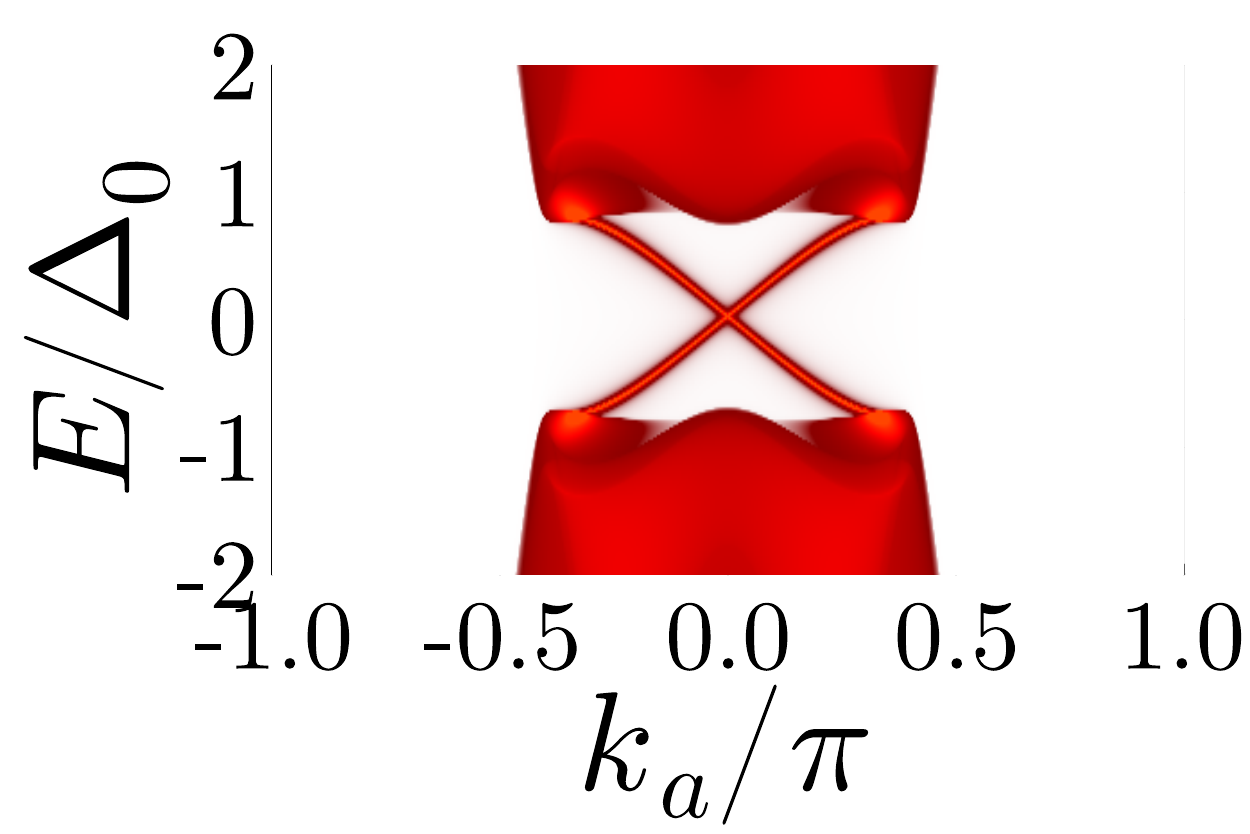}
    \end{minipage}&
    \begin{minipage}[c]{0.40\linewidth}
        \subcaption{}\vspace{-1mm}
          \includegraphics[width=\linewidth]{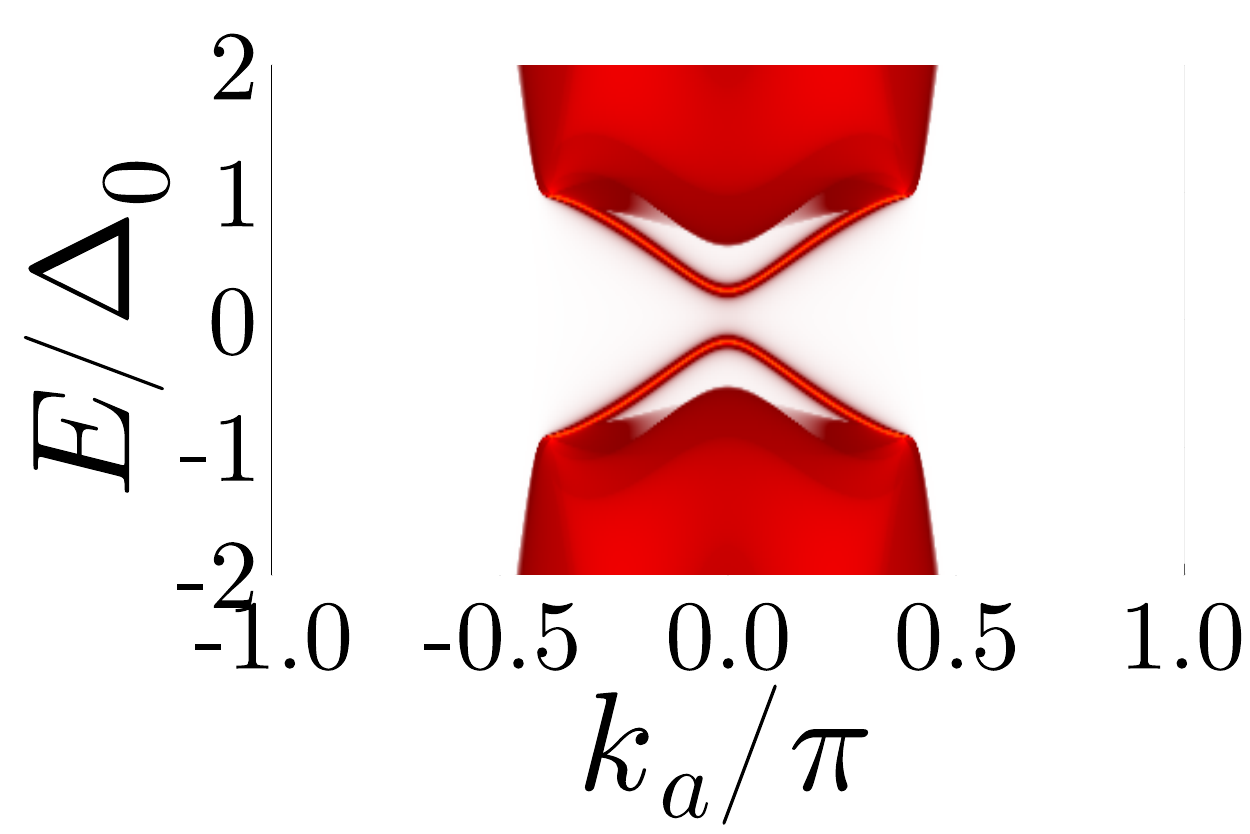}
    \end{minipage}\\
    \rotatebox[origin=c]{90}{$k_{a}=0$}&
    \begin{minipage}[c]{0.40\linewidth}
        \subcaption{}\vspace{-1mm}
          \includegraphics[width=\linewidth]{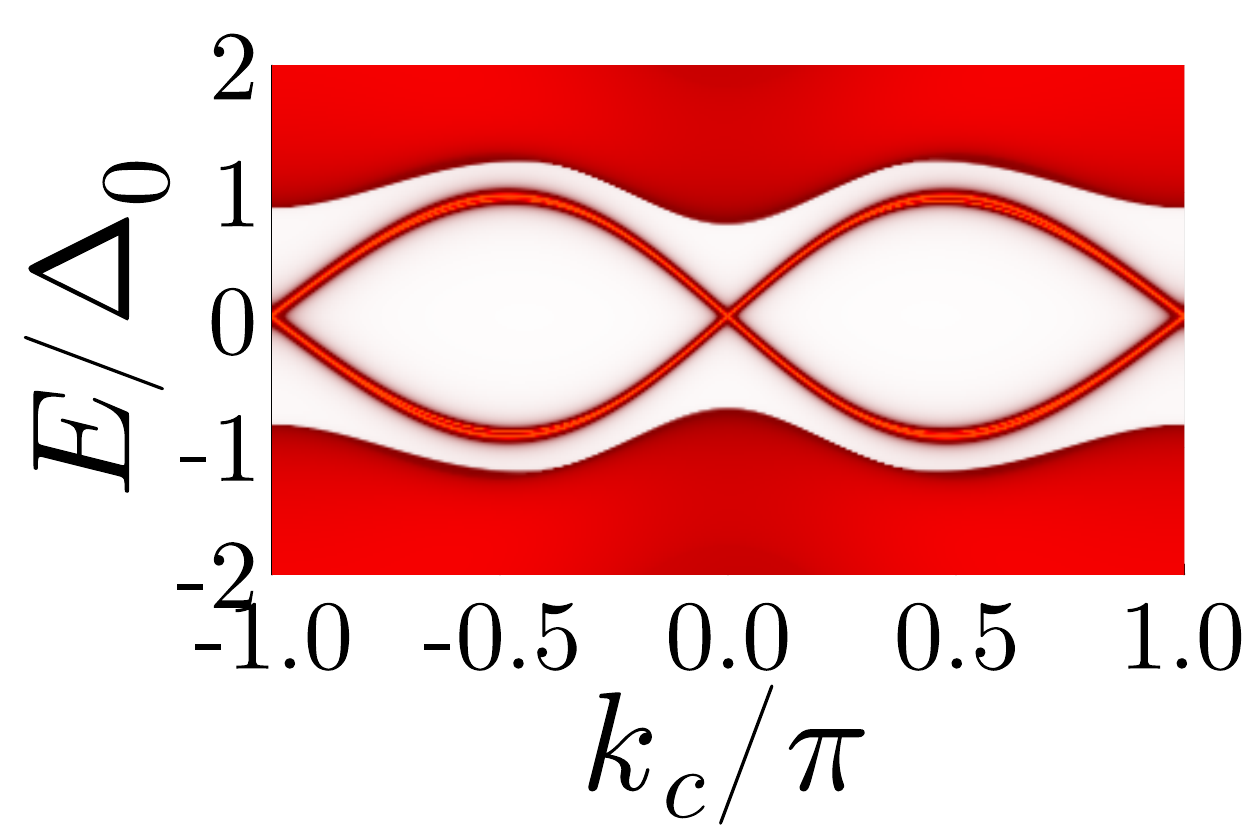}
    \end{minipage}&
    \begin{minipage}[c]{0.40\linewidth}
        \subcaption{}\vspace{-1mm}
          \includegraphics[width=\linewidth]{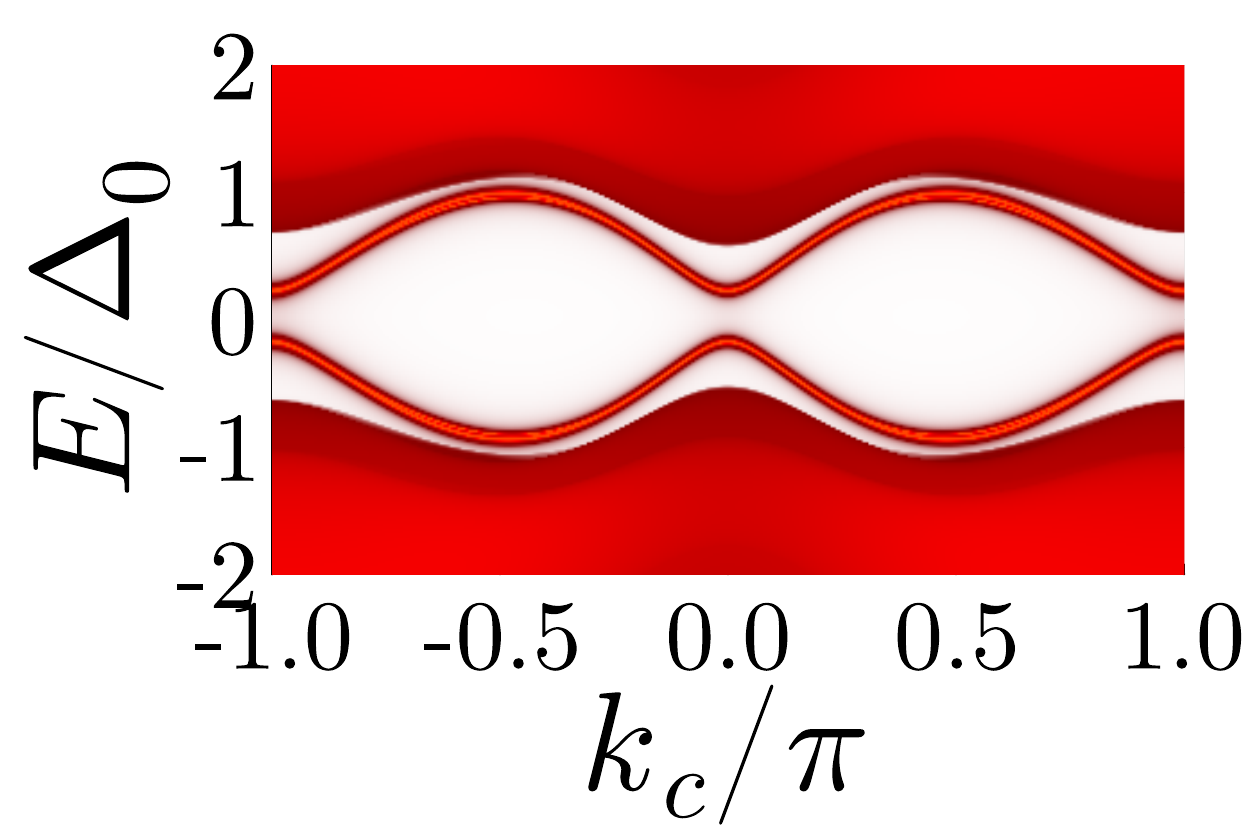}
    \end{minipage}\\
     & \multicolumn{2}{c}{
        \includegraphics[width=0.7\linewidth]{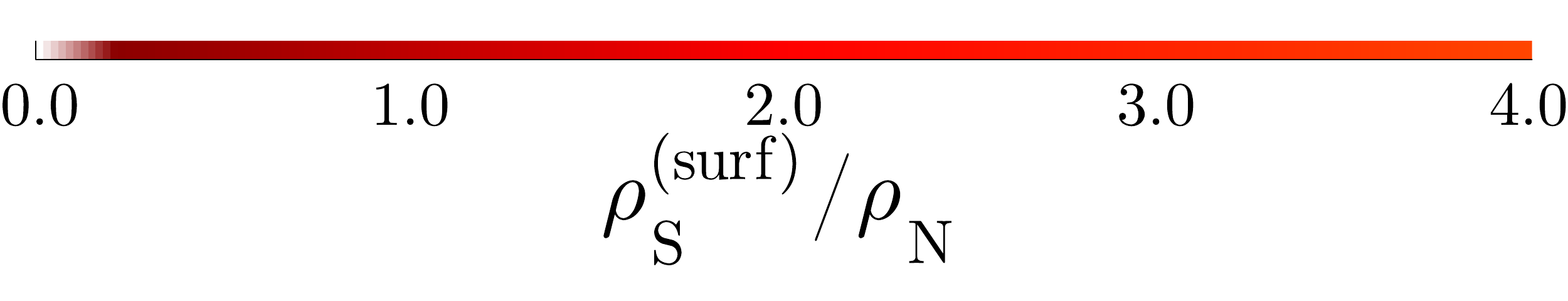}
     }
  \end{tabular}
  \caption{
    Surface LDOS for the $A_{u}$ pairing state on the (010) surface when an external magnetic field is applied along the $a$ and $b$ directions.
  }
  \label{fig: sdos_Au_Zeeman}
\end{figure}

\begin{figure}[htbp]
  \begin{tabular}{ccc}
    {\begin{tabular}{c}\fbox{\scalebox{1.5}{$B_{3u}$}}\\ (010)\end{tabular}} & $H^\mathrm{ext}_{a}≠ 0$ & $H^\mathrm{ext}_{c}≠ 0$\\
    \rotatebox[origin=c]{90}{$k_{c}=0$}&
    \begin{minipage}[c]{0.40\linewidth}
        \subcaption{}\vspace{-1mm}
          \includegraphics[width=\linewidth]{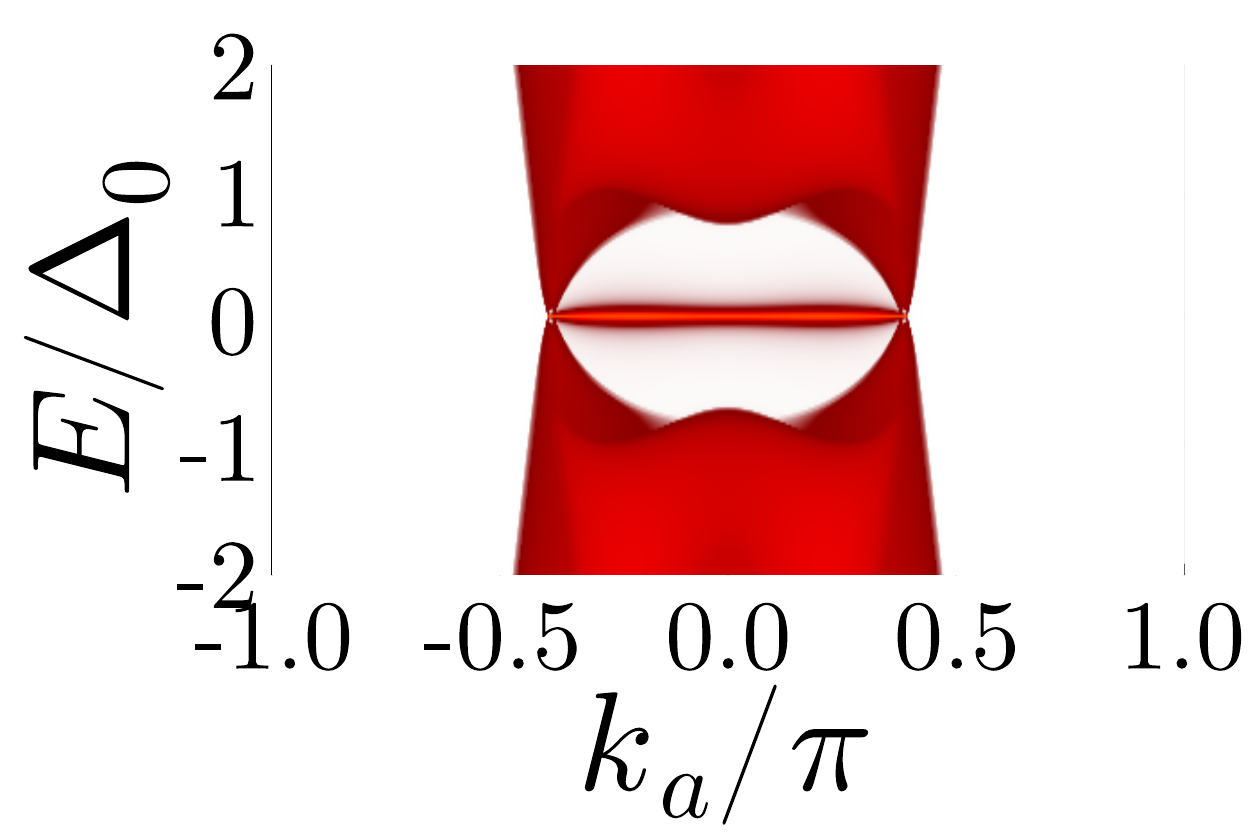}
    \end{minipage}&
    \begin{minipage}[c]{0.40\linewidth}
        \subcaption{}\vspace{-1mm}
          \includegraphics[width=\linewidth]{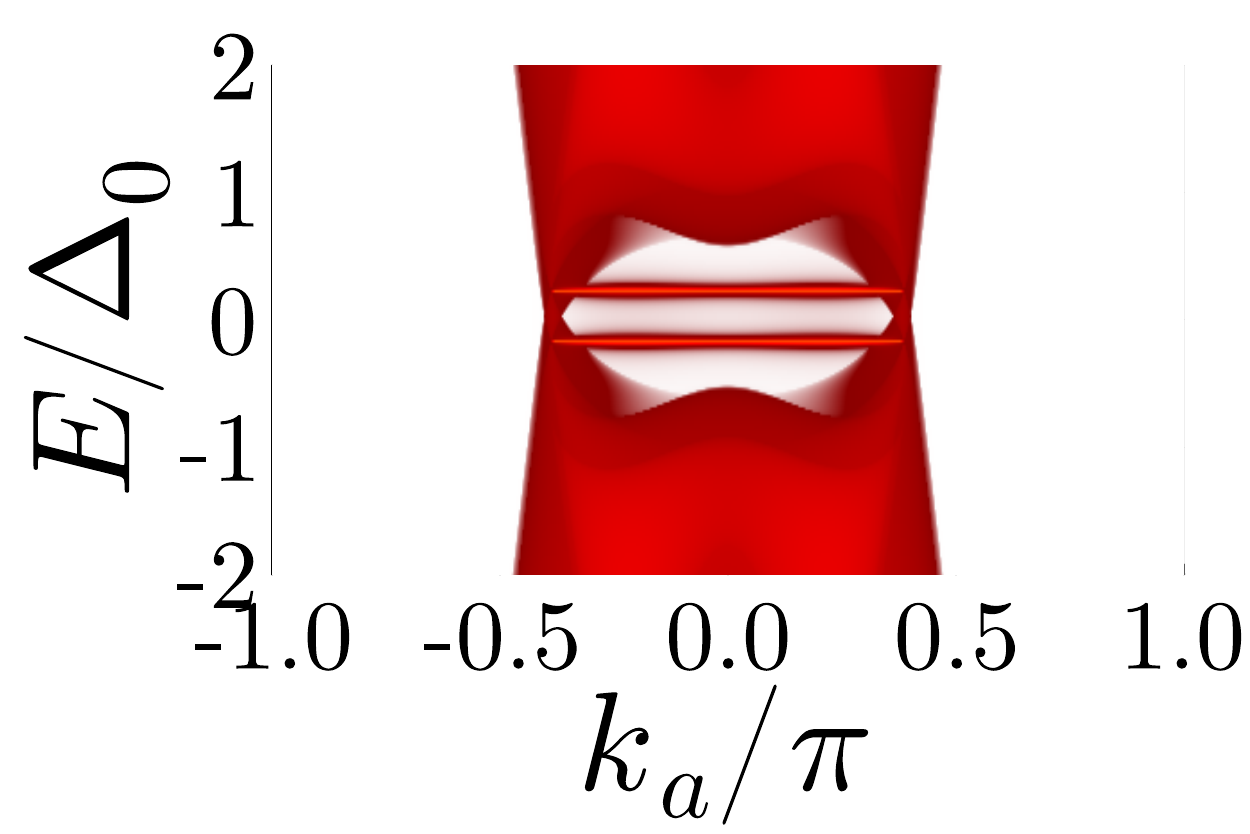}
    \end{minipage}\\
    \rotatebox[origin=c]{90}{$k_{a}=0$}&
    \begin{minipage}[c]{0.40\linewidth}
        \subcaption{}\vspace{-1mm}
          \includegraphics[width=\linewidth]{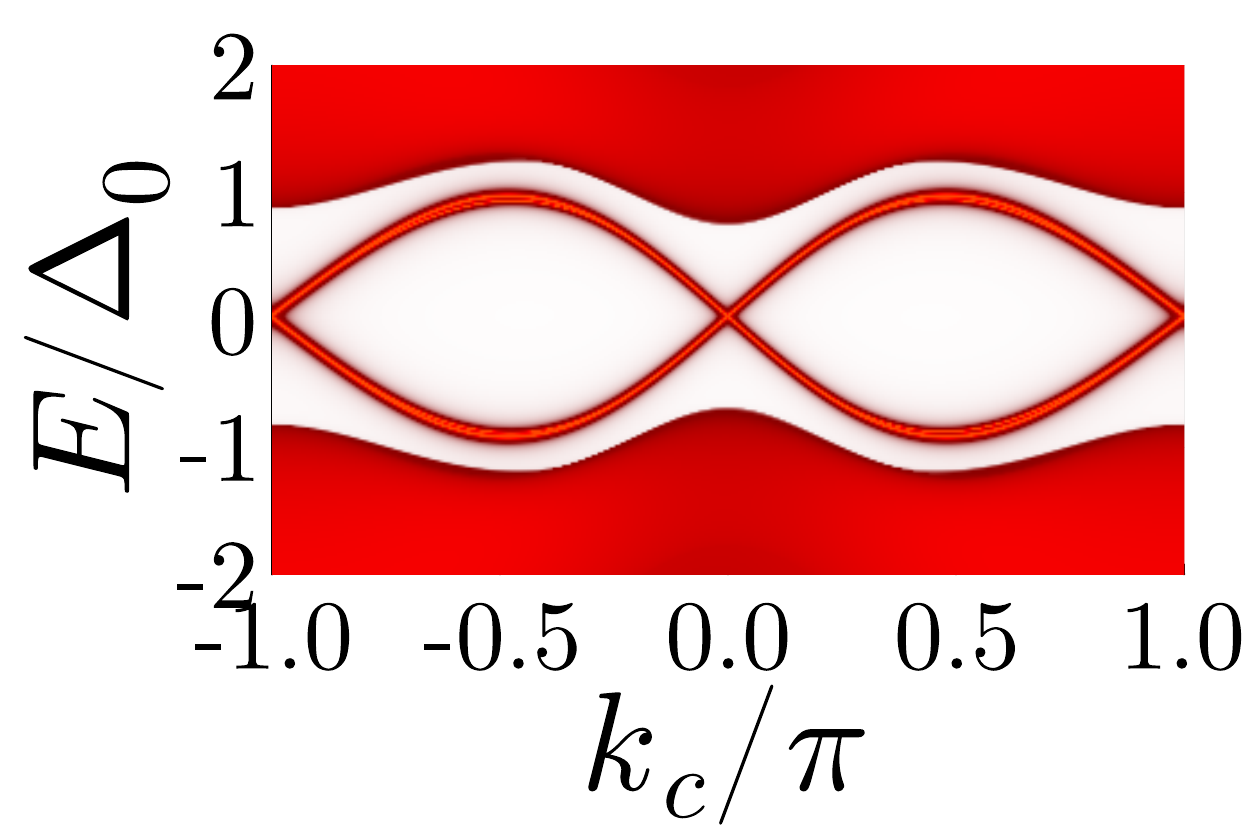}
    \end{minipage}&
    \begin{minipage}[c]{0.40\linewidth}
        \subcaption{}\vspace{-1mm}
          \includegraphics[width=\linewidth]{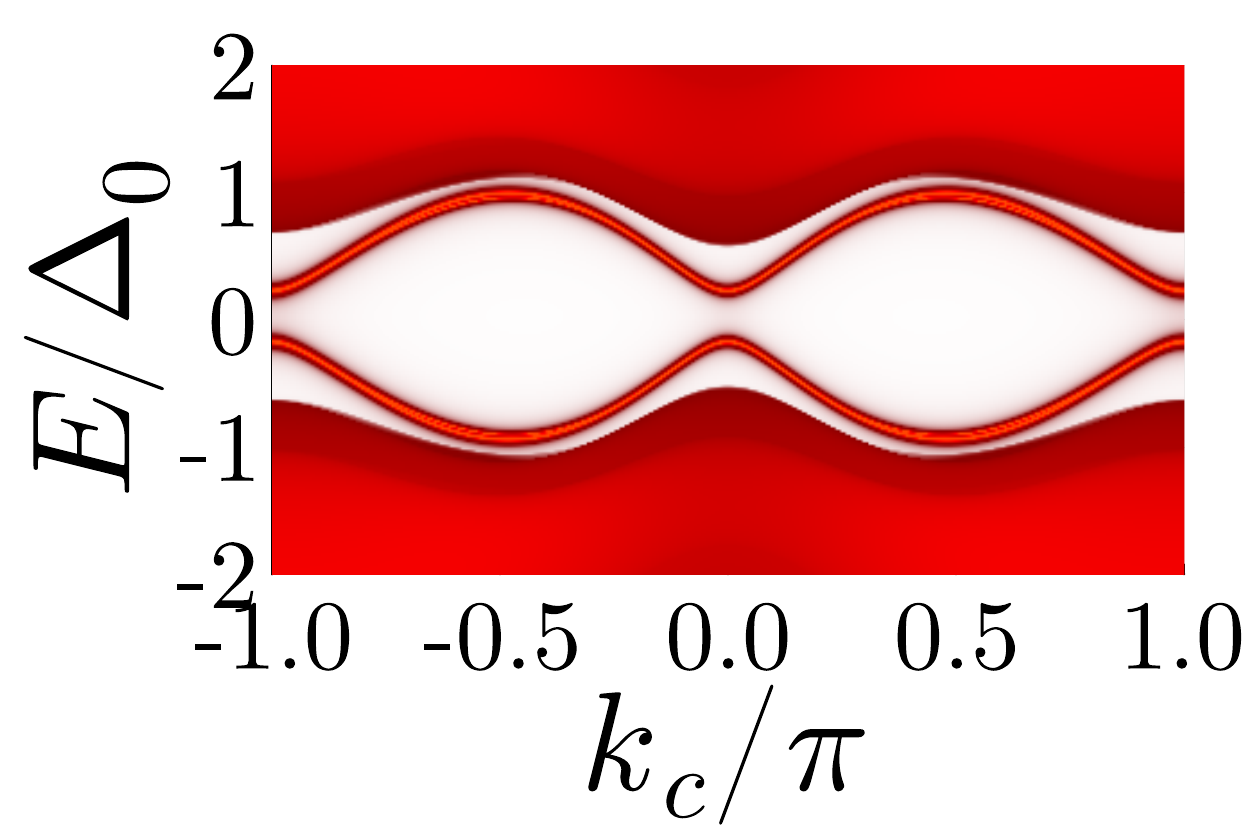}
    \end{minipage}\\
     & \multicolumn{2}{c}{
        \includegraphics[width=0.7\linewidth]{cbar-Zeeman.pdf}
     }
  \end{tabular}
  \caption{
    Surface LDOS for the $B_{3u}$ pairing state on the (010) surface when an external magnetic field is applied along the $a$ and $c$ directions.
  }
  \label{fig: sdos_B3u_Zeeman}
\end{figure}

Here we show the case of the $A_{u}$ and $B_{3u}$ pairing state on the (010) surface.
\Cref{fig: sdos_Au_Zeeman,fig: sdos_B3u_Zeeman} plot the surface LDOS in the $A_{u}$ and $B_{3u}$ pairing state on the (010) surface for each direction of the applied magnetic field, respectively.
In the case of $\bm{H^\mathrm{ext}}\parallel\bm{\hat{a}}$, as the direction of the magnetic field does not break chiral symmetry, the topological number is well-defined and the MSS remains at zero energy.
In the $\bm{H^\mathrm{ext}}\parallel\bm{\hat{c}} \; (\bm{\hat{b}})$ case for $A_{u}$ ($B_{3u}$) pairing state, the zero-energy state is maintained for the same reason.
On the other hand, in the $\bm{H^\mathrm{ext}}\parallel\bm{\hat{b}} \; (\bm{\hat{c}})$ case for $A_{u}$ ($B_{3u}$) pairing state, the direction of chiral symmetry is broken and the MSS can no longer maintain zero energy, creating an energy gap. 
The orientation of the magnetic field that opens a gap in the MSS coincides with the direction of the Ising-like anisotropic magnetic response of the MSS.

\section{Calculation method of surface Green's function}
\label{app:sdos}

We consider the surface state of \ce{UTe2} in the superconducting state.
To calculate the Green's function at the surface, we use the recursive Green's function method~\cite{umerski_1997,takagi_2020}.
Surface states are calculated for the (100), (010), and (001) surfaces.
First, we consider the finite Green's function, which is defined as follows:
\begin{align}
  G(z,\bm{k}_\parallel) = \frac{1}{z\bm{I} - H_\mathrm{BdG}^\mathrm{lattice}(\bm{k}_{∥})}
  \label{eq: Green_function_native}
\end{align}
where $\bm{I}$ is an identity matrix, $z$ is the complex number corresponding to energy, and $\bm{k}_{∥}$ is the wave number parallel to the surface (perpendicular to the open axis direction).
$H_\mathrm{BdG}^\mathrm{lattice}$ is the lattice Hamiltonian \cref{eq:lattice_hamiltonian} which is described in matrix form as follows:
\begin{align}
  H_\mathrm{BdG}^\mathrm{lattice}(\bm{k}_{∥})
  &= \mqty(
    h^{∥}_{\bm{k}_{∥}}    & t^{⟂}_{\bm{k}_{∥}}     &       & & \bm{O} \\
    \qty(t^{⟂}_{\bm{k}_{∥}})^{†} & h^{∥}_{\bm{k}_{∥}}   & t^{⟂}_{\bm{k}_{∥}} & &  \\
              & \qty(t^{⟂}_{\bm{k}_{∥}})^{†} & ⋱ & ⋱ & \\
              &           & ⋱ & ⋱ & t^{⟂}_{\bm{k}_{∥}} \\
    \bm{O}         &           &   & \qty(t^{⟂}_{\bm{k}_{∥}})^{†} & h^{∥}_{\bm{k}_{∥}} 
  ),
\end{align}
$h^{∥}_{\bm{k}_{∥}}$ is the on-site Hamiltonian and $t^{⟂}_{\bm{k}_{∥}}$ is the neighbor hopping. 
In considering the next-nearest-neighbor or longer hopping terms, we redefine these as
\begin{align}
  \begin{cases}
  h^{∥}_{\bm{k}_{∥}}
  &= \mqty(
    H^{∥}(\bm{k}_{∥})  & T^\mathrm{NN}(\bm{k}_{∥}) \\
    \qty(T^\mathrm{NN}(\bm{k}_{∥}))^{†} & H^{∥}(\bm{k}_{∥}) 
  ) \\
  t^{⟂}_{\bm{k}_{∥}}
  &= \mqty(
    T^\mathrm{NNN}(\bm{k}_{∥}) & \bm{O} \\
    T^\mathrm{NN}(\bm{k}_{∥}) & T^\mathrm{NNN}(\bm{k}_{∥}) 
  )
  \end{cases},
\end{align}
where $\bm{O}$ is a zero matrix.
Similarly, we note the matrix components of the Green's function \cref{eq: Green_function_native} as follows:
\begin{align}
  G
  &= \mqty(
    G_{1,1} & G_{1,2} & \cdots & G_{1,N} \\
    G_{2,1} & G_{2,2} &        & ⋮ \\
    ⋮      &         & ⋱      & ⋮ \\
    G_{N,1} & \cdots  & \cdots & G_{N,N} 
  ),
\end{align}
where the subscripts of $G_{i,j}$ denote the numbers of lattice site.
Two neighboring on-site Green's functions recursively follow the following recurrence formula
\begin{align}
  G_{i+1,i+1}(z,\bm{k}_{∥}) = \frac{1}{z\bm{I} - h^{∥}_{\bm{k}_{∥}} - \qty(t^{⟂}_{\bm{k}_{∥}})^{†}G_{i,i}(z,\bm{k}_{∥})t^{⟂}_{\bm{k}_{∥}}}.
  \label{eq:rec_green_func}
\end{align}
Hereafter, the notation of $\bm{k}_{∥}$ and $z$-dependence of $G$ is omitted.

Next, this recurrence formula \cref{eq:rec_green_func} is applied to a semi-infinite system.
In preparation to obtain the semi-infinite Green's function, we define the left-hand M\"{o}bius transformation as
\begin{align}
  \qty[\mqty(A_{11} & A_{12} \\ A_{21} & A_{22})] ∘ M ≡ \qty(A_{11}M + A_{12})\qty(A_{21}M + A_{22})^{-1}.
\end{align}
The following coupling law
\begin{align}
  \qty[B] ∘ \qty(\qty[A] ∘ Y) = \qty[BA] ∘ Y
\end{align}
holds for M\"{o}bius transformation.
We express \cref{eq:rec_green_func} by M\"{o}bius transformation as follows:
\begin{align}
  G_{N,N} &= \qty[X] ∘ G_{N-1,N-1},
  \label{eq:rec_green_func_semi-infinite_1}\\
  X &= \mqty(\bm{O} & \qty(t^{⟂}_{\bm{k}_{∥}})^{-1} \\ -\qty(t^{⟂}_{\bm{k}_{∥}})^{†} & \qty(z\bm{I}-h^{∥}_{\bm{k}_{∥}})\qty(t^{⟂}_{\bm{k}_{∥}})^{-1}).
  \label{eq:X}
\end{align}
By repeating this recurrence formula, we reduce \cref{eq:rec_green_func_semi-infinite_1} to the following equation
\begin{align}
  G_{N,N} =& \qty[X] ∘ G_{N-1,N-1}\notag\\
  =& \qty[X] ∘ (\qty[X] ∘ G_{N-2,N-2})\notag\\
  =& \qty[X²] ∘ G_{N-2,N-2}\notag\\
  & \quad ⋮ \notag\\
  =& \qty[X^{N-1}] ∘ G_{1,1}
  \label{eq:rec_green_func_semi-infinite_2}
\end{align}
Then, we diagonalize the non-Hermitian $2M\times 2M$ matrix $X$, where $M$ is the dimension of the matrix $(t^{⟂}_{\bm{k}_{∥}})^{-1}$. 
The eigenvalues of $X$ are $λ₁,λ₂,\cdots, λ_{2M}$ and the order of these values is defined as $\abs{λ₁}<\abs{λ₂}<\cdots<\abs{λ_{2M}}$.
Let $\bm{u}_{i}$ be the right eigenvectors of $X$ and $U$ be the matrix
\begin{align}
  U = (\bm{u}₁, \cdots, \bm{u}_{2M}).
\end{align}
Then, the matrix $X$ can be diagonalized by $U$ as
\begin{align}
  U^{-1}XU = Λ = \mqty(\dmat{Λ₁,Λ₂}).
\end{align}
By using these matrices, \cref{eq:rec_green_func_semi-infinite_2} can be transformed into the following equation
\begin{align}
  G_{N,N} &= \qty[X^{N-1}] ∘ G_{1,1} \notag\\
  &= \qty[UΛ^{N-1}U^{-1}] ∘ G_{1,1} \notag\\
  &= \qty[U]∘\qty(\qty[Λ^{N-1}]∘\qty(\qty[U^{-1}] ∘ G_{1,1})) \notag\\
  &= \qty[U]∘\qty(Λ₁^{N-1} \qty(\qty[U^{-1}] ∘ G_{1,1})\qty(Λ₂^{N-1})^{-1}).
\end{align}

Lastly, since $\abs{λ₁}<\abs{λ₂}<\cdots<\abs{λ_{2M}}$, we can take the limit of $N$ to evaluate the matrix element as follows:
\begin{multline}
  \lim_{N→∞} \qty(Λ₁^{N-1}\qty(\qty[U^{-1}] ∘ G_{1,1})\qty(Λ₂^{N-1})^{-1})_{ij} \\
  = \lim_{N→∞} \qty(\frac{λ_{i}}{λ_{M+j}})^{N-1}\qty(\qty[U^{-1}]∘G_{1,1})_{ij}\\
  = 0
\end{multline}
Therefore, we obtain the semi-infinite Green's function as follows:
\begin{align}
  G_{∞} 
  ≡ \lim_{N→∞}G_{N,N} 
  = \qty[U]∘O 
  = U_{12}\qty(U_{22})^{-1},
\end{align}
where $U_{12}$ and $U_{22}$ is the $M×M$ block matrices of $U$
\begin{align}
  U = \mqty(\xmat*{U}{2}{2}).
\end{align}
In the following numerical calculations, we take $z=E+iδ_{ϵ}$ to evaluate the retarded Green's function in \cref{eq: ARDOS_formula} and we choose a smearing factor $δ_ϵ$ as $δ_ϵ= 5.0\times 10^{-4}Δ₀$.

In our present calculation, the huge numerical error emerges to obtain the inverse of $t^{⟂}_{\bm{k}_{∥}}$ due to considering the long-range hopping term $T^\mathrm{NNN}$.
Therefore, we use the QZ (Schur) decomposition method that does not need to explicitly solve the inverse of $t^{⟂}_{\bm{k}_{∥}}$ in \cref{eq:X}~\cite{miyata_2013_Numerical,miyata_2016_Computing}.
First, we define two matrices as follows:
\begin{align}
  A₁ = \mqty(\dmat[\vb*{O}]{\bm{I},t^{⟂}_{\bm{k}_{∥}}})
  \qc{}
  A₂ = \mqty(\bm{O} & \bm{I} \\ -\qty(t^{⟂}_{\bm{k}_{∥}})^{†} & \qty(z\bm{I}-h^{∥}_{\bm{k}_{∥}}) ),
\end{align}
which can recast \cref{eq:X} into $X=A₂\qty(A₁)^{-1}$.
To diagonalize the matrix $X$, we factorize $A₁$ and $A₂$ into the QZ decomposition form as
\begin{align}
  \begin{cases}
    A₁ =& QR₁Z^{†}\\
    A₂ =& QR₂Z^{†}
  \end{cases},
\end{align}
where $Q$ and $Z$ are unitary, and $R₁$ and $R₂$ are upper triangular.
The above equation immediately shows that $A₂\qty(A₁)^{-1}$ and $R₂\qty(R₁)^{-1}$ are the same eigenvalues.
Furthermore, we denote the right eigenvectors of $R₂\qty(R₁)^{-1}$ as $\bm{v}_{i}$, $\bm{u}_{i}$ and $\bm{v}_{i}$ are connected by the following relational equation
\begin{align}
  \bm{u}_{i} = Q\bm{v}_{i}.
\end{align}
Thus, instead of computing the inverse of $t^{⟂}_{\bm{k}_{∥}}$, we use this method to diagonalize $X$ by finding the inverse of $R₁$.

\end{appendix}

\bibliographystyle{apsrev4-2}
\bibliography{main_v2.bbl}

\end{document}